\documentclass[aps,showpacs]{revtex4}
\usepackage{amsmath}
\usepackage{amssymb}
\usepackage{graphicx}
\usepackage{dcolumn}
\usepackage{bm}

\newcommand{\br}{\mathbf{r}}
\newcommand{\bk}{\mathbf{k}}
\newcommand{\prt}{\partial}
\newcommand{\bu}{\mathbf{u}}

\newcommand{\la}{\lambda}
\newcommand{\K}{\mathrm{K}}
\newcommand{\E}{\mathrm{E}}

\begin{document}

\title{
Two-dimensional supersonic nonlinear Schr\"odinger flow past an extended obstacle}

\author{
G.A. El$^{1}$,  A.M. Kamchatnov$^{2}$, V.V. Khodorovskii$^1$, E.S. Annibale$^3$ and A. Gammal$^3$ \\
$^1$Department of Mathematical Sciences, Loughborough University,\\
Loughborough,  LE11 3TU, UK \\
$^2$ Institute of Spectroscopy, Russian Academy of Sciences,
Troitsk 142190, Moscow Region, Russia \\
$^3$ Instituto de F\'{\i}sica, Universidade de S\~{a}o Paulo,
05315-970, C.P. 66318 S\~{a}o Paulo, Brazil
}

\begin{abstract}
Supersonic flow of a superfluid past a slender  impenetrable macroscopic obstacle
is studied  in the framework of  the two-dimensional defocusing nonlinear
Schr\"odinger (NLS) equation. This problem is of fundamental importance as
a dispersive analogue of the
corresponding classical gas-dynamics problem. Assuming the oncoming flow speed
sufficiently high, we asymptotically reduce the original boundary-value
problem for a steady flow past a slender body to the  one-dimensional
dispersive piston problem described by the nonstationary NLS equation, in which
the role of time is played by the stretched $x$-coordinate and the piston motion
curve is defined by the spatial body profile.
Two steady oblique spatial dispersive shock waves (DSWs) spreading from the
pointed ends of the body are generated in both half-planes. 
These are described analytically by constructing appropriate
exact solutions of the Whitham
modulation equations for the front DSW and by using a generalized
Bohr-Sommerfeld quantization rule for the oblique dark soliton fan in the rear DSW.
We propose an extension of the traditional modulation description of DSWs to include
the linear ``ship wave" pattern forming outside the nonlinear modulation region of
the front DSW. Our analytic results are supported by direct 2D unsteady numerical
simulations and are  relevant to recent experiments
on  Bose-Einstein condensates freely expanding past obstacles.
\end{abstract}


\maketitle

\section{Introduction}

In  compressible  fluid dynamics  there are two
canonical situations  in which  shock waves can be generated.
In the first case the formation of a shock occurs as a result  of breaking of an
evolving  smooth or discontinuous profile of the density (velocity)
and is described by the generalized solutions of the {\it initial-value problems}
for the ideal fluid dynamics equations.
The second type of shock waves occurs in  a supersonic fluid  flow past a body or
as a result of the motion of a piston within a tube filled with a liquid or a gas
(see, e.g., \cite{cf48, ll, wh74}) and is associated with the
{\it boundary-value  problems}.
 In a viscous fluid, the shock wave can be represented as
a narrow region within which strong dissipation processes take place
and the thermodynamic and hydrodynamic parameters of the flow
undergo a sharp change. However, if viscosity is negligibly small
compared with dispersion effects, the shock singularity is
resolved by a nonlinear wave train called a dispersive shock wave (DSW).
A remarkable feature of the DSW
is the generation of solitons at one of its boundaries  so that the
whole structure can often be asymptotically described as a
``soliton train''.

An analytical theory of one-dimensional DSWs  pioneered by
Gurevich and Pitaevskii \cite{gp74}  is based on the
assumption that the oscillatory structure of a DSW can be
asymptotically described by a modulated periodic (or, more
generally, quasi-periodic) solution of the governing dispersive
equation. The slow variations (modulations) of the travelling
periodic wave parameters such as amplitude, wavenumber etc are governed by the
so-called Whitham equations obtained by averaging of dispersive
conservation laws over the period of the travelling wave. Analyzing
the numerically observed structure of the dispersive shock wave, Gurevich
and Pitaevskii proposed a special system of nonlinear free-boundary
conditions for the KdV-Whitham system and obtained a global
self-similar modulation solution for the problem of the decay of an
initial discontinuity for the KdV equation. An analogous problem for
the defocusing nonlinear Schr\"odinger (NLS) equation was formulated and
solved in \cite{gk87,eggk95} (see also a detailed analysis in \cite{kod99,bk06}
where a different approach to the formulation of the
step problem for the Whitham equations was used). The modulation
solutions describing more general cases of breaking of  monotone and
non-monotone initial profiles were obtained in
\cite{gke92,kke92,tian93} (KdV equation) and in
\cite{ek95,kku02} (defocusing NLS equation) using Tsarev's generalized
hodograph transform method \cite{ts85}.

The modulation theory of one-dimensional unsteady expanding DSWs proved to be very effective  in different physical contexts ranging from shallow-water waves \cite{egkJFM}
to fibre optics \cite{kod99} to Bose-Einstein condensates (BECs) \cite{kgk04,ha06}.  In particular, it was successfully used in \cite{barrier09} for the analytical description of the generation of dark solitons in quasi-1D transcritical BEC flows through wide penetrable potential barriers observed recently in the experiment \cite{ea07}.

The study of two-dimensional steady DSWs occurring in the supersonic dispersive flows
past bodies was initiated in \cite{gkke} where the stationary 2D system of the governing collisionless plasma equations was asymptotically reduced to the 1D KdV equation along the
linear characteristics (Mach lines) with the stretched transverse coordinate playing the role of time (see also \cite{karpman}) and then appropriate modulation solutions were constructed and  interpreted in terms of the original steady 2D problem.

Whilst an asymptotic description of supersonic dispersive flow past body in the framework
of the weakly nonlinear KdV dynamics
captures a number of  essential  features of the wave patterns arising in the flow,
it may fail to provide an
adequate description of the waves of sufficiently large amplitude.
A different  approximation not involving small-amplitude expansions, but instead,
using the expansions in inverse Mach number,  was
proposed in \cite{ekt04} in the context of collisionless plasma dynamics.
In \cite{ekt04} the problem of the supersonic dispersive flow past slender
body was reduced to the so-called piston problem (this ``hypersonic''
transformation is known very well in classical gas dynamics---see, for instance,
\cite{ll,ovs}).
In  the Letter \cite{ek06} this transformation was applied to the problem of the
supersonic 2D NLS flow past slender obstacle, which was translated into the piston
problem for the 1D defocusing NLS equation. It was also shown how the dispersive
piston problem for the defocusing NLS equation can be asymptotically reduced to
a much better understood initial-value problem.

The present paper is devoted to a systematic study of the DSWs generated in the supersonic
flow past extended bodies in the framework of two-dimensional defocusing NLS equation.
The most relevant physical context
of this problem is the description of the flows of Bose-Einstein condensates
(BECs) past obstacles, which  is currently a subject of intensive experimental and
theoretical studies (see, for instance,  \cite{cornell05,caruso,ea07} for recent
experimental work and
\cite{wmca-99,ap04,egk06,kp08,ship1,ship2,ship3D,sh-08,two-comp,barrier09}
and references therein for some of the theoretical advances). It should be noted that
the literature on this subject is growing too rapidly to reflect all recent advances.
We also note that most of the existing theoretical work on the BEC flows past
obstacles is concerned
with the flows past small localized ``impurities" with the dimensions of order of the
healing length. In this paper, we consider an opposite situation, when the obstacle
has the size much greater than the internal coherence length of the medium. This
``slender body'' problem is  fundamentally important as a dispersive counterpart
of the classical gas dynamics problem about the supersonic flow past a ``wing"
(see, for instance,  \cite{cf48,ll}) and has an advantage of the possibility
of full analytical treatment. In addition, the solution of this problem elucidates
the macroscopic mechanisms of the generation of dark solitons and ``ship waves''
in BECs observed in the numerical and physical experiments
\cite{cornell05,caruso,egk06,ship1,ship2,ship3D,two-comp}.
Foreseeable direct physical applications could be connected with the BEC flows
in atom-chip systems (see, e.g., \cite{chip0,chip} and references therein).

The possibility of full analytical description of the 2D the supersonic
NLS flow past body problem is based on the already mentioned ``dispersive piston''
approximation \cite{ek06}.
In the recent paper \cite{ha08}, the
dispersive piston problem for 1D unsteady NLS flows was studied
for the particular case of the piston moving with
constant velocity (this corresponds to the flow past an infinite
straight concave corner in the context of the present paper---see
Section  VI.A).  In the present paper,  full analytical modulation
solutions will be constructed for this and other, more general,
cases when the piston curve is a reasonably arbitrary unimodal function, which is
necessary for the description of the supersonic NLS flow past a
finite-length body.

One of the unusual features of the NLS piston problem solution, not captured by
the single-wave KdV approximation, is the generation of a {\it nonmodulated
nonlinear periodic wave}
in the region between the piston (body surface) and the trailing edge of the
DSW for sufficiently large piston speeds.
We show that this ``transition wave'' observed in the numerical solution in
\cite{ha08} actually occurs due to the reflection of a large-amplitude DSW
from the piston surface---so that the interaction of the oncoming and reflected
modulated waves necessarily leads to the
formation of a region filled with purely periodic nonlinear oscillations.
The occurrence of  a nonmodulated nonlinear wave region in the similarity
solutions of the defocusing NLS equation was first predicted  in
\cite{eggk95} as one of the particular cases
in the general classification of the decay of an initial discontinuity.

In the recent studies \cite{egk06,ship1,ship2,ship3D} of the supersonic BEC flow
past localized obstacles  two main distinct ingredients of the generated wave
pattern have been identified and studied analytically and numerically:
the so-called ``ship waves'' corresponding to the spatial Bogoliubov modes
and generated outside the Mach cone, and oblique dark solitons generated
inside the Mach cone and stretching  behind the obstacle. An unexpected
feature of these oblique dark solitons, established first numerically in
\cite{egk06}, is their apparent stability, in striking contrast with the
well established notion of the absolute ``snake'' instability of two-dimensional
dark NLS solitons  leading to their decay into vortex-antivortex pairs
\cite{kp-1970,zakharov-1975,kuztur}, \cite{anderson}. This apparent paradox
was resolved in \cite{kp08} where it was shown that the presence of the
background BEC flow with the velocity greater than certain ``threshold''
velocity  stabilizes the dark soliton, so that it becomes only
{\it convectively} unstable, i.e. practically stable in the reference frame attached to the obstacle.

We note that in \cite{egk06,ship1,ship2,ship3D}  the ship waves and oblique dark
solitons were studied as separate  independent wave structures generated by an
idealized obstacle of small size placed in the BEC flow.  At the same time,
the process of the generation of these wave structures, as well as the connection
of their parameters with the geometry and size of the physical obstacle remained
beyond the scope of the cited studies. In this paper, by considering an analytically
tractable case of the  supersonic NLS flow past a two-dimensional slender obstacle
of finite size,
we show that the ship waves and oblique dark solitons can be described as asymptotic
far-field outcomes of the spatial ``evolution'' of two separate DSWs spreading
from the front and rear pointed ends of the body. In spite of their common origin,
the front and rear DSWs evolve in  drastically different ways: the front wave
asymptotically transforms into a
dispersing wave packet (effectively a ``ship wave'') while the rear one converts
into a fan of dark solitons. This qualitative difference occurs owing to the fact that
the front wave is developed from the compression ``hump'' forming due to the
slowing of the oncoming flow
near the increasing profile of the front part of the body while the rear wave
evolves from the density dip forming behind the body. So in terms of the
one-dimensional NLS equation,
the front wave corresponds  to the continuous spectrum  of the associated
Zakharov-Shabat linear spectral problem and the rear one---to the discrete
spectrum. This is in striking contrast with classical dissipative gas
dynamics where both shock waves spreading from the endpoints of the
wing have essentially the same structure (see, for instance, \cite{cf48,ll}).

We first develop the theory of the supersonic flow past a straight wedge by
applying the similarity modulation solutions \cite{gk87,eggk95}
to the associated 1D dispersive piston problem
(see \cite{ha08}). The comparison with full 2D numerical solution of the
NLS equation with impenetrability boundary conditions at the body surface
shows that the 1D piston approximation describes the arising wave distribution
remarkably well.

Next we analyze the flow past a slender ``wing'' by constructing asymptotic
1D analytical solutions for the front and rear DSWs and comparing them with
the full 2D numerical solutions.
We describe the front DSW behaviour by constructing an appropriate modulation
solution of the dispersive piston problem with the
piston curve corresponding to the body profile and then ``translating''
this solution in terms of the original 2D problem.

The numerically observed wave distributions of the 2D NLS flow around the corner
or the front edge of the wing, however, extend beyond the DSW region confined
to certain boundaries, $[y^-(x), y^+(x)]$. To describe the distribution of
the wave crests outside the DSW  region we extend the traditional Gurevich-Pitaevskii
type formulation of the  problem by complementing it by the modulation solution
describing the distribution in the linear wave ``packet" located outside the
external DSW boundary $y^+(x)$.  The lines of constant phase in this linear
modulation solution determine the location of the small-amplitude wavecrests
visible in numerical and physical experiments. Together with the DSW,
they form a structure which eventually transforms
into the universal Kelvin-Bogoliubov ``ship wave'' pattern \cite{ship1,ship2}.
The far-field asymptotic behaviour of our nonlinear modulation solution
describes the distributions of the wave amplitude in this ``ship wave''
as a function of the wing profile.

Finally, we consider the rear DSW, which  asymptotically decomposes into
a fan of oblique dark solitons \cite{egk06}. Instead of constructing the
full modulation solution, we describe the
asymptotic distribution of solitons in this fan using the generalized
Bohr-Sommerfeld semi-classical quantization rule for the spectral eigenvalues obtained
for the defocusing NLS equation in \cite{jl99,kku02} using the
inverse scattering transform (IST) formalism.

Our analytical solutions are compared with the full numerical simulations
of the 2D unsteady NLS flow past extended obstacles.

\section{Formulation of the problem}

We consider the supersonic NLS flow past an extended two-dimensional
body with pointed ends (a ``wing''). For simplicity we shall assume
zero attack angle.

We  describe the flow dynamics by the multidimensional defocusing
NLS equation in the canonical form
\begin{equation}\label{eq1}
   i \psi_t =-\tfrac12\Delta\psi
   +|\psi|^2\psi  .
\end{equation}
Since we
shall be interested in the potential (vortex-free) flows it is
convenient to transform Eq.~(\ref{eq1}) to a hydrodynamic form by
means of the substitutions
\begin{equation}\label{eq4}
   \psi(\br,t)=\sqrt{n(\br,t)}\exp\left(i \Theta({\bf r}, t)
   \right), \qquad  {\bf u}= \nabla \Theta \, ,
\end{equation}
where $n(\br,t)$ is the density of the ``fluid'' and ${\bf
u}(\br,t)$ denotes its potential velocity field, ${\bf r} \equiv (x,y)$. We introduce normalized
dependent variables $ \tilde{n}=n/n_0,\, \tilde{\bf u}={\bf u}/c_s,
c_s= \sqrt{n_0} $ where $n_0$ is the value of the density at
infinity and $c_s$ is the corresponding sound speed. As a result, we
obtain the system (we omit tildes for convenience of the notation)
\begin{equation}\label{eq8}
\begin{split}
 n_t+\nabla \cdot (n\bu)=0,\\
  \bu_t+(\bu\cdot\nabla)\bu+\nabla n+\nabla\left[\frac{(\nabla n)^2}{8n^2}
   -\frac{\Delta n}{4n}\right]=0\, , \\
    \nabla \times {\bf u} =0 \,
       \end{split}
\end{equation}
(here $\nabla \equiv (\prt_x,\prt_y)$).

We assume the uniform oncoming flow with constant density $n=1$ and
the velocity ${\bf u} = (M,0)$ directed parallel to $x$ axis. Here $M>1$
is the Mach number of the oncoming supersonic flow. The system
(\ref{eq8}) then should be solved with the boundary conditions at
infinity,
\begin{equation}\label{eq10}
   n \to 1,\quad \bu \to (M,0)\quad\text{as}\quad |{\bf r}| \to \infty
\end{equation}
and the impenetrability condition at the body surface $S$:
\begin{equation}\label{eq11}
   \left.\bu\cdot{\bf N}\right|_S=0 \, ,
\end{equation}
where ${\bf N}$ denotes a unit vector of outer normal to the body
surface.  Similar to classical gas dynamics theory of supersonic
flows (see \cite{ll} for instance) we shall be interested in an established, steady,
wave pattern. Hence we confine ourselves to stationary
solutions of the problem (\ref{eq8})--(\ref{eq11}) and replace
Eqs.~(\ref{eq8}) by their time-independent versions for $n(x,y)$,
 $\bu=(u(x,y),v(x,y))$:
\begin{equation}\label{eq12}
\begin{split}
   (nu)_x+(nv)_y=0,\\
   uu_x+vu_y+n_x+\left(\frac{n_x^2+n_y^2}{8n^2}-
   \frac{n_{xx}+n_{yy}}{4n}\right)_x=0,\\
   uv_x+vv_y+n_y+\left(\frac{n_x^2+n_y^2}{8n^2}-
   \frac{n_{xx}+n_{yy}}{4n}\right)_y=0, \\
    u_y -v_x=0 \, .
   \end{split}
\end{equation}
Let the shape of the body in the upper half-plane be given by a unimodal (one-hump)
function: $y=F(x) > 0$ for $x \in (0,L)$,
$F(0)=F(L)=0$  and $F(x)=0$ for $x \notin [0,L]$, $L$ being the body length in dimensionless
units (see Fig.~1).  Thus we have  ${\bf
N}\propto(F'(x),-1)$, and the boundary conditions (\ref{eq10}),
(\ref{eq11}) are transformed to
\begin{equation}\label{eq14a}
   n=1,\quad u=M,\quad v=0\quad\text{at}\quad x^2 + y^2 \to \infty,
\end{equation}
\begin{equation}\label{eq14b}
   v=uF'(x)\quad\text{at}\quad y=F(x).
\end{equation}
\begin{figure}[ht]
\centerline{\includegraphics[width=8cm]{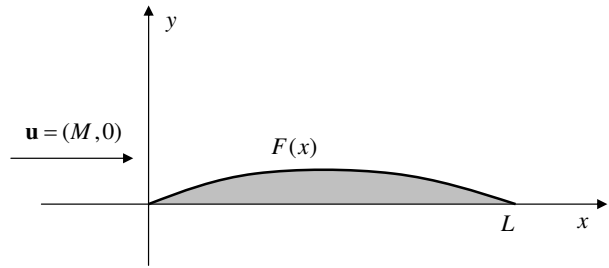}}
\caption{Flow past a wing}
\label{fig1}
\end{figure}
The flow in the lower half-plane can be considered independently in a completely analogous way.

\section{Piston problem approximation and qualitative description of the wave pattern}

The system  (\ref{eq12})--(\ref{eq14b}) is still too complicated for
a direct analytical treatment. However, when the flow can be
considered as highly supersonic the steady
problem of the two-dimensional flow past slender body can be asymptotically
transformed to a
much simpler problem of 1D ``unsteady'' flow along $y$ axis with the
scaled $x$ coordinate playing the role of ``time'' \cite{ekt04}. To
this end, we substitute into Eqs.~(\ref{eq12}) the new variables
\begin{equation}\label{eq15}
   u=M+u_1+O(1/M),\quad T=x/M,\quad Y=y,
\end{equation}
assuming $M^{-1} \ll 1$. Then to leading order we obtain
\begin{equation}\label{eq16}
\begin{split}
    n_T+(nv)_Y=0,\\
 v_T+vv_Y+n_Y+\left(\frac{n_Y^2}{8n^2}
   -\frac{n_{YY}}{4n}\right)_Y=0,
   \end{split}
\end{equation}
\begin{equation}\label{eq18}
 u_{1}=0.
\end{equation}
Equations (\ref{eq16}) represent the hydrodynamic form of the 1D
defocusing NLS equation
\begin{equation}\label{nls-1D}
   i\Psi_T+\tfrac12\Psi_{YY}-|\Psi|^2\Psi=0
\end{equation}
for a complex field variable
\begin{equation}
\Psi=\sqrt{n}\exp\left(i\int^Y v(Y',T)dY'\right),
\end{equation}
 and we can apply the
well-developed analytical methods to its study. It is remarkable that in the
case of a slender body, for which $M \alpha =O(1)$, where
$\alpha=\mathrm{max}|F'(x)|$, the boundary condition (\ref{eq14b})
reduces (to leading order in $M^{-1}$) to the classical piston
conditions (see \cite{ll} for instance)
\begin{equation}\label{eq19}
   v=v_p= df/dT\quad\text{at}\quad Y=f(T),
\end{equation}
where the piston motion is described by the function $f(T)=F(MT)$.
Condition (\ref{eq10}) at infinity transforms into
\begin{equation}\label{eq19a}
n=1,  \  \ v = 0 \quad \hbox{as} \ \  Y \to \infty \, .
\end{equation}

Thus, we have reduced the problem of the flow past slender body to the
piston problem for 1D flow along a tube with a piston
moving inside it according to the law (\ref{eq19}). In contrast to the
classical gas dynamics, the piston problem is now posed for
dispersive equations (\ref{eq16}).

The piston reduction for 2D hypersonic dispersive dissipationless flows was
first introduced
in \cite{ekt04} in a rather general form and in \cite{ek06} it was formulated
in the  present NLS context.  In \cite{ha08}  the
dispersive piston problem for 1D defocusing  NLS equation was studied
for the simplest case of the piston moving with
constant velocity (this corresponds to the flow past an infinite
straight concave corner in the context of the present paper---see
Section VI). In the subsequent sections, analytical modulation
solutions will be constructed for this and other, more general,
cases when the piston curve is a non-monotone function, which is
necessary for the description of the supersonic NLS flow past a
finite-length body (a ``wing'').
\begin{figure}[ht]
\centerline{\includegraphics[width=8cm]{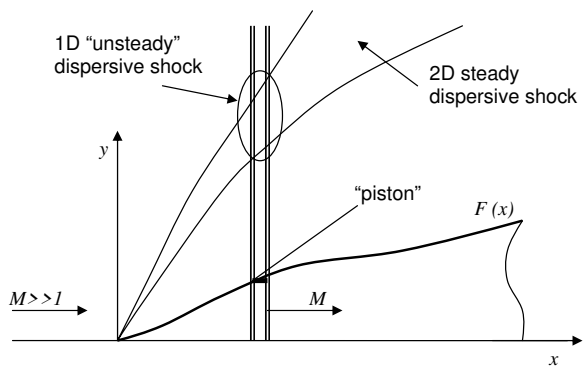}}
\caption{Piston
analogy in the problem of supersonic flow of dispersive fluid past
body}\label{fig2}
\end{figure}
In classical viscous gas dynamics, the supersonic flow past a wing
leads to the generation of two spatial shock waves (oblique jumps of
compression) spreading from the front and the rear edges of the wing
(see \cite{ll} for instance). In terms of the piston problem this
corresponds to the formation of two shocks during two different
phases of the piston motion: forward and reverse.

Before we proceed with the quantitative analysis  of this problem we
briefly outline the
qualitative structure of the dispersive flow past finite-length body
using the theoretical results of \cite{karpman,gkke,ek06}.
We assume that the length of the body is much greater than
typical dispersive (coherence) length of the
medium. Then in dispersive hydrodynamics, both shocks spreading from the body
edges resolve into expanding nonlinear oscillatory zones,  the {\it oblique spatial
dispersive shock waves}. At finite distances from the body surface
these two spatial dispersive shocks have similar structure (see
\cite{gp74}): each represents a modulated nonlinear wave acquiring a
form close to a chain of oblique solitons at one edge of the
oscillatory zone and degenerating into a linear wave at the opposite edge.
However, as was indicated above, at large distances from the body the two dispersive
shocks demonstrate drastically different behaviour: in the present case of the NLS
hydrodynamics the
dispersive shock spreading from the rear edge of the body
transforms into the oblique soliton train while the dispersive shock
forming at the front end of the body completely degenerates into a
vanishing amplitude dispersing linear wave packet.

\section{Modulation theory for the defocusing NLS equation: account of results}

The theory of DSWs is based on the study of a certain nonlinear
free-boundary problem for the modulation (Whitham) equations---the so-called
Gurevich-Pitaevskii problem. In this section we make a brief review
of the relevant results of the modulation theory for the defocusing
NLS equation which are necessary for the analysis of spatial DSWs
generated in the steady supersonic NLS flow past slender body. A
detailed derivation of the single-phase  NLS modulation system can be found in
\cite{kamch2000}.

\subsection{Travelling wave solution and modulation equations}

The periodic travelling wave solution of the defocusing NLS equation
(\ref{eq16}) can be expressed in terms of the Jacobi elliptic function ``${\rm sn}$''
and is characterized by four constant parameters $\la_1\leq\la_2\leq\la_3\leq\la_4$,
\begin{equation}\label{eq013}
n =\frac14(\la_4-\la_3-\la_2+\la_1)^2+ (\la_4-\la_3)
(\la_2-\la_1)\,{\rm sn}^2\left(\sqrt{(\la_4-\la_2)(\la_3-\la_1)}\,
\theta,m\right) \, ,
\end{equation}
\begin{equation}\label{v}
v=U - \frac{C}{n} \, ,
\end{equation}
where $C=\frac{1}{8} (-\lambda_1 - \lambda_2 + \lambda_3 +
\lambda_4) (-\lambda_1 + \lambda_2 - \lambda_3 + \lambda_4)
(\lambda_1 - \lambda_2 - \lambda_3 + \lambda_4)$,
\begin{equation}\label{eq016}
\theta=Y-UT-\theta_0,\qquad U=\frac12 \sum_{i=1}^4\la_i,
\end{equation}
$U$ being the phase velocity of the nonlinear wave and $\theta_0$ initial phase.
The modulus $0 \le m \le 1$  is defined as
\begin{equation}\label{eq015}
m=\frac{(\la_2-\la_1)(\la_4-\la_3)}{(\la_4-\la_2)(\la_3-\la_1)},
\end{equation}
and the wave amplitude is
\begin{equation}\label{amp}
a= (\la_4-\la_3)
(\la_2-\la_1) \, .
\end{equation}
The wavelength is equal to
\begin{equation}\label{eq017}
\mathfrak{L}= \int \limits_{\la_3}^{\la_4}
\frac{d\lambda}{\sqrt{(\la-\la_1)(\la-\la_2)(\la-\la_3)(\la_4-\la)}}=
\frac{2{\K}(m)}{\sqrt{(\la_4-\la_2)(\la_3-\la_1)}},
\end{equation}
${\K}(m)$ being the complete elliptic integral of the first kind.

In the limit   as $m \to 1$ (i.e. as $\la_3  \to \la_2$) the travelling wave
solution (\ref{eq013}) transforms into a dark soliton riding on a ``pedestal''
$n_0$:
\begin{equation}\label{sol}
n=n_0 - \frac{a_s}{\hbox{cosh}^2 (\sqrt{a_s}(Y-U_sT - \theta_0))}\, ,
\end{equation}
where the background density $n_0$, the soliton amplitude $a_s$  and the
soliton velocity $U_s$  are expressed in terms of $\lambda_1, \lambda_2, \lambda_4$ as
\begin{equation}\label{22}
n_0=\frac{1}{4}(\la_4 - \la_1)^2,  \quad a_s=(\la_4 - \la_2)(\la_2 - \la_1) , \quad    U_s=\frac{1}{2}(\la_1+2\la_2+\la_4)\, .
\end{equation}
Allowing the parameters $\la_j$ to be slowly varying functions of $Y$ and $T$,
one arrives at a  modulated nonlinear periodic wave  in which the evolution
of $\la_j$'s is governed by the Whitham  modulation equations in the diagonal
Riemann form \cite{fl86,pavlov87}
\begin{equation}
\label{eq018} \frac{\partial \la_i}{\partial
T}+V_i(\la)\frac{\partial \la_i}{\partial Y}=0, \qquad i=1,2,3,4,
\end{equation}
which are obtained via the averaging of the NLS conservation laws over the period of
the travelling wave solution (\ref{eq013}) (see \cite{wh74,kamch2000}
for the detailed description of the Whitham method).
The characteristic velocities can be calculated using the formula
\cite{gke92,kamch2000}
\begin{equation}
\label{eq019}
V_i(\la)=\left(1-\frac{\mathfrak{L}}{\partial_i\mathfrak{L}}\partial_i\right)U , \quad i=1,2,3,4 \, , \quad \hbox{where} \quad
\partial_i\equiv\partial/\partial \la_i \, .
\end{equation}
Substitution of Eq.~(\ref{eq017}) into Eq.~(\ref{eq019}) gives the
explicit expressions
\begin{equation}\label{vi}
\begin{split}
V_1&=\tfrac12 \sum \la_i
-\frac{(\la_4-\la_1)(\la_2-\la_1)\K}{(\la_4-\la_1)\K-(\la_4-\la_2)\E},\\
V_2&=\tfrac12 \sum \la_i
+\frac{(\la_3-\la_2)(\la_2-\la_1)\K}{(\la_3-\la_2)\K-(\la_3-\la_1)\E},\\
V_3&=\tfrac12 \sum \la_i
-\frac{(\la_4-\la_3)(\la_3-\la_2)\K}{(\la_3-\la_2)\K-(\la_4-\la_2)\E},\\
V_4&=\tfrac12 \sum \la_i
+\frac{(\la_4-\la_3)(\la_4-\la_1)\K}{(\la_4-\la_1)\K-(\la_3-\la_1)\E},
\end{split}
\end{equation}
where  $\E=\E(m)$ is the complete elliptic integral
of the second kind.
The characteristic velocities  (\ref{vi})  are real for all values of the Riemann
invariants therefore system (\ref{eq017})
is hyperbolic. Moreover, it is not difficult to show using (\ref{eq019}) that
$\partial_i V_i >0$ for all $i$  so the NLS-Whitham system (\ref{eq018}), (\ref{vi})
is {\it genuinely nonlinear} \cite{jl99}.

An asymptotic modulated wave solution is obtained by substituting the solution
of the modulation equations (\ref{eq018}) back into the travelling wave (\ref{eq013}).
We stress that initial phase $\theta_0$ in (\ref{eq016}) is ``erased'' in the averaging
procedure so the resulting modulated wave is defined with the accuracy to an arbitrary
shift within the
wave spatial period.

For the DSW analysis in the subsequent sections we shall need the reductions of the formulae (\ref{vi}) for the limiting cases when $m=0$ and $m=1$.

The harmonic limit $m=0$ can be achieved in one of two possible ways:
one sets either $\la_2=\la_1$ or $\la_3=\la_4$.
Then:
\begin{eqnarray}
\hbox{When} \ \  \la_2=\la_1:   \quad    V_2=V_1=\lambda_1 + \frac{\lambda_3 + \lambda_4}{2} +
\frac{2(\lambda_3-\lambda_1)(\lambda_4 - \lambda_1)}{2\lambda_1 - \lambda_3-\lambda_4} \, ,&&
\nonumber \\
&& \label{m01} \\
V_3= \frac{3}{2} \la_{3} + \frac12{\la_{4}}\, , \quad
V_4= \frac{3}{2} \la_{4} + \frac12{\la_{3}} \, .&& \nonumber
\end{eqnarray}
\begin{eqnarray}
\hbox{When} \ \  \la_3=\la_4:   \quad    V_3=V_4=\lambda_4 + \frac{\lambda_1 + \lambda_2}{2} +
\frac{2(\lambda_4-\lambda_2)(\lambda_4 - \lambda_1)}{2\lambda_4 - \lambda_2-\lambda_1} \, ,&&
\nonumber \\
&& \label{m02} \\
V_1= \frac{3}{2} \la_{1} + \frac12{\la_{2}}\, , \quad V_2= \frac{3}{2} \la_{2} + \frac12{\la_{1}}\, . && \nonumber
\end{eqnarray}
In the soliton limit we have $m=1$.
This can happen only if $\la_2=\la_3$, so we obtain:
\begin{eqnarray}
\hbox{When} \ \  \la_2=\la_3:   \quad    V_2=V_3=\frac{1}{2}(\lambda_1+ 2\lambda_2 + \lambda_4)\, ,&&
\nonumber \\
&& \label{m1} \\
V_1= \frac{3}{2} \la_{1} + \frac12{\la_{4}}\, , \quad V_4= \frac{3}{2} \la_{4} + \frac12{\la_{1}}\, . && \nonumber
\end{eqnarray}

Thus in both  harmonic ($m \to 0$) and soliton  ($m \to 1$)  limits
the fourth-order modulation system (\ref{eq018}),
(\ref{vi}) reduces to the system of three equations, two of which
are decoupled.  Moreover, one can see that in all considered limiting cases
the decoupled equations agree with the {\it dispersionless limit} of the NLS
equation (\ref{eq16}). Indeed, the dispersionless limit of the NLS equation
is the ideal shallow-water system
\begin{equation}\label{eq21}
 n_T+(nv)_Y=0,\quad v_T+vv_Y+n_Y=0\, ,
\end{equation}
which can be represented in the diagonal form by introducing Riemann invariants
\begin{equation}\label{eq20}
 \la_\pm=\frac12{v}\pm\sqrt{n}
\end{equation}
\begin{equation}\label{er}
\frac{\partial \la_{\pm}}{\partial T} + V_{\pm}(\la_+, \la_-)
\frac{\partial \la_{\pm}}{\partial Y}=0\, ,
\end{equation}
where
\begin{equation}\label{V}
V_+=\frac{3}{2} \la_++ \frac12{\la_-} \, , \qquad V_-=\frac32
{\la_-}+ \frac12{\la_+} \, ,
\end{equation}

\subsection{Hodograph transform and reduction to the Euler-Darboux-Poisson equation}

We fix two Riemann invariants,
\begin{equation}\label{}
\la_1=\la_{10}=\hbox{constant}\, , \quad
\la_2=\la_{20}=\hbox{constant}
\end{equation}
to reduce (\ref{eq018}) to the system of two equations
\begin{equation}\label{34}
\frac{\partial \la_3}{\partial T}+V_3(\la_3, \la_4)\frac{\partial
\la_3}{\partial Y}=0\, , \qquad \frac{\partial \la_4}{\partial
T}+V_4(\la_3, \la_4)\frac{\partial \la_4}{\partial Y}=0\, ,
\end{equation}
where $V_{3,4}(\la_3, \la_4)  \equiv  V_{3,4}(\la_{10}, \la_{20}, \la_3, \la_4)$.
Applying the hodograph transform to system (\ref{34}) one arrives at
a linear system for $Y(\la_3, \la_4)$, $T(\la_3, \la_4)$
\begin{equation}\label{hod1}
\frac{\partial Y}{\partial \la_3}-V_4(\la_3, \la_4)\frac{\partial T}{\partial
\la_3}=0\, , \qquad \frac{\partial Y}{\partial
\la_4}-V_3(\la_3, \la_4)\frac{\partial T}{\partial \la_4}=0 \, .
\end{equation}
Now we make in (\ref{hod1}) the change of variables
\begin{equation}\label{Ts1}
 Y - V_j T = W_j  \, , \qquad j=3,4\, ,
\end{equation}
which reduces it to a symmetric system for $W_3(\la_3,\la_4)$, $W_4(\la_3,\la_4)$:
\begin{equation}\label{Ts2}
 \frac{\partial _i W _j}{W_i - W_j} = \frac{\partial_i
V_j}{V_i - V_j} \ ; \quad  \ i,j = 3, 4 \, , \quad  i \ne j ; \quad
\partial_i\equiv\partial/\partial \la_i \, .
\end{equation}
The symmetry between $V_j$ and $W_j$ in (\ref{Ts2}) and the ``potential'' structure (\ref{eq019}) of the functions $V_j$ implies the possibility of introducing
a single scalar function $g(\la_3, \la_4)$ instead of the vector  $(W_{3}, W_4)$:
\begin{equation}
\label{scalar}
W_i=\left(1-\frac{\mathfrak{L}}{\partial_i\mathfrak{L}}\partial_i\right)g \, ,\quad
i=3,4,
\end{equation}
or, which is the same,
\begin{equation}\label{scalar1}
W_i=g+ 2(V_i-U) \frac{\partial g}{\partial \la_i}\, .
\end{equation}
Then substituting (\ref{eq019}), (\ref{scalar1}) into (\ref{Ts2}) we arrive, taking into account (\ref{eq017}),  at the Euler-Darboux-Poisson (EDP) equation for $g(\la_3, \la_4)$
first  obtained in the present NLS context in \cite{gke92} (see also \cite{ek95})
\begin{equation}\label{edp}
2(\la_4-\la_3)\frac{\partial^2 g}{\partial \lambda_3 \partial \la_4}
=\frac{\partial g}{\partial \la_4} - \frac{\partial g}{\partial \la_3} \, .
\end{equation}
The general solution of the EDP equation (\ref{edp}) can be represented in
the form (see, for instance, \cite{tricomi})
\begin{equation}
\label{gs} g=\int \limits _{0} ^{\la_3} \frac{\phi_1(\la) d
\la}{\sqrt{(\la - \la_3)(\la_4-\la)}}
+
\int \limits _{0} ^{\la_4} \frac{\phi_2(\la) d \la}{\sqrt{(\la- \la_3)(\la_4-\la)}} \, ,
\end{equation} where $\phi_{1,2}(\la)$ are arbitrary (generally,
complex-valued) functions.

As a matter of fact, the same construction can be realized for any
pair of the Riemann invariants while the remaining two invariants
are fixed. Moreover, equations (\ref{Ts1})--(\ref{edp}) turn out
to be valid even when all four Riemann invariants vary \cite{gke92,ek95}.
This becomes possible for two reasons. Firstly, the
NLS modulation system (\ref{eq18}), (\ref{eq19}) is integrable via
the generalized hodograph transform \cite{ts85} which converts it
into overdetermined consistent system (\ref{Ts2}) where
$i,j=1,2,3,4$, $i \ne j$. Secondly, the ``potential'' structure of the
characteristic speeds (\ref{eq019}) makes it possible to use the same
substitution (\ref{scalar}) for all $i=1,2,3,4$ which results in the
 consistent system of six EDP equations (\ref{edp})
involving all pairs $\lambda_i, \lambda_j$, $i \ne j$.

Thus, the problem of integration of the nonlinear Whitham system (\ref{eq018})
with complicated coefficients (\ref{vi}) is essentially reduced to solving the
classical linear EDP equation
(\ref{edp}) so practically one needs to express the functions $\phi_{1,2}(\la)$
in the general solution (\ref{gs}) in terms of the initial or boundary conditions
for the NLS equation (\ref{eq1}).

One should note, that  classical hodograph solutions do not include the special
family of the simple-wave solutions as the latter correspond to the vanishing
of the Jacobian
of the hodograph transform $(\la_i, \la_j) \mapsto (Y,T)$ (see, for instance,
\cite{wh74}). However, the similarity solution can be formally included in the
hodograph solutions in the generalized form (\ref{Ts1}). Indeed, putting one
of $W_k=0$ and setting constant all the Riemann invariants $\la_j$ with
$j \ne k$ one arrives at the  similarity solution, in which $\la_k=\la_k(Y/T)$
is implicitly specified by the equation $V_k=Y/T$.

\subsection{Free-boundary matching conditions for the modulation equations}

In the description of the DSW,  the Whitham
equations (\ref{eq018}) must be equipped with certain boundary
conditions for the Riemann invariants $\lambda_i$ \cite{gk87}. These
conditions are the NLS analogs of the Gurevich-Pitaevskii conditions
\cite{gp74} formulated for the KdV dispersive shock waves.
To be specific,  we formulate  boundary conditions for
the right-propagating DSW,  which corresponds to the spatial DSW generated
in the upper-half plane in the
problem of the supersonic NLS flow past body. Without loss of generality
we assume that the formation of the
DSW starts at the origin of the $(Y,T)$-plane. In the Gurevich-Pitaevskii
setting the upper $(Y,T)$-half plane is split into three regions (see Fig.~2):
$(-\infty, Y^-(T))$, $[Y^-(T), Y^+(T)]$ and $(Y^+(T), + \infty)$.
\begin{figure}[ht]
\centerline{\includegraphics[width=8cm]{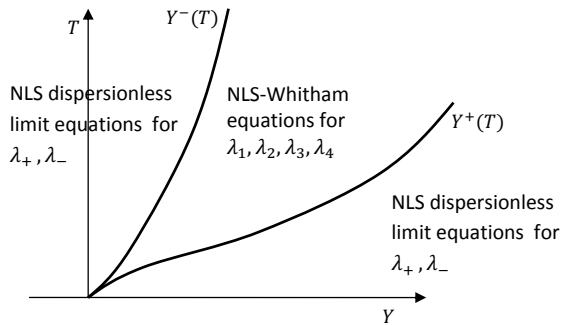}}
\caption{
Splitting of the $YT$-plane in the Gurevich-Pitaevskii problem for the
defocusing NLS equation.
}
\label{fig3}
\end{figure}

In the ``outer'' regions $(-\infty, Y^-(T))$ and $(Y^+(T), + \infty)$ the flow is
governed by the
dispersionless limit of the NLS equation, i.e. by the shallow-water system
(\ref{er}), (\ref{V}) for the Riemann invariants $\la_{\pm}$. In the DSW region
$[Y^-(T), Y^+(T)]$ the averaged oscillatory flow is described by four Whitham
equations (\ref{eq018}) for the Riemann invariants $\lambda_j$
with the following matching conditions at the trailing $Y^-(T)$ and leading
$Y^+(T)$ edges of the DSW
(see \cite{gk87,ek95} for details):
\begin{equation}
\begin{array}{l}
\hbox{At} \ \ Y=Y^-(T):\qquad \la_3=\la_2\, , \  \ \la_4 = \la_+ , \  \   \la_1 = \la_- \, , \\
\hbox{At} \ \ Y=Y^+(T):\qquad \la_3=\la_4\, , \ \  \la_2 = \la_+ ,  \  \  \la_1 = \la_- \, .
\end{array}
\label{bc}
\end{equation}
Here $\la_{\pm}(Y,T)$ are the Riemann invariants of the dispersionless limit of
the NLS equation in the
hydrodynamic form (\ref{er}), (\ref{V}).
The free boundaries $Y^{\pm}(T)$ are defined by the kinematic conditions
\begin{equation}\label{mult}
\frac{dY^-}{dT}=V_ 2(\la_1, \la_2, \la_2, \la_4)=V_3(\la_1, \la_2, \la_2, \la_4)\, , \qquad \frac{dY^+}{dT}=V_3(\la_1, \la_2, \la_4, \la_4)=V_4(\la_1, \la_2, \la_4,  \la_4)
\end{equation}
and so are  the multiple characteristics of the Whitham system. The multiple
characteristic velocities $V_2=V_3$  and
$V_3=V_4$ in (\ref{mult}) are explicitly given  by (\ref{m1}) and  (\ref{m02})
respectively. Determination of $Y^{\pm}(T)$
is an inherent part of the construction of the full modulation solution.
We also emphasize that matching conditions (\ref{bc})
are consistent with the limiting structure of the Whitham system
(\ref{eq018})  at $m=0$ and $m=1$  (see (\ref{m02}), (\ref{m1})) and
reflect the spatial oscillatory structure of the DSW
in the defocusing NLS hydrodynamics (as is known very well, such a DSW
has a dark soliton $(m=1)$ at the trailing edge and degenerates into the
vanishing amplitude
harmonic wave ($m=0$) at  the leading edge---see
\cite{gk87,eggk95,kgk04,ha06}).

One should mention that if one is interested only in the  class of $Y/T$
similarity modulation solutions arising in  the decay of an initial discontinuity problem
one can use, instead of (\ref{bc}),  a reformulation of the modulation problem
as an {\it initial-value problem} for $\la_j$,
where three of the invariants are constant at $T=0$ and for the forth one the
so-called ``regularized'' initial initial condition is used
(see \cite{kod99,bk06,ha06,ha07}). The resulting
initial-value problem has the global expansion fan  solution. This type of
the problem formulation, however,  seems to be less natural when one is
interested in a more general (not self-similar) class of solutions when the
integration of the modulation equations involves the  hodograph transform
(\ref{Ts1}), (\ref{Ts2}) (the poor compatibility of the initial-value problems
with the hodograph method is known very well in classical hydrodynamics
(see, for instance,  \cite{wh74})).
The free-boundary Gurevich-Pitaevskii type formulation (\ref{bc}), on the contrary,
is ideally compatible with the generalized hodograph transform as in any of
the hodograph space coordinate planes $(\la_i, \la_j$) it transforms into the
classical Goursat-type characteristic boundary problem
for the EDP equation \cite{gke92,ek95}.

\section{Asymptotic reformulation of the NLS piston problem as an initial-value problem}

The general dispersive piston problem (\ref{eq19}), (\ref{eq19a}) for the defocusing
NLS equation (\ref{eq16}) is difficult to tackle directly.
It is, therefore, desirable to reformulate it in terms of a much better explored
initial-value problem. The key in this reformulation is the possibility to
use the semi-classical Whitham description  which is applicable when  the characteristic
piston displacements are  much greater than unity while the piston speed is
$O(1)$ (this formally corresponds to the supersonic flow past a slender body with
the length $L \gg M$ in our original setting formulated in Section II, however,
one can expect that the results will be relevant to moderate body lengths as well).
We now assume the qualitative
picture of the flow described in the end of Section III and divide the upper part of the $(Y,T)$-plane in the
piston problem into five distinct regions (see Fig.~4).  In the regions
$I$ and $V$ the flow is undisturbed so we have $n=1$, $v=0$ there.
The corresponding ``dispersionless'' Riemann invariants (\ref{eq20}) are
$\la_{\pm}= \pm 1$.
In the region $III$ for $Y>f(T)$, the ``gas'' is put into
``motion'' by the ``piston'' moving according to Eq.~(\ref{eq19})
(we shall omit the the quotation marks for the terms related to unsteady gas flows henceforth)
and  near the piston the gas motion can be described
by the  dispersionless limit of the defocusing NLS equation (\ref{er}), (\ref{V}).
However, the formal solution of the
nonlinear hydrodynamic-type equations (\ref{er}), (\ref{V})
cannot be extended to the whole $(Y,T)$-plane because
the $Y$-derivatives blow up along certain lines in this plane so the region $III$,
where the flow is smooth, is separated from the
constant flow regions $I$ and $V$ by two DSW regions $II$ and $IV$ which spread
from the points $(0,0)$ and $(0, L/M)$, corresponding to the endpoints of
the ``wing'' (strictly speaking, one should impose a certain restriction on
the behaviour of $f(T)$ near $T=0$ to have the front DSW emanating strictly
from the point $(0,0)$---this restriction will be explained in the end of this Subsection).
The qualitative structure of these DSWs was described in Section II.
We denote the leading (outer, i.e. facing the oncoming flow)  and trailing
(inner, i.e. facing the body surface) edges
of the front DSW (region $II$) as $Y^{+}_f(T)$ and $Y^{-}_f(T)$ respectively,
and, similarly, for the rear DSW (region $V$) edges, we use the notations
$Y^{\pm}_r(T)$.

Now, the plan is to determine the flow parameters $n_p$ and $v_p$ at the piston
surface and then to trace them back to $T=0$ using the solution of the dispersionless
equations (\ref{er}), (\ref{V}). The kinematic condition (\ref{eq19})
defines $v_p=df/dT$ so we just need to find the flow density at the piston.
This can be done by considering
the data transfer along the characteristics in the Gurevich-Pitaevskii
setting of the problem where the entire wave pattern is asymptotically
described by hyperbolic equations of hydrodynamic type (the NLS-Whitham
system (\ref{eq018}) in the regions $II$ and $IV$ and the dispersionless
limit of the NLS equation (\ref{er}) in the regions $I$, $III$ and $V$).

We formulate the matching conditions for both DSWs  using the general rule
(\ref{bc}). For the front DSW we have:
\begin{eqnarray}
\hbox{At} \ \ Y&=&Y^-_f(T):\qquad \la_3=\la_2\, , \  \ \la_4 = \la_+ , \  \
\la_1 = \la_- \, ,  \nonumber \\
&& \label{bc1} \\
\hbox{At} \ \ Y&=&Y^+_f(T):\qquad \la_3=\la_4\, , \ \  \la_2 = 1 ,  \  \
\la_1 = -1 \, . \nonumber
\end{eqnarray}
Similarly, for the rear DSW:
\begin{eqnarray}
\hbox{At} \ \ Y&=&Y^-_r(T):\qquad \la_3=\la_2\, , \  \ \la_4 = 1 , \  \
\la_1 = -1\, ,  \nonumber \\
&& \label{bc2} \\
\hbox{At} \ \ Y&=&Y^+_r(T):\qquad \la_3=\la_4\, , \ \  \la_2 = \la_+  ,  \  \
\la_1 =  \la_- \, . \nonumber
\end{eqnarray}
It then follows that to satisfy the governing equations (\ref{eq018}), (\ref{er}) and the matching conditions (\ref{bc1}) and (\ref{bc2})
one has to put
\begin{equation}\label{la-1}
\lambda_1=\lambda_-=-1
\end{equation}
within the respective domains of definitions of $\la_1$ (regions $II$ and $IV$) and $\la_-$ (regions $I$, $III$ and $V$).
This condition (\ref{la-1}) of transfer of the Riemann invariant of the dispersionless system across the DSW replaces the traditional shock jump conditions for classical viscous shocks
(see \cite{el05} for a detailed discussion of  transition conditions across DSWs).
\begin{figure}[ht]
\centerline{\includegraphics[width=7cm,height=
9cm,clip]{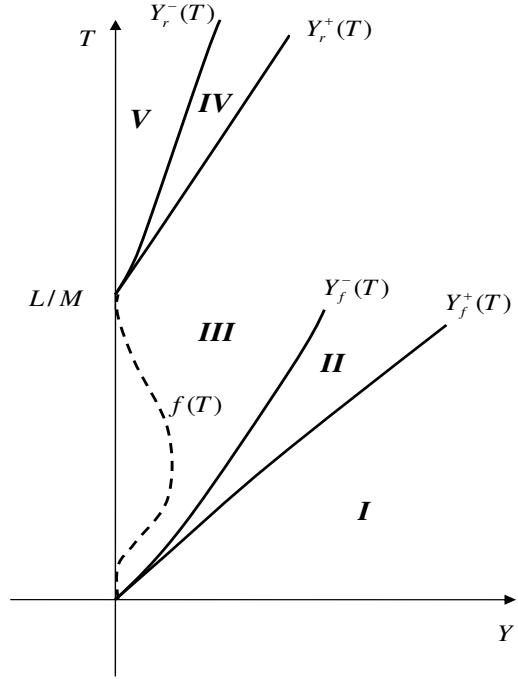}}\vspace{0.3 true cm}
\caption{$(Y,T)$-plane of
the NLS piston problem.  Dashed line:
the piston trajectory $Y=f(T)$. The lines $Y^{\pm}_{f}(T)$,
$Y^{\pm}_{r}(T)$ are the edges of the front and rear dispersive shocks
respectively.} \label{fig4}
\end{figure}
Hence, we have at the ``piston''
$v_p/2-\sqrt{n_p}=-1$ which yields the gas density
\begin{equation}\label{np}
 n_p=(v_p+2)^2/4
\end{equation}
in the region between the piston and DSW.
Then using (\ref{eq19}) we get
\begin{equation}\label{eq23}
   \la_- =-1,\quad \la_+=
   df/dT+1\quad\text{at}\quad Y=f(T).
\end{equation}
We are now able to translate these boundary conditions at the ``piston'' into the
equivalent initial conditions at $T=0$. This
problem for the system (\ref{eq21}) can  be easily solved using
characteristics. Indeed, we have $\la_-=-1$, hence $\la_+$ obeys the
simple wave equation following from (\ref{er}) (see, e.g.,
\cite{kku02})
\begin{equation}\label{eq24}
   \frac{\prt\la_+}{\prt T}+{\tfrac{1}{2}(3\la_+-1)}\frac{\prt\la_+}{\prt Y}=0 \, .
\end{equation}
Solution of (\ref{eq24}) with boundary condition (\ref{eq23}) is readily  found  using characteristics:
\begin{equation}\label{char}
 \lambda_+=f'(\xi)+1\, , \quad Y= f(\xi) + (\tfrac32
f'(\xi)+1)(T-\xi)\, ,
\end{equation}
where $\xi$ is a parameter along the piston curve $Y=f(T)$.
Then, setting in  (\ref{char}) $T=0$ we arrive at a parametric form of the
equivalent initial distribution of the Riemann invariant $\la_+$:
\begin{equation}\label{eq27}
\la_+(Y, 0): \qquad   \la_+=  f'(\xi)+1,\quad Y=f(\xi)-(\tfrac32
f'(\xi)+1)\xi\, .
\end{equation}
Distribution (\ref{eq27})  together with the initial condition
\begin{equation}\label{eq28}
\la_-(Y, 0) = -1
\end{equation}
define, via (\ref{eq20}), initial conditions for
the NLS equation in the hydrodynamic form (\ref{eq16}).
\begin{figure}[ht]
\centerline{\includegraphics[width=8cm]{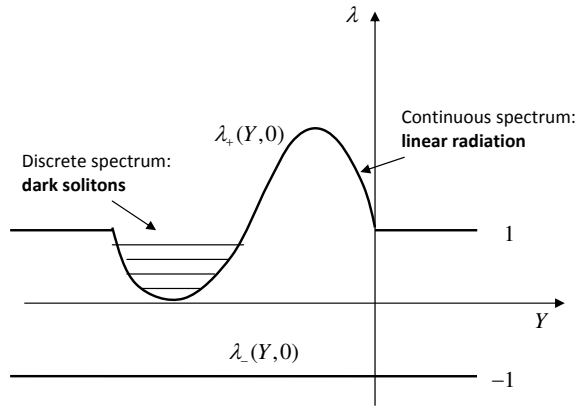}}
\caption{Sketch of asymptotically equivalent initial conditions for
$\lambda_{\pm}$ in the problem of supersonic NLS flow past  slender profile
with $f'(0)=f'(l)=0$.}
\label{fig5}
\end{figure}
It  is important to emphasize that initial conditions (\ref{eq27})--(\ref{eq28})
and the piston boundary conditions (\ref{eq19}), (\ref{eq19a}) are equivalent only
asymptotically,
as our reformulation is made within the conditions of applicability of
the Whitham modulation approach. One should also stress that in the context
of the flow past  body problem, the solution to the initial value problem
(\ref{eq16}), (\ref{eq27})--(\ref{eq28}) is defined only in the
region $Y  \ge f(T)$, i.e. outside the body (piston).

Now one can make some qualitative predictions about the asymptotic structure of
the flow in  the problem of the 2D NLS flow past slender body. First we assume that
$f(0)=f(l)=f'(0)=f'(l)=0$, where $l=L/M$. Also to avoid unnecessary complications
at this stage we assume that $\la_+(Y,0)$,
specified parametrically by (\ref{eq27}), is a single-valued function (this restriction
is not essential but it makes our analysis more transparent). Then it is not difficult
to see from (\ref{eq27})  that the ``translated'' initial profile for the hydrodynamic
Riemann invariant $\la_+$ corresponding to the flow past a wing has the shape of
a large-scale ``bi-polar'' pulse (see Fig.~\ref{fig5}) while the invariant $\la_-$
is constant. Note that the pulse is supported on the interval $[-l,0]$. Then,
the semi-classical approach to the inverse scattering transform for the defocusing
NLS equation developed in \cite{jl99,kku02} enables one to associate
the ``well'' part of the  initial profile of $\la_+$ with certain
distribution of dark solitons in the rear far-field asymptotic
while the front ``barrier'' part is responsible for the linear dispersing
radiation in the region $II$ as $T \to \infty$. The asymptotic formula for the
amplitude distribution in the dark soliton fan generated out of the rear DSW
will be presented in Section VII. The modulation solution for the front wave
gradually transforming, via the nonlinear DSW stage, into the
Kelvin-Bogoliubov ``ship-wave'' pattern (see \cite{ship1,ship2})
will be constructed in Section VI.
 \begin{figure}[ht]
\centerline{\includegraphics[width=8cm]{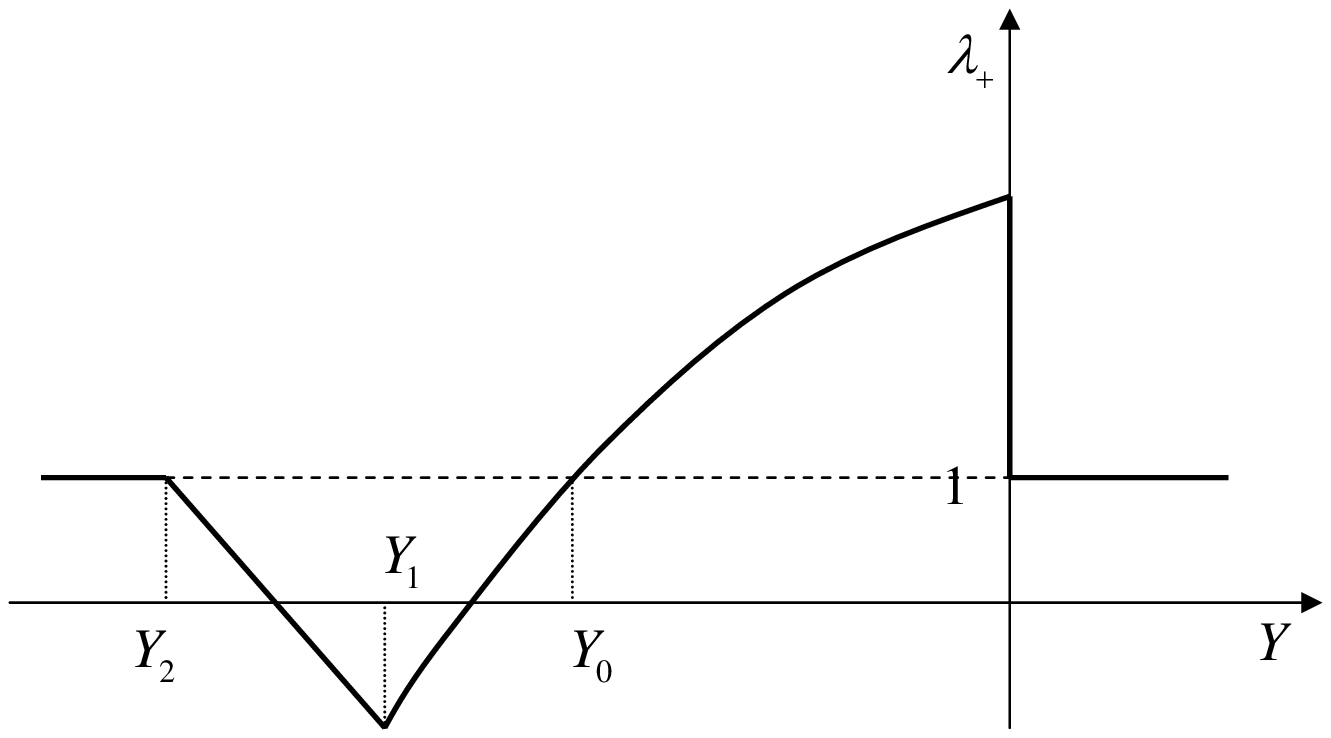}}
\caption{Sketch of an asymptotically
equivalent initial condition for $\lambda_+$ in the problem of supersonic NLS
flow past  slender wing with $f'(+0) \ne 0, f'(l-0) \ne 0$.}
\label{fig6}
\end{figure}

For a more ``realistic'' wing shape (as in Fig.~1) we have $f(0)=f(l)=f'(-0)=f'(l+0)=0$
but $f'(+0) \ne 0$, $f'(l-0) \ne 0$, where $f'(a + 0)$ and $f'(a - 0)$ denote the
right and left derivatives of $f(x)$ at $x=a$ respectively.  The qualitative behaviour
of the solution remains the same but the the quantitative description undergoes
some technical modification. Indeed, one can see from (\ref{eq27}) that  the
discontinuity of the derivative $f'(\xi)$ at $\xi=l$ implies that the rear endpoint
of the wing  maps back to an interval $[Y_2,Y_1]$, where
\begin{equation}\label{Y2}
Y_2=-l \, , \quad Y_1=-(\tfrac{3}{2}f'(l-0)+1)l\, .
\end{equation}
At these points the function $\la_+(Y,0)$ assumes the values
\begin{equation}\label{Y1}
\lambda_+(Y_2,0)=1 \, , \quad \lambda_+(Y_1,0)=1+f'(l-0) \, .
\end{equation}
On the interval $[Y_1,Y_2]$ the function $\lambda_+(Y,0)$ is linear:
\begin{equation}\label{lapar2}
\la_+(Y,0) = 1-\frac{2}{3}\left(1+{Y}/{l} \right)  \quad \hbox{for} \quad  Y_2<Y<Y_1 \, ,
\end{equation}
and  $\la_+(Y_2,0)=1$ for $Y<Y_2$. One can readily see that the function $\la_+(Y,0)$
is continuous everywhere except
for the point $Y=0$ where the profile $\la_+(Y,0)$ has a discontinuity:
\begin{equation}\label{}
\la_+(0,0)= 1+f'(+0)>1, \quad \hbox{and} \quad \la_+(Y,0)=1 \quad \hbox{for} \quad Y>0 \, .
\end{equation}
We also note that one can see from (\ref{eq27}) that the point $x_0$ of the maximum of
the body profile $F(x)$ maps to the point $Y_0=f(x_0/M)-x_0/M$ on the $Y$-axis
so that $\la_+(Y_0, 0)=1$.
A typical profile of the function $\la_+(Y,0)$ is shown in Fig.~\ref{fig6}.

For convenience of  the presentation we shall assume that $\la_+(Y,0)$ is a
single-valued function on the interval $[Y_1,0]$. This implies a restriction
\begin{equation}\label{}
0 \le \frac{d \la_+(Y,0)}{d Y} < \infty \qquad \hbox{for} \quad Y_1 < Y  < 0 \, .
\end{equation}
Then from (\ref{eq27}) we obtain the corresponding condition for the function $f(\xi)$
(i.e. for the body profile $F(x)$)
\begin{equation}\label{restrict}
0 \le \frac{-f''(\xi)}{1+ \tfrac{1}{2} f'(\xi) + \tfrac{3}{2}\xi f''(\xi) } < \infty \qquad \hbox{for} \quad 0 < \xi < l\, .
\end{equation}
One should stress, that actually there is no need for the function $\la_+(Y,0)$
to be one-valued as it is a formal projection, along the characteristics of
the Riemann-Hopf equation (\ref{eq24}), of the given physical distribution
(\ref{eq23}) of $\la_+$  specified on the piston curve. So our resulting
formulae will not be restricted exclusively to the profiles satisfying the
inequality (\ref{restrict}).

We also formulate the condition necessary for the front DSW be generated exactly from
the edge of the body at $(0,0)$. This is obtained from the condition that the profile
$\la_+(Y,T)$ breaks exactly at the initial moment $T=0$ as in Fig.~6. This is clearly
the case if $f'(+0) \ne 0 $. However, if $f'(+0)=0$ then $\la_+$ can tend to unity at $Y=0$
according to the square root law, $\la_+(Y)\propto\sqrt{-Y}$ as $Y\to-0$. This means that
$\left.dY/d\la_+\right|_{T=0}=0$ at $\xi=0$ and this condition can be satisfied if
$f^{\prime\prime}(\xi)\to\infty$ but $f^{\prime\prime}(\xi)\xi\to\mathrm{const}$
where $\mathrm{const}$ can be equal to zero. In particular, such a behaviour takes place
for $f^{\prime\prime}(\xi)\propto\xi^{-\beta},\,\,0<\beta\leq1$.
If $\left.dY/d\la_+\right|_{T=0}\neq0$
then the wave breaking occurs at a later moment $T=T_b$ at the point $Y=Y_b$ which are
determined by the equations
$$
\left.\frac{dY}{d\la_+}\right|_{T_b}=0,\qquad
\left.\frac{d^2Y}{d\la_+^2}\right|_{T_b}=0.
$$
Simple calculation with the use of Eqs.~(\ref{char}) yields the equation for $\xi_b$:
$$
(f'(\xi_b)+2)f^{\prime\prime\prime}(\xi_b)=4(f^{\prime\prime}(\xi_b))^2.
$$
When $\xi_b$ is found, then $T_b$ and $Y_b$ are calculated according to the formulae
$$
T_b=\xi_b+\frac{4f^{\prime\prime}(\xi_b)}{3f^{\prime\prime\prime}(\xi_b)},\qquad
Y_b=f(\xi_b)+\frac{2(3f'(\xi_b)+2)f^{\prime\prime}(\xi_b)}{3f^{\prime\prime\prime}(\xi_b)}.
$$
In spatial $x,y$-terms this means means that the point of generation of the DSW
is detached from the obstacle.

\subsubsection*{Example: parabolic profile---a ``wing''}

We illustrate the described mapping of the body profile onto the initial profile of the Riemann invariant $\la_{+}$
by considering a parabolic ``wing'' with an opening angle $\alpha$ above the $x$-axis and the length $L$
so that the function $F(x)$ in (\ref{eq14b}) is given by
\begin{equation}\label{Fpar}
F(x)=\alpha x(1 - x/L) , \qquad 0 \le x \le L\, .
\end{equation}

Then the piston function  is
$f(T)=F(MT)=\alpha M T(1 - T/l)$, where $l=L/M$. Now, the corresponding initial distributions of the ``dispersionless'' Riemann invariants $\la_\pm $ are given by (\ref{eq27}), (\ref{eq28}), (\ref{lapar2}), that is we have the following specification for $\la_+(Y,0)$
\begin{equation}\label{lapar1}
\la_+(Y,0):  \qquad    \la_+=1+ \alpha M(1 - 2{\xi}/{l})\, , \quad Y= -\xi \left( 1 + \frac{\alpha M}{2} \left( 1
-  4\xi/l\right) \right) \quad \hbox{for} \quad Y_1<Y<0\, ,
\end{equation}
\begin{equation}\label{lalin}
\la_+(Y,0) = 1-\frac{2}{3}\left(1+{Y}/{l} \right)  \quad \hbox{for} \quad  Y_2<Y<Y_1 \, ,
\end{equation}
\begin{equation}\label{lapar3}
\la_+(Y,0)=1 \quad \hbox{for} \quad  -\infty < Y<Y_2 \ \ \hbox{and} \ \ Y>0 \, .
\end{equation}
Here (see (\ref{Y1}), (\ref{Y2}))
\begin{equation}\label{}
Y_1=-l\left(1 - \frac{3}{2} \alpha M\right)  \, , \qquad Y_2=-l \, ,
\end{equation}
so that the minimal value of $\la_+$ is $\la_+(Y_1,0)=1-\alpha M$.
Also we have
\begin{equation}\label{}
Y_0=f(l/2)-l/2= -l/2(1-\alpha M/2)\, , \qquad  \ \la_+(Y_0, 0)=1\, .
\end{equation}
and  $\la_+(0,0)=1+\alpha M$.
We note that condition (\ref{restrict}) is satisfied if the denominator
in it is negative as $\xi  \to l$  which implies
a simple inequality $\alpha M < 2/7$.

\section{Flow past straight corner}

\subsection{Analytical theory}

We first consider a model problem of the flow past an infinite straight
corner specified by the function
\begin{equation}\label{concave}
  F(x)=0, \quad \hbox{for} \quad
 x<0 ; \qquad F(x)=\alpha x \quad \hbox{for} \quad x \ge 0\, ,
\end{equation}
\begin{figure}[ht]
\centerline{\includegraphics[width=8cm]{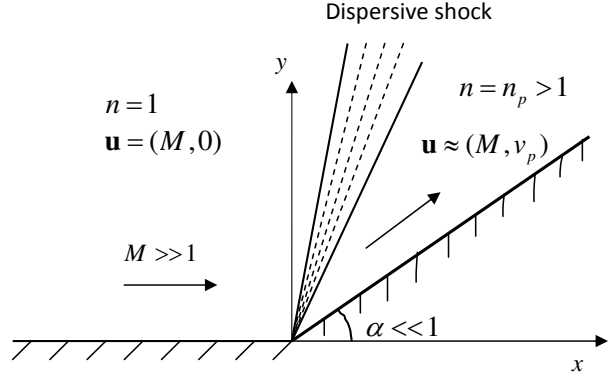}}
\caption{Supersonic dispersive flow past concave corner.
The flow speed and density in the region between
the corner and DSW are $v_p=\alpha M$, $ n_p=(v_p+2)^2/4$.}
\label{fig7}
\end{figure}
where $\alpha >0$ is some constant  (see Fig.~\ref{fig7}).
To apply the piston approximation we
need to assume that $\alpha \sim M^{-1} \ll 1$ so the piston curve in
(\ref{eq23}) is $f(T)=\alpha M T$, i.e. the piston speed is
\begin{equation}\label{}
v_p=\frac{df}{dT}=\alpha M,
\end{equation}
Using (\ref{np}) we obtain that in the piston approximation the flow parameters
in the region between the  body surface and the DSW
are simply
\begin{equation}\label{intermediate}
u=M\, , \qquad v=v_p=\alpha M\, , \qquad n=n_p=(M\alpha +2)^2/4 \, .
\end{equation}
Now, using (\ref{eq27})--(\ref{eq28}) we obtain the asymptotically equivalent
initial conditions for
the NLS equation (\ref{eq16}) in terms of
$\la_{\pm}$ (see (\ref{eq20}))
\begin{equation}\label{init1}
\begin{split}
T=0: \quad \la_-=-1\, ; \\ \qquad \la_+=A^+ \ge 1 \quad \hbox{for}
\quad Y \le 0 \, ,  \quad \hbox{and} \quad  \la_+=1, \quad \hbox{for} \quad Y>0\, .
\end{split}
\end{equation}
Here
\begin{equation}\label{A+}
A^+=1+ \alpha M\, .
\end{equation}
Of course, in the context of the flow past body problem, the solution is
defined only for $ Y \ge Y_p=\alpha M T$.
Thus, the problem essentially reduces to the much studied problem of
the decay of an initial discontinuity for the defocusing NLS
equation (see \cite{gk87,eggk95}) with some restrictions for
the domain of the solution.

The relevant  modulation solution has the form of a centered characteristic fan
\begin{equation}\label{mod1}
\la_1 = -1\, , \quad \la_2=1 \, , \quad \la_4=A^+ \, ,
\end{equation}
\begin{equation}\label{mod2}
\frac{Y}{T}=V_3(-1, 1, \la_3, A^+)
\end{equation}
or explicitly (see (\ref{vi}))
\begin{equation}\label{sss}
\frac{Y}{T}=\frac12 (\la_3+1+\alpha M)
-\frac{(1+\alpha M -\la_3)(\la_3-1)\K(m)}{(\la_3-1)\K(m)- \alpha M \E(m)} \, ,
\end{equation}
where
\begin{equation}\label{ml}
m=\frac{2(1+\alpha M - \la_3)}{\alpha M (\la_3+1)},
\end{equation}
\begin{figure}[ht]
\centerline{\includegraphics[width=8cm]{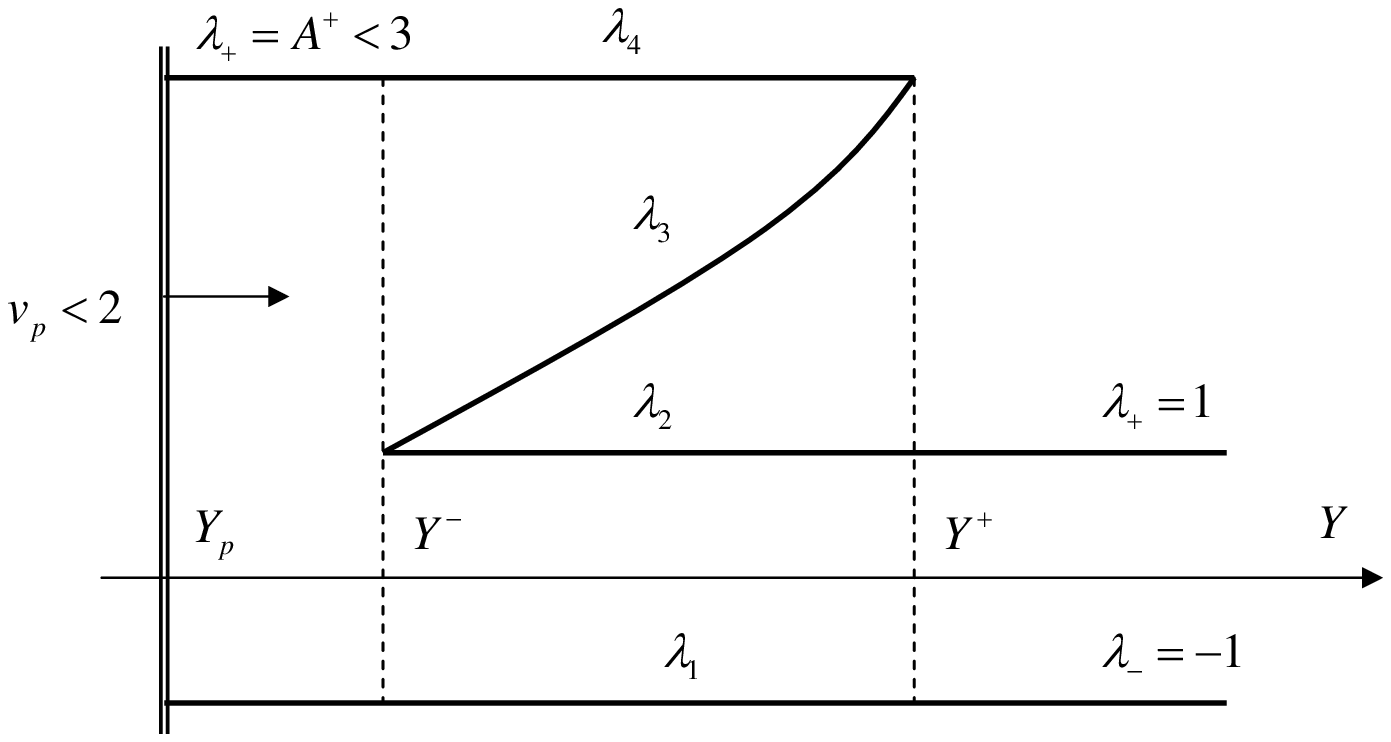} \quad
\includegraphics[width=7.5cm]{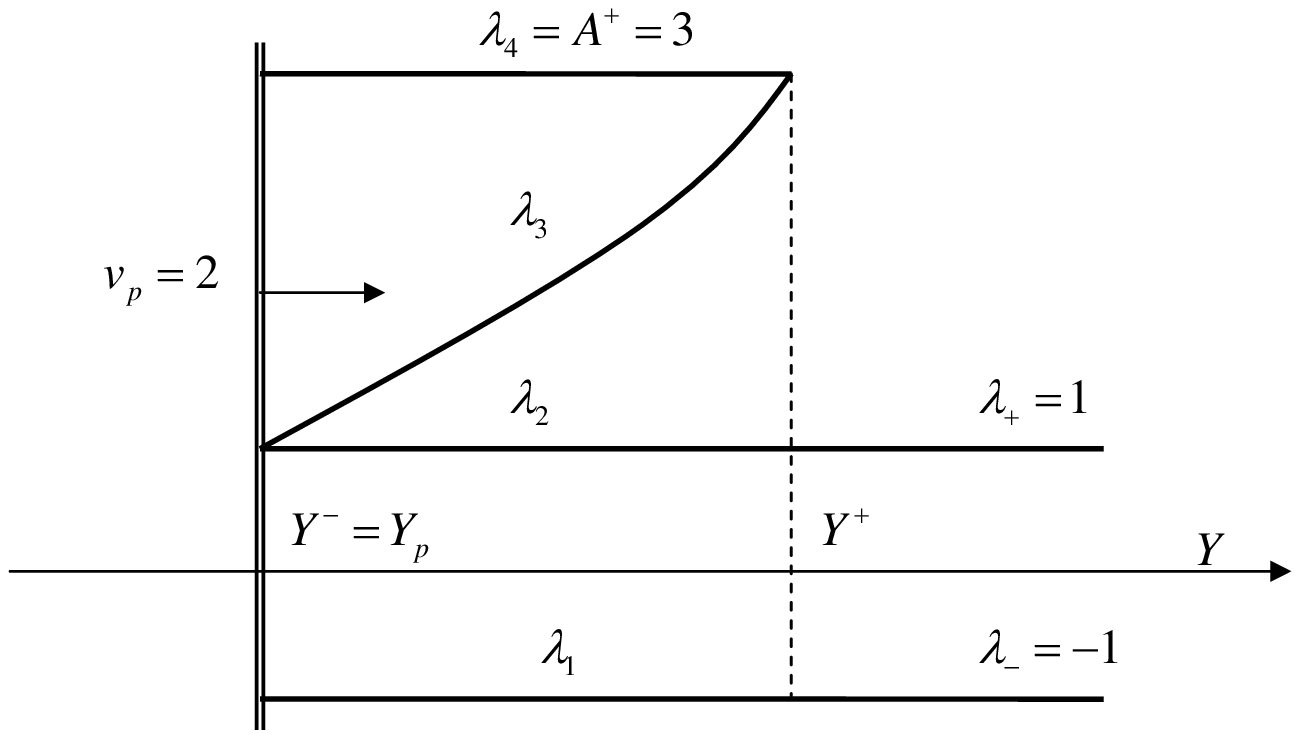}}  \caption{Behaviour
of the Riemann invariants in the similarity DSW at some $T>0$. The
vertical double line at $Y=Y_p$ marks the ``piston'' (local body surface)
position. Left: subcritical piston speed, $v_p <2$;
Right: critical piston speed $v_p=2$---formation
of a ``black'' soliton at the trailing edge $Y^- = Y_p$}
\label{fig8}
\end{figure}

The DSW is confined to an expanding region
$\tau ^- T \le Y \le  \tau ^+ T$, where the ``speeds'' (or rather slopes)
$\tau^{\pm}$ of the edges  are calculated from (\ref{sss}) as the boundary
values of the similarity variable
$\tau = Y/T$:
 \begin{equation}\label{tr1}
\tau^-=\tau(m=1)= 1+\frac{\alpha M}{2} \, ,
\end{equation}
\begin{equation}\label{lead1}
\tau^+= \tau(m=0)= \frac{2(\alpha M)^2 +4(\alpha M) + 1}{1 + \alpha M} \, .
\end{equation}
We note that the trailing edge speed $\tau^-$ is translated into the slope of
the oblique dark soliton forming at the DSW edge facing  the body surface
\begin{equation}\label{s-}
s^-=\tau^-/M=1/M+ \alpha/2 \, .
\end{equation}
The amplitude of this soliton is
\begin{equation}\label{a-}
a^-=(\la_4-\la_3)(\la_2-\la_1)=2(A^+-1)=2 \alpha M \, .
\end{equation}
The density profile in the soliton is defined by formula (\ref{sol})
in which one substitutes  the pedestal $n_0=n_p=(2+ \alpha M)^2/4$,
the amplitude  $a= 2\alpha M$ and the `velocity' $U_s=\tau^-$.
We note that it follows from (\ref{s-}) that in the flow past corner problem
the oblique dark soliton is formed {\it outside} the ``conventional''
Mach cone defined by the slope
$1/\sqrt{M^2-1} \approx 1/M$. This is in an apparent contrast with the wave
pattern described in \cite{egk06,egk07} where the oblique dark
solitons were shown to be
necessarily formed  {\it inside the Mach cone}. However,
in \cite{egk06,egk07} the oblique dark solitons were considered to be
generated by the {\it point-like} obstacle. In that case  the background flow density and, correspondingly, the sound speed were equal to unity so that  the Mach number
in the background flow was everywhere equal $M$.  In the present case
of the flow past corner, the oblique dark soliton is generated on a non-unity
background  which results in a different value of the local sound speed and,
therefore, in the changed definition of the Mach cone which is now specified
by the local Mach number $M_l=M/\sqrt{n_p}$. As a result, the adjusted Mach angle
becomes $1/\sqrt{M_l^2 -1} \approx 1/M+\alpha/2$ which coincides with the oblique
soliton slope (\ref{s-}). Thus, in the supersonic NLS flow past a corner an oblique
dark soliton is formed along the
actual Mach line. This agrees with the result in \cite{gk87} where it was
shown that the trailing dark soliton in the DSW moves with the sound speed.
Since $s^+=\tau^+/M>s^-$ the implication of this fact is that the  DSW is
located entirely outside the Mach cone.

The schematic behavior of the Riemann invariants in the modulation solution
(\ref{mod1}), (\ref{sss}) is shown in Fig.~8.
As was already mentioned, in the context of the supersonic flow past body
(or the piston) problem  the solution (\ref{sss}) is defined only for
$Y \ge v_p T=\alpha M T$.
Then from the condition  $\tau^-=\alpha M$ we obtain the critical value
$v_p=\alpha M = 2$ for which the greatest dark soliton in the DSW is generated
right at the
body surface (see Fig.~8, right panel). Incidentally, this value of $v_p$ also
implies that the density at the minimum of this greatest soliton turns zero
which constitutes the
appearance of a {\it vacuum point} at the trailing edge of the DSW \cite{eggk95}.
Indeed, from (\ref{eq013}), the minimum of the density in the travelling
wave solution is $n_{min} =\frac14(\la_4-\la_3-\la_2+\la_1)^2 $.
Substituting (\ref{mod1}) we obtain the distribution for the local minima of $n$ in
the DSW,
\begin{equation}\label{nmin}
n_{\min}(m) =\frac14(\alpha M -\la_3(m)-1)^2 \, ,
\end{equation}
where the dependence $\la_3(m)$ is given by (\ref{ml}). Then the requirement
that $n_{min}(1)=0$ immediately yields $\alpha M =2$.
Generally, setting in (\ref{nmin})  $\alpha M >2$ one gets from $n_{\min}(m)=0$
\begin{equation}\label{m*}
 m^*=\frac{4}{(\alpha M)^2} <1\, ,
\end{equation}
i.e. the vacuum point occurs inside the DSW.  Since at the vacuum point we
have $\lambda_1=-1$, $\lambda_2 = 1$, $\lambda_3=\alpha M -1$, $ \lambda_4=1+\alpha M$
the phase velocity (\ref{eq016}) at the vacuum point is $U^*=\alpha M$, i.e.
is equal to the piston velocity.
This seems to imply that the DSW gets attached to the piston and is realized
only partially with the modulus ranging from $0$
to $m^*$.
\begin{figure}[ht]
\centerline{\includegraphics[width=7.5cm]{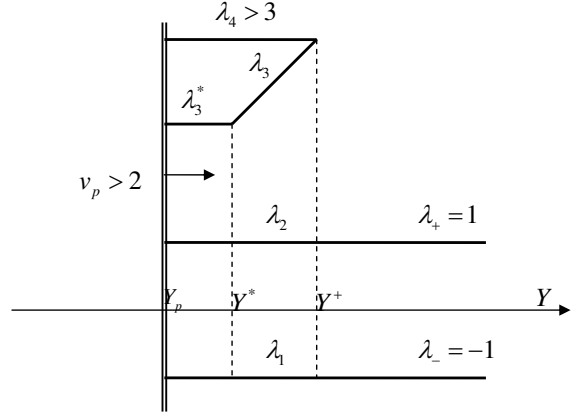}}
\caption{Riemann invariants for the supercritical
``piston velocity'', $v_p>2$. The periodic transition wave occupies the
region $[Y^p, Y^*]$ where $Y^*=\tau^* T$ (see (\ref{taus})).}
\label{fig9}
\end{figure}
However,  it turns out that one cannot attach the partial DSW directly to the
piston, instead, one should introduce an additional periodic ``transition wave''
with $m=m^*$ between the  DSW and the piston \cite{ha08}. As a matter of fact,
the vacuum point is present at each period of this transition wave.
The generation of a {\it nonmodulated} periodic nonlinear wave in the piston
problem can be explained in the
following way. The vacuum point phase velocity $U^*$  coincides with the
{\it nonlinear group
velocity} only when $m^*=1$, i.e. for the dark soliton, when  the multiple
characteristic velocity $V_2=V_3=U$ is nothing but the soliton speed.  Generally,
for $m^* \ne 1$
one has $V_2 \ne V_3 \ne v_p$ and, therefore, one should introduce a reflected wave.
As a result of the DSW reflection  from the piston (body surface) one generally would get
a two-phase wave region characterized by six Riemann invariants (see,
for instance \cite{bk06,ha07} for the corresponding Whitham equations),
with two of them
changing. However, the requirement of self-similarity of the modulation
solution in the problem of the supersonic flow past {\it straight} corner
imposes the restriction that only one Riemann invariant can change.
This implies that the two varying Riemann invariants in the general
two-phase modulated solution must coincide with each other with the
consequence that there is only one oscillating phase described by
{\it four} distinct {\it constant} Riemann invariants  (the varying
multiple Riemann invariant
can be ignored as it essentially describes the propagation of the
vanishing amplitude linear wave packet against the cnoidal wave background).
So, as a result of nonlinear wave interaction one effectively gets a
non-modulated finite-amplitude periodic wave, which in the present
context can be viewed as a nonlinear standing wave. The behaviour of
the Riemann invariants in the described ``supercritical''
modulation solution is schematically shown in Fig.~9. The region of the
intermediate periodic wave expands with the speed
\begin{equation}\label{taus}
\tau^*=V_3(-1, 1, \alpha M -1, 1+ \alpha M)=\alpha M -
\frac{2(\alpha M -2)\K(m^*)}{(\alpha M - 2)\K(m^*) - \alpha M \E (m^*)}\, ,
\end{equation}
where $m^*= 4/ (\alpha M)^2$ (see (\ref{m*})). The transition wave amplitude is
(see (\ref{amp}))
\begin{equation}\label{}
a^*=(\la_4 - \la_3)(\la_2 - \la_1)=4\, ,
\end{equation}
and it does not depend on the value of $\alpha M >2$. The latter only
affects the transition wave width $\tau^* T$ (and the local wave shape, via $m^*$).
This is in striking contrast with the classical dissipative piston problem,
where the density jump across the shock  increases without limitations as
the piston velocity grows.
We also note that in the physical $xy$-plane the transition wave is located
between the body surface and the centered line with the slope $s^*=\tau^*/M$.
An explicit expression for the oscillating density profile in the transition
wave in physical $xy$ plane is
\begin{equation}\label{transition}
n = 4\,{\rm sn}^2\left(\alpha M (y-\alpha x)
,m^*\right) \, ,
\end{equation}
and the wavelength of the transition wave (\ref{transition}) in any $x$-section
is calculated as $\mathfrak{L}$ (see (\ref{eq017}))  and is given by
\begin{equation}\label{}
\mathfrak{L}=\frac{2}{\alpha M} \K(m^*)\, .
\end{equation}
In conclusion we note that the described transition wave solution
actually represents part of the special similarity solution obtained in
\cite{eggk95} as a particular case in the decay of an initial discontinuity
problem (see the case 6 in the full classification of \cite{eggk95}).
The transition wave was also very recently observed in \cite{ha08} in the
numerical simulations of
the dispersive piston problem.

\subsection{Comparison with numerical solutions}

\begin{figure}[ht]
\centerline{\includegraphics[width=6cm]{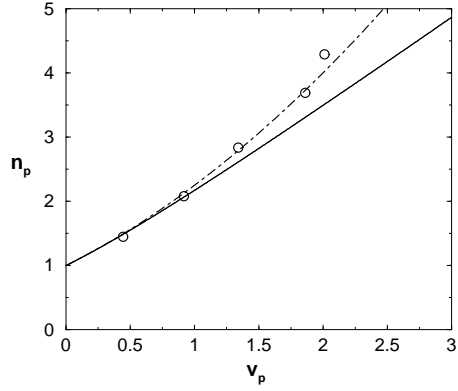}}
\caption{Dependence of the
flow density $n_p$ near the corner surface on the vertical velocity component
$v_p$ for the flow
with $M=10$. Dashed line: dependence (\ref{intermediate}) obtained in the
dispersive piston approximation. Circles: data obtained from the full 2D numerical
solution. Solid line:
$n_p(v_p)$ - dependence for the classical dissipative piston problem}
\label{fig10}
\end{figure}

We have performed two series of numerical simulations. Firstly, we constructed
the full  unsteady numerical solutions for the 2D NLS flow past corner for $M=10$ and
different values of the corner angle $\alpha$.   Secondly, we performed parallel numerical simulations of the associated 1D piston problem (\ref{nls-1D})--(\ref{eq19a})
for the corresponding values of the piston velocity $v_p=M\alpha$.
In both cases we have used finite-difference codes with the impenetrability
condition $\psi=0$ at the body
(piston) surface. The results of the numerical simulations have been then
compared with the analytical modulated solution obtained in the previous subsection.
We note that from
the numerical point of view it is more convenient to perform simulations for a
symmetric wedge with the opening angle $\alpha$ above the $x$-axis and to use the
wave pattern in the upper half-plane for the comparison.

First, we have made a comparison for the DSW transition condition, which in our
case is expressed by the formula (\ref{intermediate}) specifying the parameters
of the constant flow between the DSW and the corner surface provided the oncoming
flow parameters are $u=M$, $v=0$, $n=1$.
The comparison of the dependence  $n_p(v_p)$ for $M=10$ with the numerical data
for the density in the 2D flow past a wedge  is shown in Fig.~\ref{fig10}.
This is also compared with the dependence $n_p(v_p)$ for the classical dissipative
piston problem  for polytropic gas with $\gamma = 2$, corresponding to the
dispersionless limit of the NLS equation. The corresponding classical piston
jump condition is specified by the equation $v_p = (n_p - 1)\sqrt{\frac{1+n_p}{2n_p}}$.
The comparison is made for $M=10$. One can see excellent agreement between the
analytical dispersive piston curve and the numerical data obtained from the full
2D simulations of the flow past corner problem. At the same time one can see
noticeable departure of the dispersive piston curve from the classical piston
curve. The numerical simulations data and the dispersive piston curve split at
$v_p=2$ (i.e. at $\alpha = v_p/M=0.2$, which also agrees with our solution as
for $M\alpha>2$ the theory predicts the formation of a transition wave so that
the region of a constant flow between the corner and DSW disappears.

\begin{figure}[ht]
\centerline{\includegraphics[width=8cm]{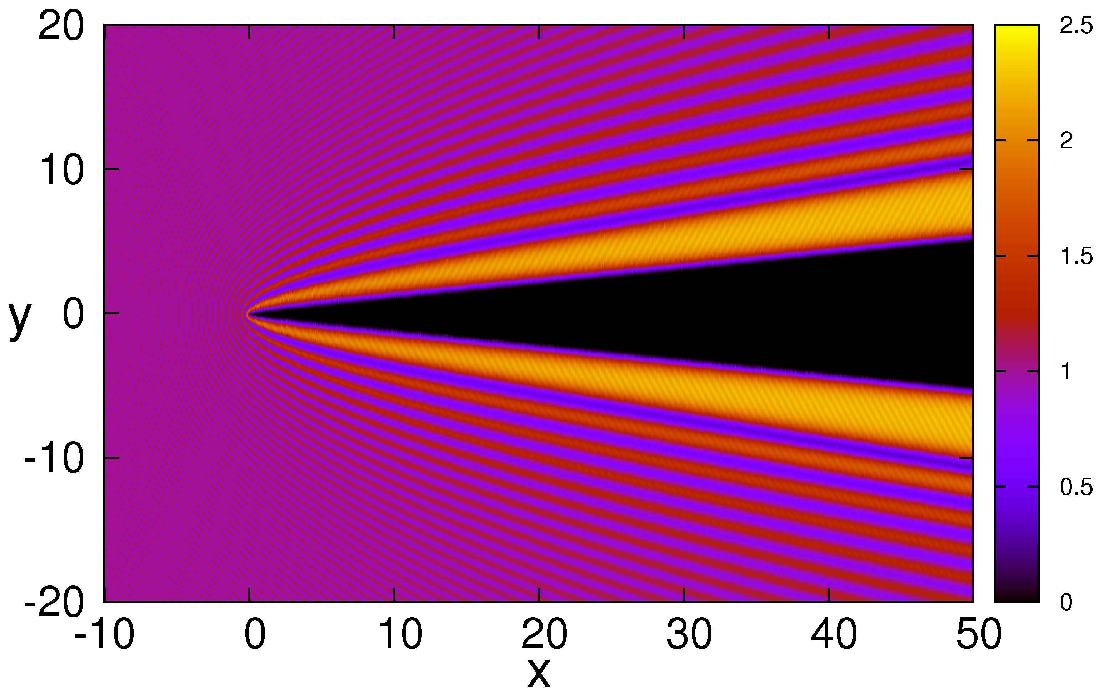} \qquad
\includegraphics[width=7cm]{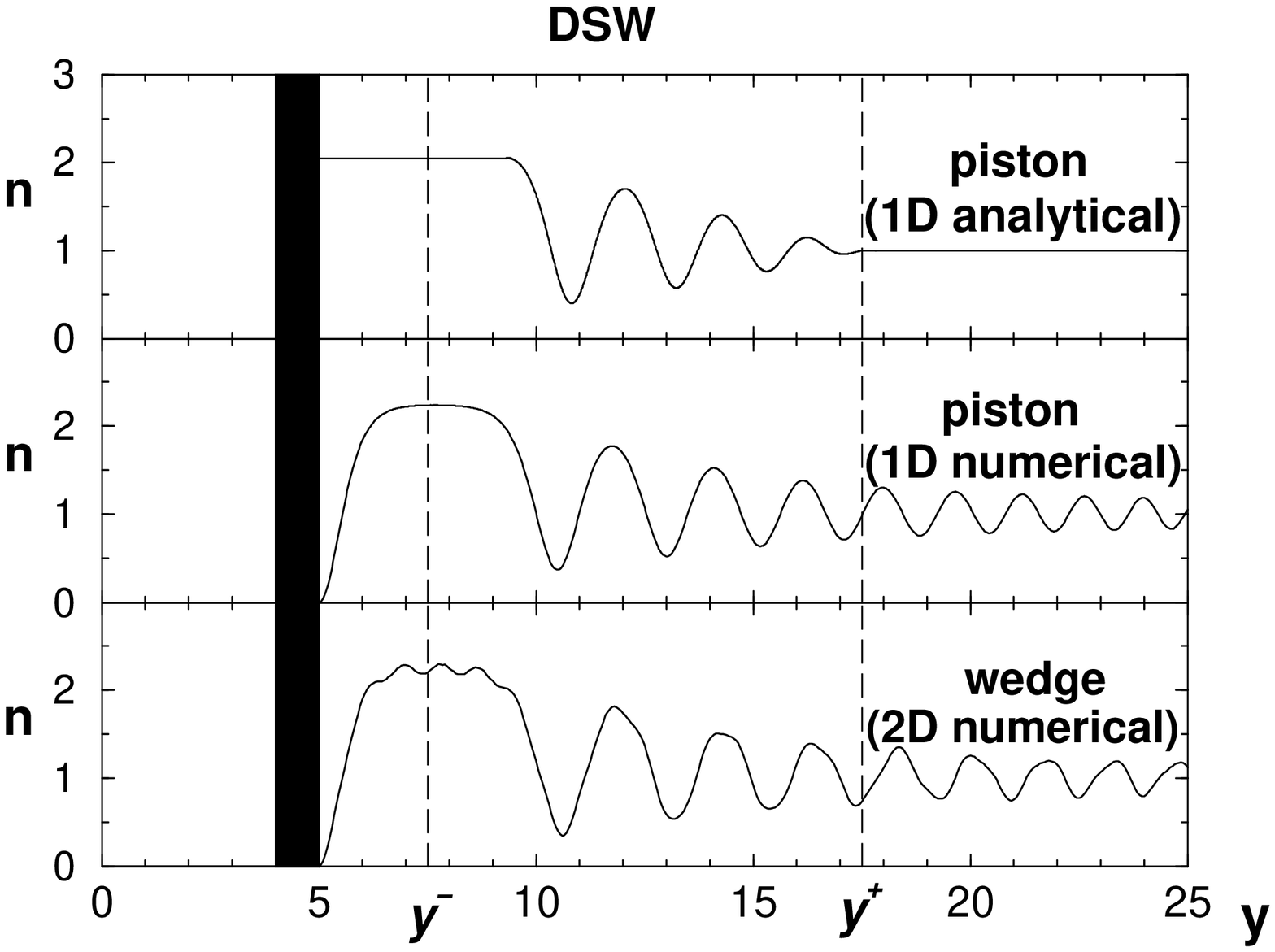}}
\caption{Left (color online): 2D density plot for of the supersonic ($M=10$)
NLS flow past a wedge with the opening angle above the $x$-axis $\alpha=0.1$.
Right: density profile $n(y)$ at  $x=50$. Top: analytical modulated solution;
middle: numerical solution of the associated 1D piston problem; bottom: 1D cut of the full
2D solution at $t=15$. Points $y^-$ and $y^+$ mark the boundaries of the
DSW predicted by the modulation solution. The body surface (piston) is
located at about $y = 5$.}
\label{fig11}
\end{figure}
\bigskip
\begin{figure}[ht]
\centerline{\includegraphics[width=8cm]{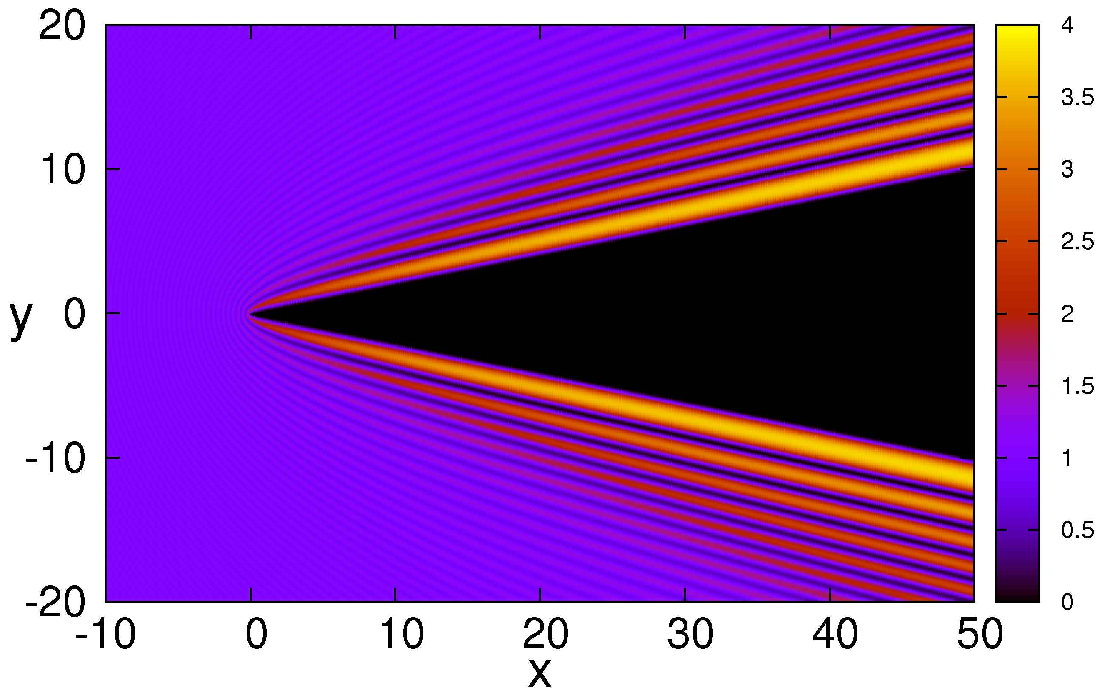} \quad
\includegraphics[width=7cm]{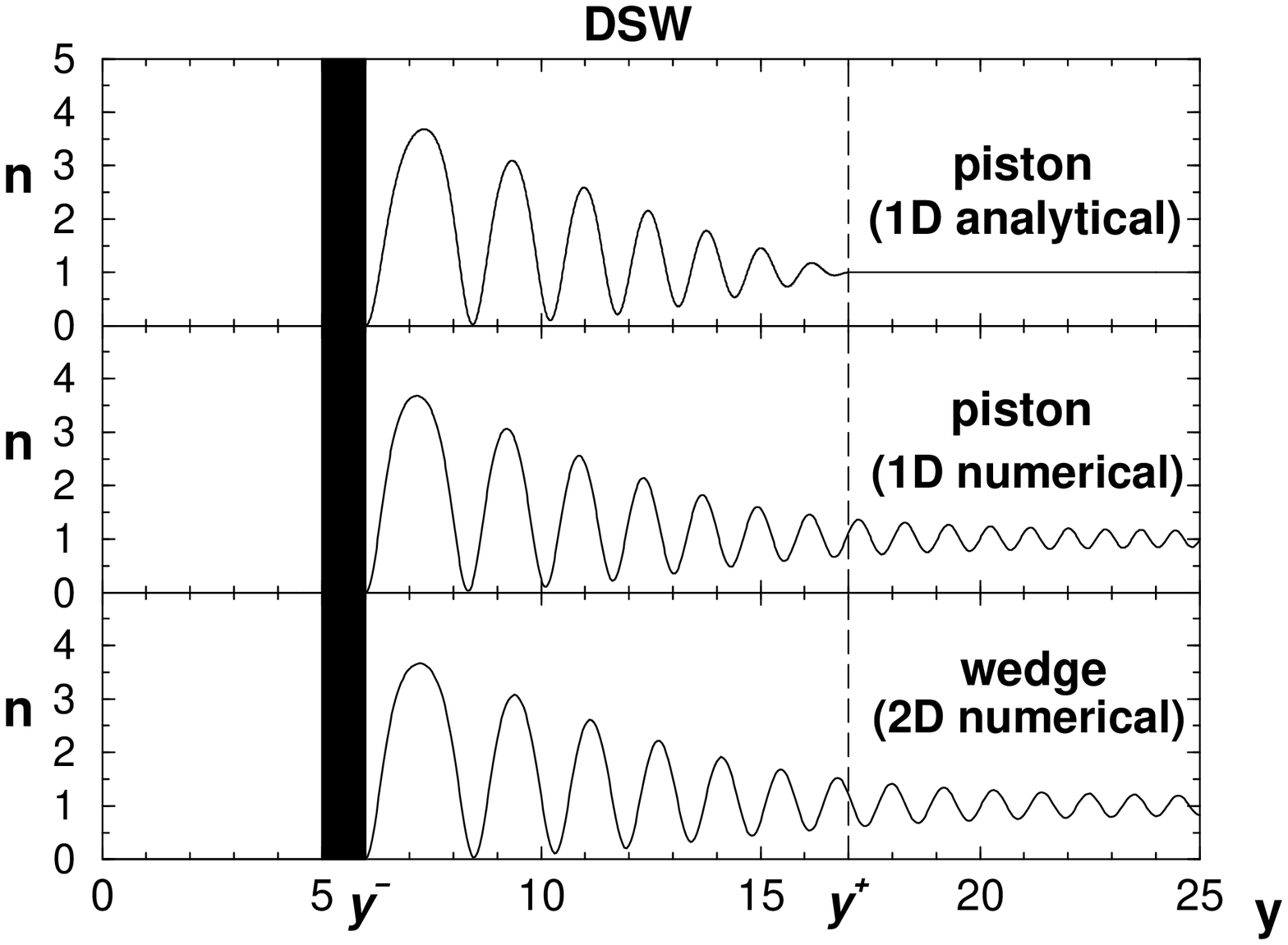} }
\caption{Left (color online): 2D density plot for  the supersonic ($M=10$) NLS flow
past a wedge with the opening angle above the $x$-axis $\alpha=0.2$.
Right: density profile $n(y)$ at  $x=30$; top: analytical modulated solution;
middle: numerical solution of the associated 1D piston problem; bottom:
1D cut of the full
2D solution at $t=15$. Points $y^-$ and $y^+$ mark the boundaries of the
DSW predicted by the modulation solution. The body surface (piston) is located
at about $y = 6$.}
\label{fig12}
\end{figure}
\bigskip
\begin{figure}[ht]
\centerline{\includegraphics[width= 9cm]{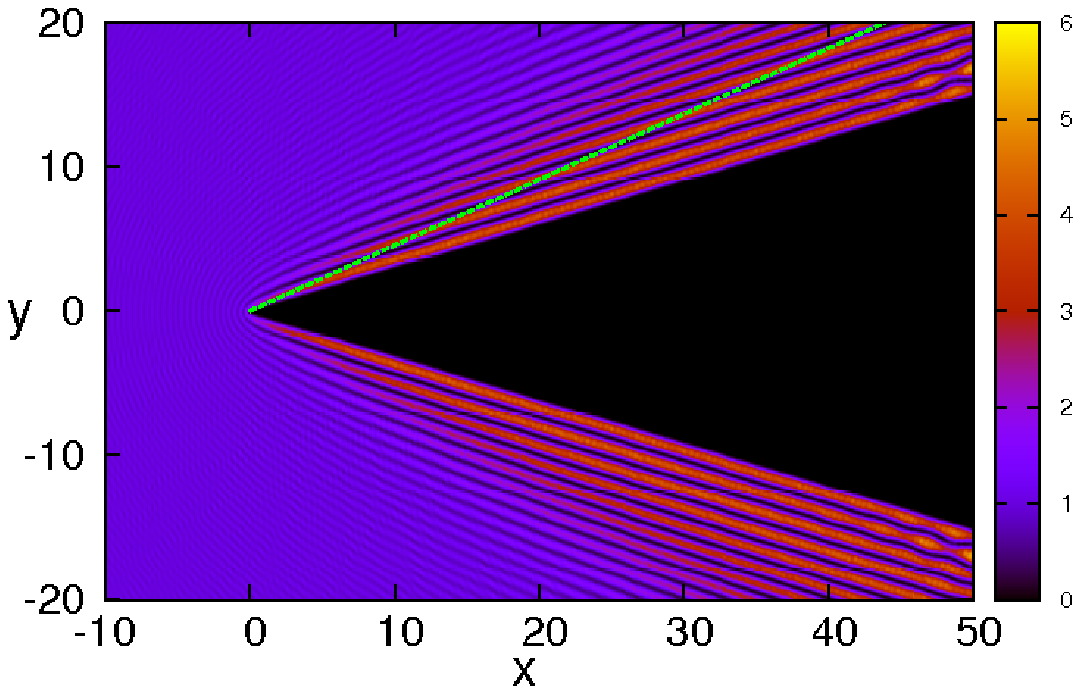} 
\includegraphics[width=7cm]{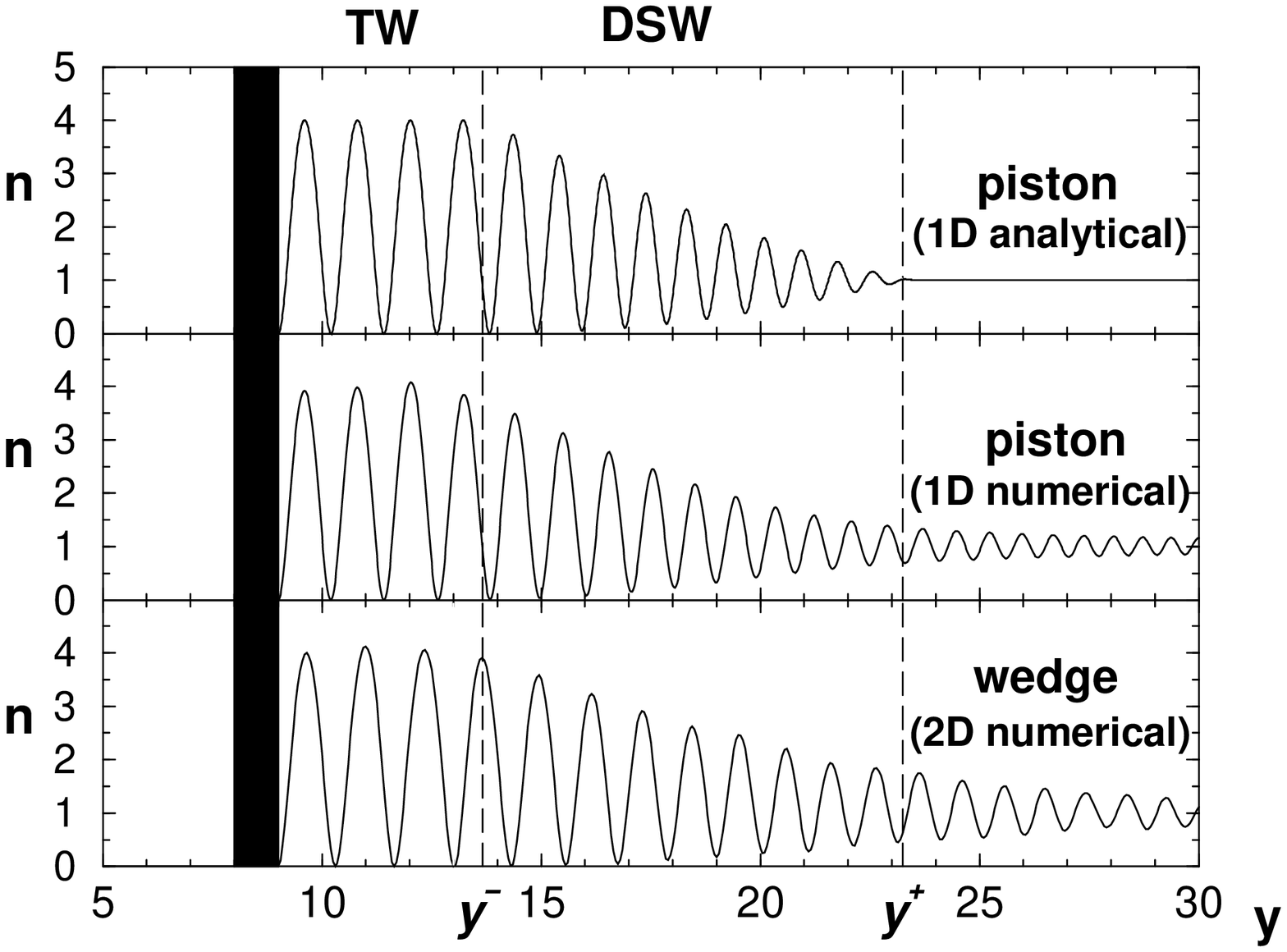}}
\caption{Left (color online): 2D density plot for the supersonic $M=10$ NLS flow
past a wedge with the opening angle above the $x$-axis $\alpha=0.3$. The dashed line marks the end
of the transition wave predicted by the theory.
Right: density profile $n(y)$ at  $x=30$; top: analytical modulated solution;
middle: numerical solution of the associated 1D piston problem; bottom: 1D cut of the full
2D solution at $t=15$. Points $y^-$ and $y^+$ mark the boundaries of the DSW
specified by the modulation solution. The body surface (piston) is located at
about $y = 9$.}
\label{fig13}
\end{figure}
\bigskip
\begin{figure}[ht]
\centerline{\includegraphics[width=5cm]{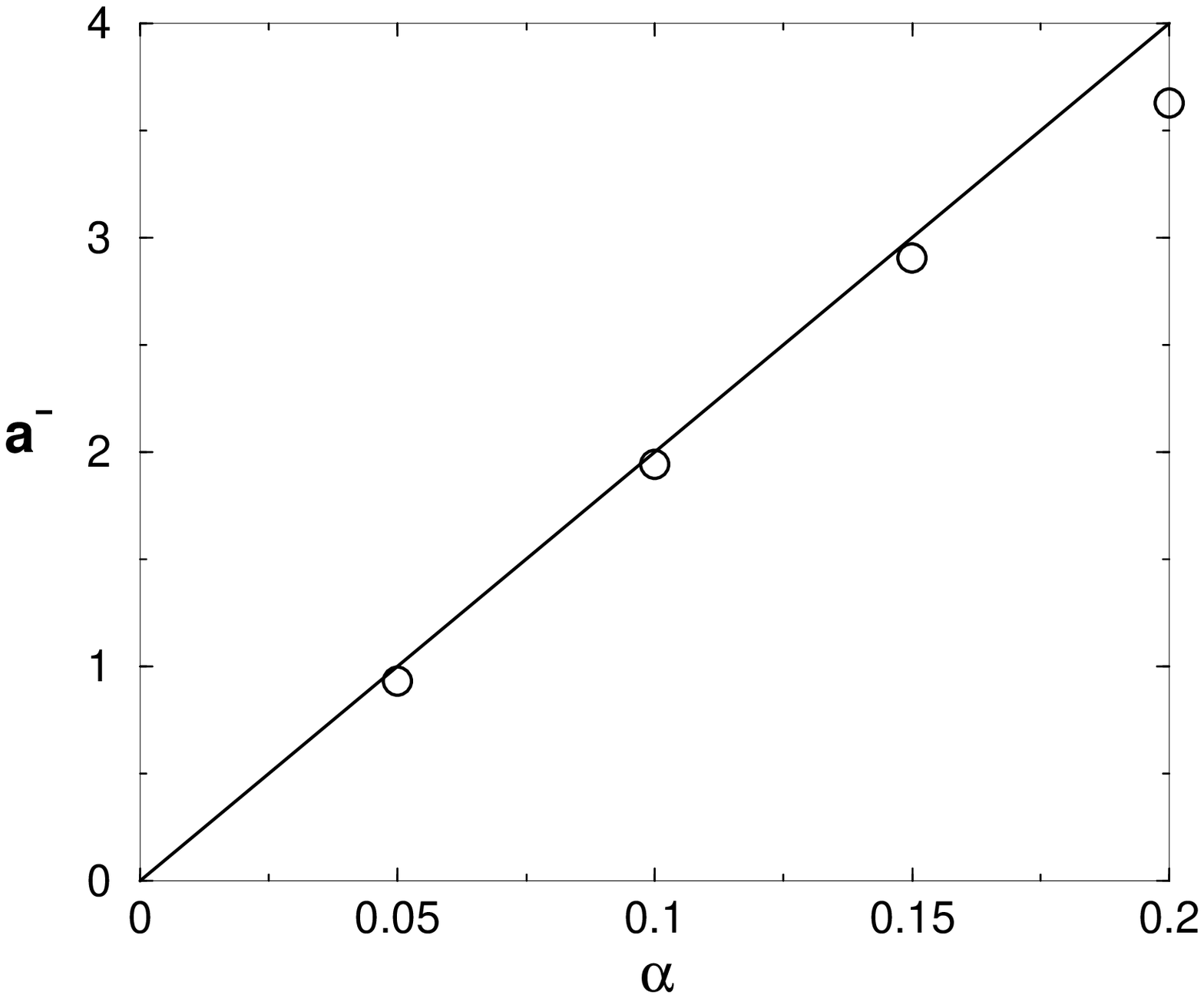} \qquad  \quad \includegraphics[width=5cm]{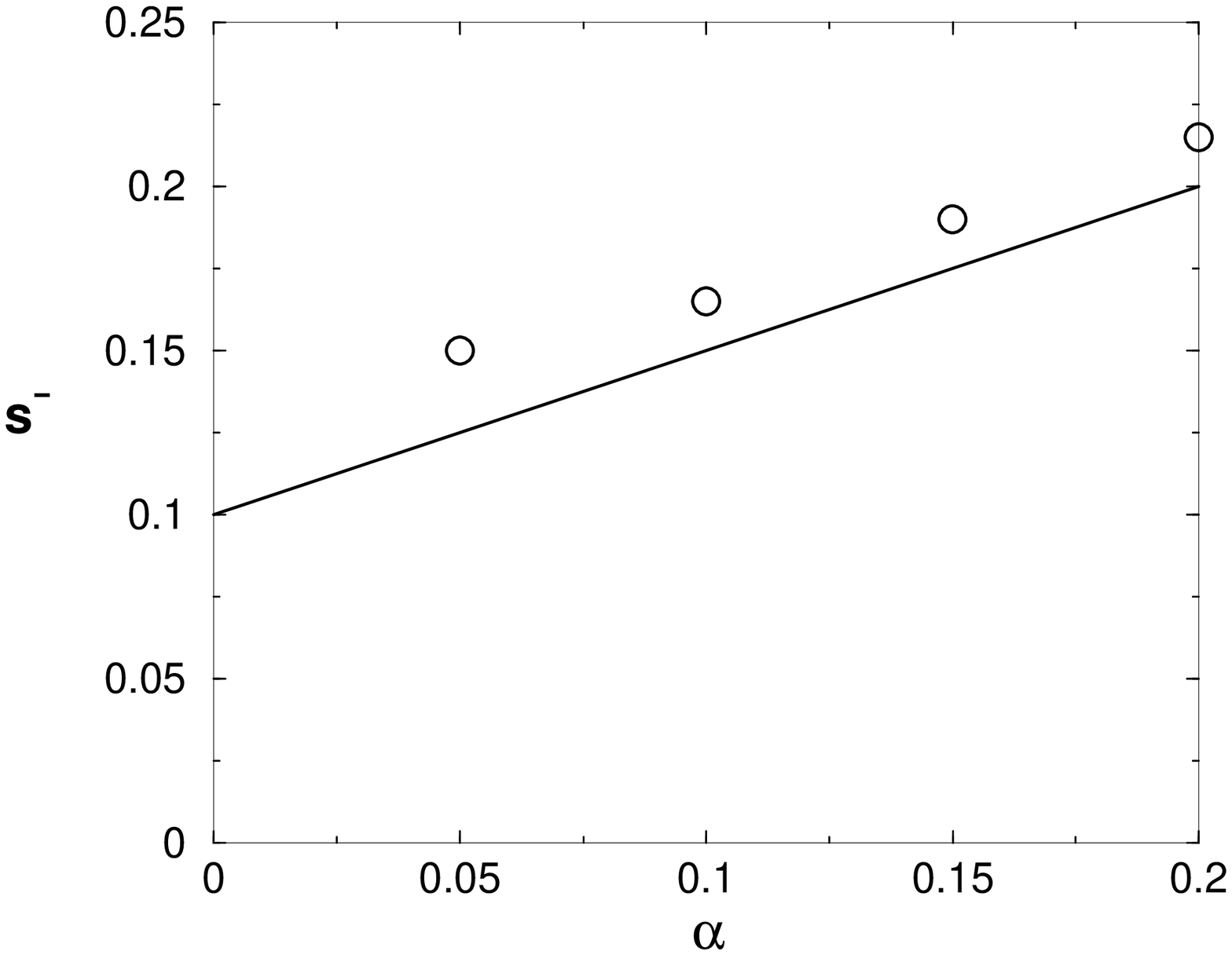}}
\caption{Parameters of the first dark soliton in the DSW as functions of the corner angle.
Left: the soliton amplitude $a^-$; Right: the soliton slope $s^-$.
The numerical values are taken at $x=50$}
\label{fig14}
\end{figure}

In Figs.~\ref{fig11}--\ref{fig13} the 2D density plots (left) and 1D
cross-section density profiles (right) are presented for the flows with
$M=10$ past corners with $\alpha = 0.1, 0.2$ and $0.3$ respectively.
The analytical solutions (top panel) are compared with the numerical
solution of the asymptotically equivalent 1D dispersive piston problem
(middle panel) and with the $x$-section of full 2D solution. One can
see that 1D numerical dispersive piston solutions agree remarkably well
with the results of the full 2D simulations.
The agreement between the analytical solutions and numerical simulations is
also very good in the DSW region (we note that the exact position of the
wave in the analytical solution is determined up to a characteristic coherence
length (soliton half-width) as the initial phase $\theta_0$ in (\ref{eq013})
is not defined by the modulation theory). The predicted occurrence of the
vacuum point at the body surface at $\alpha M=2$ and the generation of the
non-modulated transition wave for $\alpha M >2$ are seen very well in
Figs.~\ref{fig12}, \ref{fig13}. The predicted position $y^*=\tau^* x/M$
of the right boundary of the transition wave (see (\ref{taus})) also
agrees very well with the numerical simulations---see Fig.~\ref{fig13}.

Next, in Fig.~\ref{fig14} the comparisons for the amplitude (\ref{a-})
and slope (\ref{s-}) of the first dark soliton in the DSW as functions
of the corner angle are presented. One can see that the agreement is
excellent for the soliton amplitude and quite good for the slope.
One should emphasize that the  accuracy inherent the hypersonic approximation
(\ref{eq16}) implies that the amplitude formula  (\ref{a-}) is defined with
the accuracy $O(1/M)$ while for the slope $s^-$ given by (\ref{s-}) the
accuracy is $O(1/M^2)$. Since the slope formula is $s^-=1/M + \alpha/2$
one can expect that  a noticeable discrepancy between analytical and numerical
values for of $s^-$ may be the case for small angles, say  for
$\alpha \lesssim 0.1$. Of course, this will contribute, on level of
$O(x^*/M^2)$, to the error in the analytical determination the spatial
$y$-location of the oblique dark soliton at some $x$-cross-section made
at $x=x^*$ (as in Figs.~\ref{fig11}--\ref{fig13}). We note that
the analytically predicted soliton location is also subject to an
arbitrary, up to a typical wavelength, shift  inherent in the modulation theory.

One should also note some important feature of the wave pattern that is not captured
by the modulated solutions as seen in the right upper panels of
Figs.~\ref{fig11}--\ref{fig13}. Indeed, one can see noticeable small-amplitude
oscillations beyond the outer harmonic edge $y^+$ of the DSW (as defined by
the modulation theory). In the  theory of one-dimensional DSWs these linear
oscillations are usually ignored. However, in the considered here 2D problem
these linear oscillations represent an essential part of the observable wave pattern
(see the left panels in Figs.~\ref{fig11}--\ref{fig13}) and should be taken
into account. A similar wave distribution was considered recently in
\cite{ship1,ship2,ship3D}  in connection with the Bogoliubov-Kelvin
``ship waves'' generated by a point-like obstacle placed in the supersonic
BEC flow (see also in \cite{caruso} the discussion of the experimentally
observed patterns).
An extended modulation solution describing the combined
wave pattern including both the DSW and the linear ``ship-wave'' distribution
will be constructed in the next section.

It is worth noting that in the strongly nonlinear region
near the the wedge boundary at large $x$ one can see the oscillations of the
dark soliton crest lines (see the density plot in Fig.~\ref{fig13} (left panel)). This is the manifestation of the so-called ``snake'' instability of dark solitons
with respect to bending disturbances \cite{kp-1970,
zakharov-1975,kuztur}. However, for large enough oncoming flow velocity these
unstable disturbances are convected by the flow along solitons and hence
they become just convectively unstable in the reference frame related with
the obstacle \cite{kp08}. Therefore, for the considered here large Mach numbers, the DSW structure can be regarded as
effectively stable and thus can be treated as a modulated stationary solution of the 2D
NLS equation.

\section{Flow past wing}

\bigskip
\begin{figure}[ht]
\centerline{\includegraphics[width=10cm]{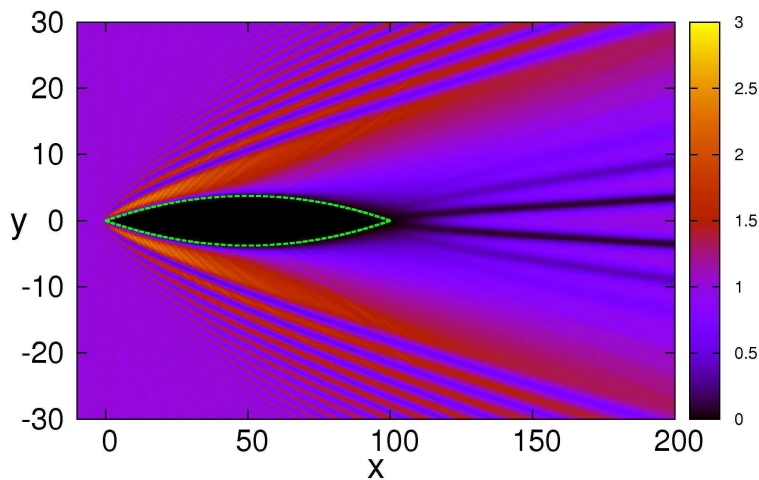}}
\caption{(color online)
Supersonic NLS flow past a wing: density plot. The oncoming (from the left)
flow speed is $M=10$. The green dashed line shows the wing profile specified
by Eq.~(\ref{Fpar})}
\label{fig15}
\end{figure}

Now we consider  supersonic flow past an extended slender finite body---a wing.
From the very beginning we assume zero attack angle so without loss of generality
we shall consider the wave pattern only in the upper half-plane. The density plot
for supersonic ($M=10$) flow past the wing having a symmetric parabolic form
specified by the function (\ref{Fpar}) with
$\alpha=0.15$ and $L=100$ is shown in Fig.~\ref{fig15}. One can see that the
wave pattern agrees with the qualitative predictions made in Section V using
the inverse scattering transform reasoning
applied to the asymptotically equivalent initial data of the type shown in
Fig.~\ref{fig5}. Indeed, one can see the front DSW, similar to that in the
straight wedge case described in Section VI and the fan of oblique dark
solitons spreading from the rear edge of the wing. Unlike the straight
wedge case, though, the front DSW is not characterized by a constant jump
of density $n$ and velocity $v$ across it,
so the depth of the oscillations decreases as the distance from the generation
point at $(0.0)$ increases. As a result, the front wave degenerates into a
small-amplitude dispersing wave with the distribution of wavecrest wave
having the form  similar to the ``ship-wave'' pattern described in
\cite{ship1,ship2}.
The length of the wing used in the simulations is not sufficiently large
to identify the details of the intermediate front and rear DSWs.
However, we shall construct full modulation solution for the front DSW and,
by considering its asymptotic behavior for large $x,y$ will derive the amplitude
and wavelength distributions applicable
to the ship-wave pattern. For the rear DSW, instead of constructing full modulation
solution, we shall take advantage of the semi-classical Bohr-Sommerfeld type
distribution \cite{jl99,kku02} for the distribution of eigenvalues in
the Zakharov-Shabat scattering problem.

The crucial difference between our consideration in this paper and the results
obtained in earlier papers \cite{egk06,ship1,ship2} on dark solitons
and ship waves is that here we
asymptotically solve the {\it boundary value problem} for the 2D NLS equation
and express the parameters of the resulting wave distributions in terms of
the initial profile, while the previous papers were concerned with the study
of certain particular solutions of the 2D NLS equation.

\subsection{Flow past  front edge of a wing}

\subsubsection{Formulation of the problem}

To model the  flow of a superfluid past the front edge of a wing we consider
the function $F(x)$ of the type shown in Fig.~16 (left) so that
$F=0$ for $x\le0$; $F'(+0)=\alpha \ll 1$, $F'(x)>0$ for $0<x \le x_0$ and
$F'(x)=0$ for $x \ge x_0$.
Then one can readily see from (\ref{eq27}) that  for highly supersonic flows
the asymptotically equivalent initial condition for
$\lambda_{+}=\tfrac{1}{2}v+\sqrt{n}$ has the
shape shown in Fig.~16 (right). The other Riemann invariant $\lambda_-=-1$
(see (\ref{eq28})). Also $Y_0=f(\xi_0)-\xi_0$ where $\xi_0=x_0/M$.
In terms of the piston problem
this corresponds to the forward motion  of the piston. The initial piston
velocity is $v_p=M\alpha$ (as in the problem of  the flow past straight corner
with the angle $\alpha$) but then the motion of the piston slows down until it
eventually stops at $T=\xi_0$. To avoid unnecessary complications connected
with the formation of the transition wave we shall assume that $\alpha M <2$.
\begin{figure}[ht]
\centerline{\includegraphics[width=6cm]{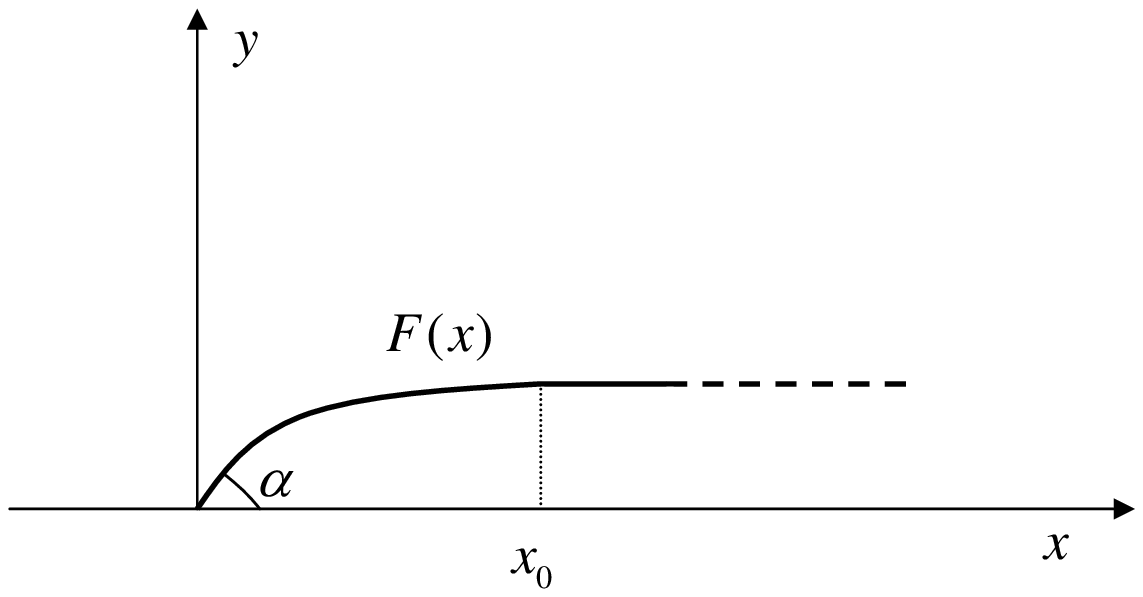} \qquad  \quad \includegraphics[width=5.5cm]{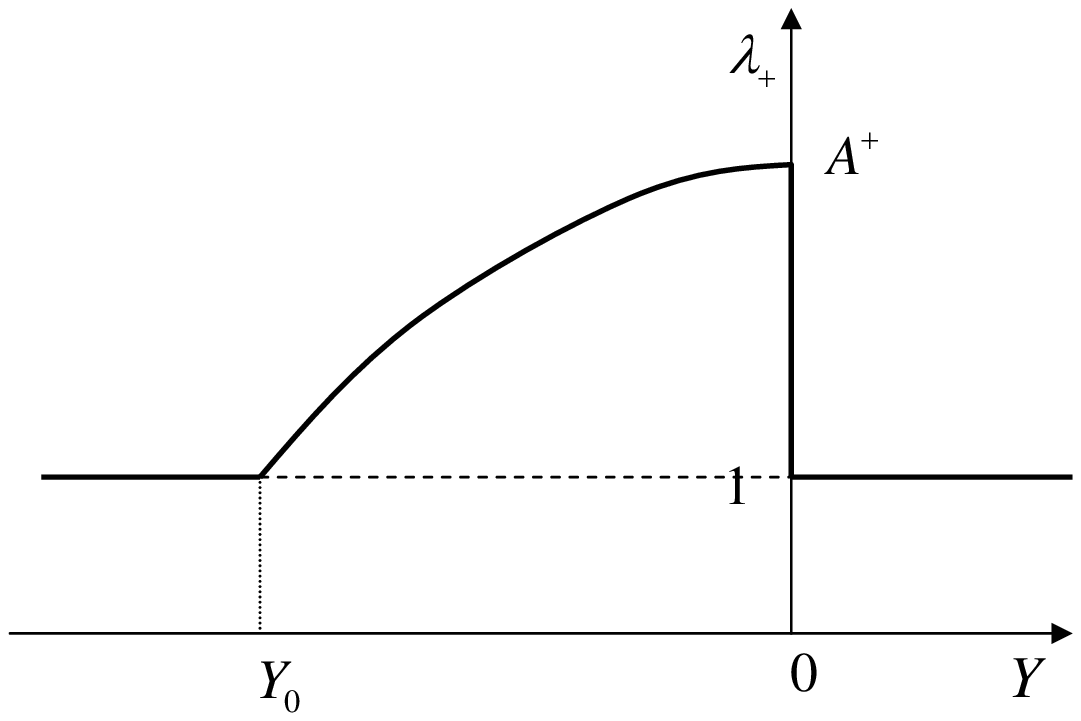}}
\caption{Left: front edge of a wing in an upper half-plane.
Right: asymptotically equivalent initial condition for $\lambda_+$.}
\label{fig16}
\end{figure}

As was explained in Section V, it is clear from the IST-based reasoning
that the disturbance caused by the front edge of the
wing in the supersonic NLS flow  will eventually (for $T \gg 1$)
transform into a linear dispersive wave radiation. However, for an intermediate
values of $T$
the spatial ``evolution'' of this disturbance leads to the formation of a
DSW having the structure similar to that generated in the  flow past straight
corner described in the previous Section. Thus, remarkably, even in this
``solitonless'' configuration, the DSW and dark solitons still  form,
albeit as an intermediate wave pattern. While in the evolutionary problems
this wave pattern is transient,
in our 2D stationary problem the intermediate front DSW
exists for all times and transforms into linear waves only at large
distances from the body.
The essential difference is that, due to the presence of the spatial
scale $x_0$, the corresponding modulation dynamics is no
longer self-similar resulting in the wave parameter variations along
the wave crest lines which now have curved geometry.
In particular, one can expect
that the oblique dark soliton forming  at the trailing edge of the DSW
will initially (i.e. at  $x=0, y=0$) have in the Whitham approximation
the slope $ s^-=\alpha/2+1/M$
and the amplitude $a^-=2 \alpha M$
as in the corresponding straight corner case, but as the distance from
the body increases, its amplitude and slope will both decrease and asymptotically one can
expect that $a^- \to 0$ and $s^- \to 1/M$ as $x \to \infty$.

\subsubsection{Modulation solution}

We use (\ref{bc1}), (\ref{la-1})  to formulate matching conditions for
the Riemann invariants in the front DSW (we shall omit the subscript
in $Y^{\pm}_f$):
\begin{equation}
\begin{array}{l}
\hbox{At} \ \ Y=Y^-(T):\qquad \la_3=\la_2\, , \ \  \la_4 = \la_+ , \  \   \la_1 = -1 \, , \\
\hbox{At} \ \ Y=Y^+(T):\qquad \la_3=\la_4 \, , \ \  \la_2 = 1,  \  \  \  \  \la_1 = -1 \, ,
\end{array}
\label{match1}
\end{equation}
where  $\la_+(Y,T)$ is the solution of  the simple-wave equation (\ref{eq24})
with the initial condition $\lambda_+(Y,0))$ defined by (\ref{eq27}).
Thus we have  for $\la_+$ an implicit representation
\begin{equation}\label{simple}
Y-{\tfrac{1}{2}(3\la_+-1)}T= w(\la_+) \,,
\end{equation}
where $w(\la_+)$ is the inverse function to $\la_+(Y,0)$ (note that for
general non-monotone initial profile $\la_+(Y,0)$ one would need to consider
two monotone branches of $w(\la_+)$ separately, but in our case of the
profile shown in Fig.~16, right,  there is only one branch specified
by (\ref{eq27})).
Schematic behavior of the Riemann invariants corresponding to the matching
conditions (\ref{match1}) is shown in Fig.~17.
\begin{figure}[ht]
\centerline{\includegraphics[width=6cm]{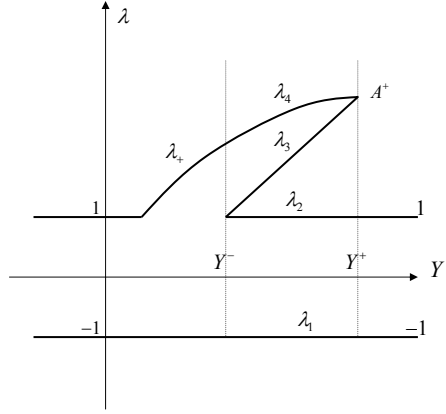} }
\caption{Schematic behavior of Riemann invariants  in the modulation
solution for the front DSW.}
\label{fig17}
\end{figure}

Importantly, the modulation problem
(\ref{eq018}), (\ref{bc3}) is no longer self-similar so one should use the
hodograph transform to solve it (see Section IV.B). First, since
$\la_1=-1$ and $\la_2=1$ satisfy both modulation equations (\ref{eq018})
and matching conditions (\ref{match1}) we have $\la_1=-1$ and $\la_2=1$
everywhere, hence there are only two modulation equations
for $\la_3$ and $\la_4$ left to solve. These transform via the
substitution (see (\ref{Ts1}))
\begin{equation}\label{Ts11}
Y-V_{3}(-1, 1,\la_3, \la_4)T=W_{3}(\la_3, \la_4)\, , \qquad Y-V_{4}(-1, 1,\la_3, \la_4)T=W_{4}(\la_3, \la_4)
\end{equation}
into a system of two linear partial differential equations for
$W_{3,4}(\la_3, \la_4)$
\begin{equation}\label{Ts21}
\frac{1}{W_4 - W_3} \frac{\partial  W _3}{\partial \la_4}= \frac{1}{V_4 - V_3} \frac{\partial  V_3}{\partial \la_4}\, , \qquad
\frac{1}{W_3 - W_4} \frac{\partial  W _4}{\partial \la_3}= \frac{1}{V_3 - V_4} \frac{\partial  V _4}{\partial \la_3}\, .
\end{equation}
The boundary conditions for equations (\ref{Ts21}) are obtained by
considering  the hodograph solution (\ref{Ts11}) at the free boundaries
$Y^{\pm}$ and applying
to it the matching conditions (\ref{match1}).  At the trailing edge
$Y=Y^-(T)$ we have $\la_3=1$ and $V_4(-1,1,1,\la_4) =\tfrac{1}{2}(3\la_4-1) $
(see (\ref{m1})) so that the
second equation (\ref{Ts11}) becomes
\begin{equation}\label{TsTrail}
 Y- \tfrac{1}{2}(3\la_4-1) T=W_{4}(1, \la_4) \, .
\end{equation}
Comparing (\ref{TsTrail})  with the simple-wave solution (\ref{simple})
at $Y=Y^-$, where $\la_4=\la_+$ (see (\ref{match1})) we obtain the
boundary condition for (\ref{Ts21})
\begin{equation}\label{bp1}
W_4(1, \la_4)=w(\la_4) \, .
\end{equation}
At the leading edge $Y=Y^+$ we have $\la_3=\la_4 $ and (see (\ref{m02}))
\begin{equation}\label{mult1}
 V_3(-1, 1, \la_4, \la_4)= V_4(-1, 1, \la_4, \la_4 )=2\la_4-1/\la_4
 \equiv V^*(\la_4) \, .
\end{equation}
The multiple characteristic velocity $V^*(\la_4)$ determines the speed of
the leading edge (see (\ref{mult})).  Note that since $\la_4>1$ we always
have $\partial_4 V^*>0$
and for the initial data of the type shown in Fig.~16 (right) the
speed of the leading edge at $T=0$ is $V^*(A^+)$ which is the greatest
characteristic  speed in the system. Therefore
the characteristic $dY/dT=V^*(A^+)$  is not intersected by other
characteristics of the family $dY/dT=V_4$ so the equation of
the leading (harmonic, $m=0$) edge of the DSW is simply
\begin{equation}\label{leadline}
Y- (2A^+-1/A^+) T=0\, .
\end{equation}
Substituting $A^+=1+\alpha M$  we obtain the slope of the outer
(facing the oncoming flow) edge of the DSW in the physical $x,y$ plane  as
\begin{equation}\label{slopelead}
s^+=\frac{y}{x}=\frac{2(\alpha M)^2 +4(\alpha M) + 1}{M(1 + \alpha M)} \, .
\end{equation}
Thus, the outer edge of the spatial DSW is determined  by the opening
angle $\alpha$ alone and does not depend on the specific
body contour (indeed, (\ref{slopelead}) coincides with the expression
for the  slope of the outer edge of  the DSW generated by the flow
past straight infinite corner with the angle $\alpha$ (see (\ref{lead1})).

The obtained leading edge equation (\ref{leadline}) should be consistent
with the hodograph solution (\ref{Ts11}) considered at $m=0$.
Then comparing (\ref{Ts11}) for $\la_3=\la_4=A^+$  with (\ref{leadline})
we get
\begin{equation}\label{bp2}
W_3(A^+, A^+)=W_4(A^+, A^+)=0\, .
\end{equation}
Equations (\ref{bp1}) and (\ref{bp2}) provide boundary conditions for
the linear system (\ref{Ts21}).
Using  transformation (\ref{scalar1}), which in our case  is
explicitly represented as
\begin{equation}\label{scalar11}
W_i(\la_3, \la_4)=g+2 [V_i(-1, 1, \la_3, \la_4) -  \tfrac12 (\la_3+\la_4) ] \frac{\partial g}{\partial \la_i}\, , \qquad i=3,4\, ,
\end{equation}
the system (\ref{Ts21}) is further reduced to the EDP equation (\ref{edp})
for the potential
function $g(\la_3, \la_4)$ (see Section IV.B for details). Now we need to translate
boundary conditions (\ref{bp1}) and (\ref{bp2})  for the hodograph equations
(\ref{Ts21}) into the boundary conditions for the EDP equation.

Substituting (\ref{scalar11})  into (\ref{bp1})   we obtain
\begin{equation}\label{odeg}
g(1, \la_4)+2(\la_4-1)\frac {\partial g(1, \la_4) }{ \partial \la_4}=w(\la_4).
\end{equation}
Ordinary differential equation (\ref{odeg}) is readily integrated to give
\begin{equation}\label{bc3}
g(1, \la_4)=-\frac{1}{2\sqrt{\la_4- 1}}\int \limits _{\la_4} ^{A^+}
\frac{w(z)}{\sqrt{z - 1}}dz \, ,
\end{equation}
where we have chosen the constant of integration such that $g(1, A^+)=0$
(this requirement is not essential).
Now, without loss of generality we  take the  general solution (\ref{gs})
of the EDP equation (\ref{edp}) in an equivalent form
\begin{equation}
\label{gs1} g(\la_3, \la_4)=\int \limits _{1} ^{\la_3} \frac{\phi_1(\la) d
\la}{\sqrt{(\la_4- \la)(\la_3-\la)}} + \int \limits _{\la_4} ^{A^+}
\frac{\phi_2(\la) d \la}{\sqrt{(\la - \la_4)(\la- \la_3)}} \, ,
\end{equation}
 where $\phi_{1,2}(\la)$ are arbitrary (generally,
complex-valued) functions. From (\ref{bp2}) we obtain $\phi_1(\la)
\equiv 0$. Next, applying boundary condition (\ref{bc3}) we arrive
at the integral Abel equation (see, for instance,  \cite{abramowitz}) for
$\phi_2(\lambda)$
\begin{equation}\label{abel}
\int \limits _{\la_4} ^{A^+} \frac{\phi_2(\la) d \la}{\sqrt{(\la -
\la_4)(\la- 1)}}=-\frac{1}{2\sqrt{\la_4- 1}}\int \limits _{\la_4}
^{A^+} \frac{w(z)}{\sqrt{z - 1}}dz \, .
\end{equation}
The solution to equation (\ref{abel}) (obtained via the inverse  Abel
transform for $\phi_2(\la)/\sqrt{\la - 1}$) is
\begin{equation}\label{phi2}
\phi_2(\la)=\frac{1}{2\pi\sqrt{\la- 1}}\int \limits _{\la} ^{A^+}
\frac{-w(z)}{\sqrt{z - \la}}dz \, .
\end{equation}
Substituting $\phi_1=0$ and $\phi_2(\lambda)$ given by (\ref{phi2}) into the general
solution (\ref{gs1}) and changing the order of integration we obtain a compact
representation for
the solution to the EDP equation for the problem of the flow past front
part of the wing
\begin{equation}\label{gs2}
g(\la_3, \la_4)=\frac{1}{\pi\sqrt{\la_4- 1}}\int \limits _{\la_4}
^{A^+} \frac{-w(z)}{\sqrt{z -
\la_3}}\K\left(\frac{(z-\la_4)(\la_3-1)}{(z-\la_3)(\la_4-1)}\right)dz \, ,
\end{equation}
where $\K(z)$ is the complete elliptic integral of the first kind.
Now, formulae (\ref{gs2}), (\ref{scalar11}) and (\ref{Ts11}) provide the
exact implicit modulation solution to the NLS initial value problem with the
initial profile of the type
shown in Fig.~16 (right). Strictly speaking, one should now show  that
the obtained solution is global, i.e. the mapping $(\la_3, \la_4) \mapsto  (Y,T)$
specified by
(\ref{Ts11}),  (\ref{scalar11}) and (\ref{gs2}) is invertible for all $T$. However,
instead of giving full mathematical proof of the invertibility of the hodograph
transform (\ref{Ts11})  for our solution,  it seems to be more instructive just
to show that the obtained modulation solution has a physically meaningful
asymptotic behaviour for $T \gg 1$,
which, apart from providing us with the useful information about distributions
of physical parameters at large distances from the body, will  be a convincing
enough indication that the solution  is valid for all $T$.

To study the long-time behaviour of the obtained solution we  express $T$
from the hodograph formulae (\ref{Ts11}), (\ref{scalar11}) as
\begin{equation}\label{T}
T=\frac{W_3-W_4}{V_4-V_3}=2\frac{[V_3 -  \tfrac12 (\la_3+\la_4) ]
\tfrac{\partial g}{\partial \la_3} - [V_4 -  \tfrac12 (\la_3+\la_4) ]
\tfrac{\partial g}{\partial \la_4} }{V_4 - V_3}\, ,
\end{equation}
where we have denoted $V_j \equiv V_j(-1,1,\la_3, \la_4)$, $W_j
\equiv W_j(\la_3, \la_4)$ for brevity.
Next, substituting the solution (\ref{gs2}) into (\ref{T}) we obtain an
explicit expression for $T$ in terms of $\la_3$, $\la_4$. Analysis of this
expression shows that
$T \to \infty$ implies $\la_3 \to \la_4$. Since the wave amplitude
$a=2(\la_4-\la_3)$ and
the modulus $m=2a/[(\la_4-1)(\la_3+1)]$ (see (\ref{eq015})) we obtain
that $a \to 0$, $m \to 0$ as $T \to \infty$ everywhere except for a small
vicinity of the trailing edge point where $\la_3 \to 1$, so one has $a\to 0$
but $m \to 1$. That means that the front DSW  asymptotically transforms into
a vanishing amplitude linear wave packet (the asymptotic behaviour of the
trailing soliton will be considered separately).

Indeed,  a straightforward analysis shows that  for the obtained solution
$W_{3,4}( \la_3, \la_4)/T  \to 0$  as $\la_3 \to \la_4$ (i.e. $T \to  \infty$).
Then we have
from the hodograph solution (\ref{Ts11}) to leading order in $1/T$
\begin{equation}\label{group}
T \gg 1 : \qquad \frac{Y}{T} \cong V_3(-1,1,\la_4,\la_4)=2\la_4 -\frac{1}{\la_4}\, .
\end{equation}
Next, expanding (\ref{T}) for small $\la_4 - \la_3 \ll 1$ we obtain, after
some algebra,  the leading order asymptotic behaviour (provided $\lambda_4$
is not too close to $1$)
\begin{equation}\label{aT}
a \cong \frac{1}{T^{1/2}} A(\la_4)\, ,
\end{equation}
where
\begin{equation}\label{A}
A(\la_4) = 4\sqrt{\frac{\la_4(\la_4+1)}{\pi(2\la_4^2+1)}\ } \ (\la_4-
1)^{1/4} \left(\int \limits _{\la_4} ^{A^+}\frac{-w(z)}{\sqrt{z - \la_4}}dz \right)^{1/2}.
\end{equation}
Asymptotic behaviour (\ref{group}), (\ref{aT}) is consistent with the
modulation theory for linear waves (see for instance \cite{wh74}). Indeed,
using the definitions of the phase velocity $U$ (\ref{v}) and the wavenumber
$k=2\pi/\mathfrak{L}$ where $\mathfrak{L}$ is the wavelength (\ref{eq017})
one can see that in the linear limit $\la_3 \to \la_4$
one has
\begin{equation}\label{kU}
\la_3=\la_4: \qquad k=\frac{2\pi}{\mathfrak{L}(-1,1,\la_4,\la_4)}=2\sqrt{\la_4^2-1}\, , \ \ \hbox{so}\quad U= \la_4=\sqrt{1+k^2/4}\, ,
\end{equation}
the latter being the linear dispersion relation of the NLS equation (\ref{nls-1D}), $\omega_0(k)=kU=k\sqrt{1+k^2/4}$.
Then the right-hand side of (\ref{group}), $2\la_4 - 1/\la_4 = (1+k^2/2)/\sqrt{1+k^2/4}$,
is nothing but the linear group velocity $\omega'_0(k)$ so (\ref{group})
is simply the similarity solution of the kinematic modulation equation
$k_T+\omega_0'(k) k_Y=0$ for the linear wave packet.

The asymptotic  behaviour (\ref{aT}) of the amplitude is also consistent
with the linear wave energy conservation law
$\partial_T a^2 + \partial_Y (\omega'_0(k) a^2)=0$. However, the function
$A(\la_4)$ defining the relation of the
asymptotic wave amplitude distribution with the body profile cannot be determined
within the linear theory and requires the full nonlinear analysis presented here.
One should mention that the eventual transformation of the front DSW into a linear
radiation also agrees with the
general reasoning of the inverse scattering transform method as the initial
conditions of the type described in the beginning of this section (see also Fig.~10)
represent a ``solitonless potential'' having only a continuous spectral
component. Indeed, formulae (\ref{group})--(\ref{A})
could also be obtained via the inverse scattering transform formalism but the
employed here method via the solution of the Whitham equations appears to be
more direct and efficient for the purpose.

Finally, using (\ref{group})--(\ref{kU}) we represent the asymptotic amplitude
and wavenumber distributions  explicitly in terms of the original spatial
variables $x,y$ to perform later a comparison with the numerical simulations
of the 2D NLS flow past slender obstacle:
\begin{equation}\label{frontas}
x,y \gg 1: \qquad  a \cong \left(\frac{M}{x} \right)^{1/2} A \left(\frac{\tau+\sqrt{\tau^2+8}}{4}\right),
\quad k \cong \frac{1}{2}\sqrt{\left(\tau+\sqrt{\tau^2+8}\right)^2-16},
\quad \hbox{where} \ \ \tau=M\frac{y}{x} \, .
\end{equation}
One can see that $k \to 0$ as $\tau \to 1$, the latter being the Mach line
in the hypersonic approximation.
This will also emerge in the next section where the trailing (soliton) edge
of the front DSW will be shown to asymptotically
approach the Mach line as $x \to \infty$. A remarkable feature of the asymptotic
wavenumber $k$ distribution in (\ref{frontas})
is that it  does not depend on the shape and size of the body (provided the
conditions of applicability of the piston approximation are satisfied).
This will allow us to construct an analytic description of the universal ``ship-wave''
pattern generated in the supersonic NLS flow past slender bodies.
\begin{figure}[ht]
\centerline{\includegraphics[width=5cm]{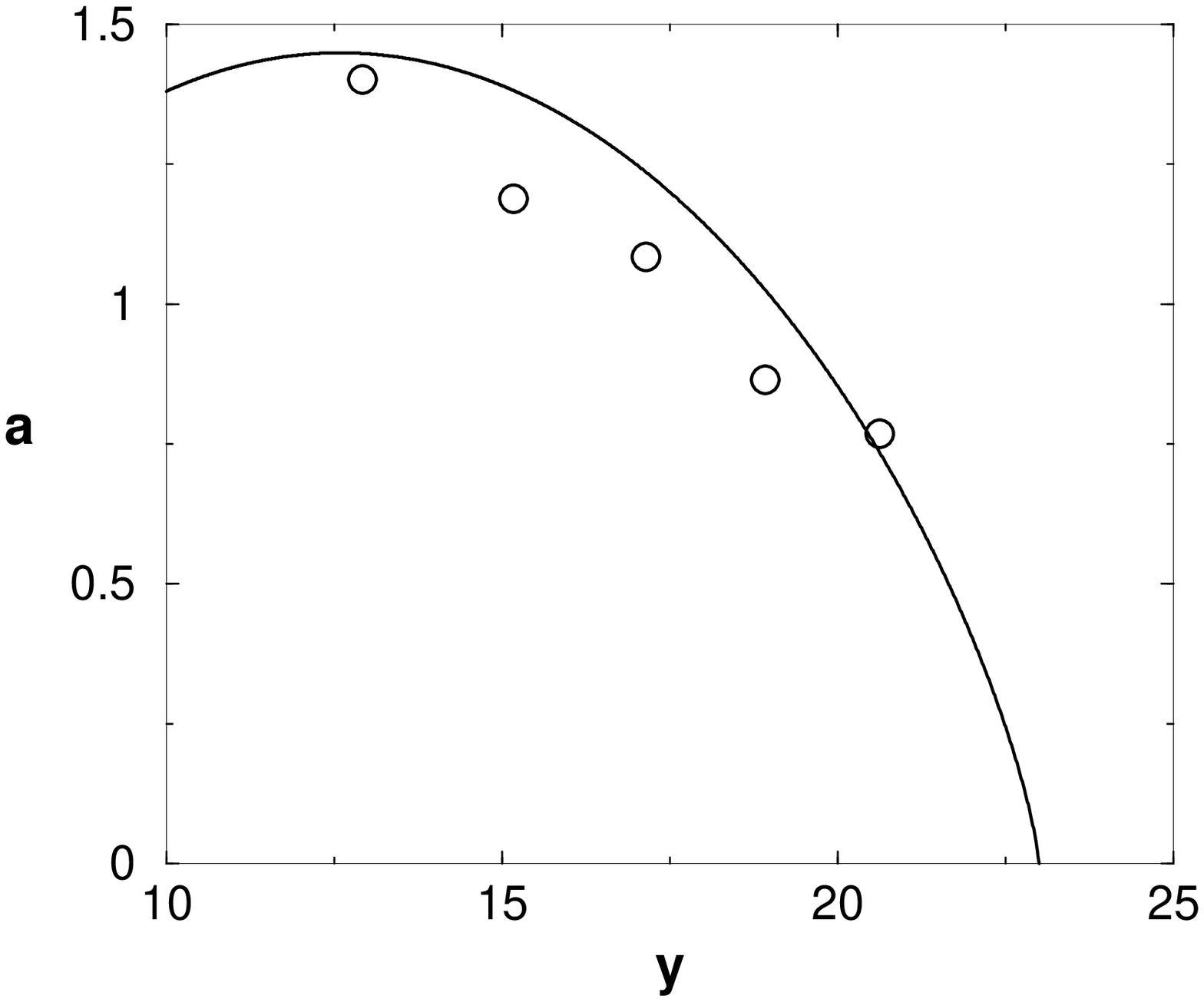} \qquad  \qquad \includegraphics[width=5cm]{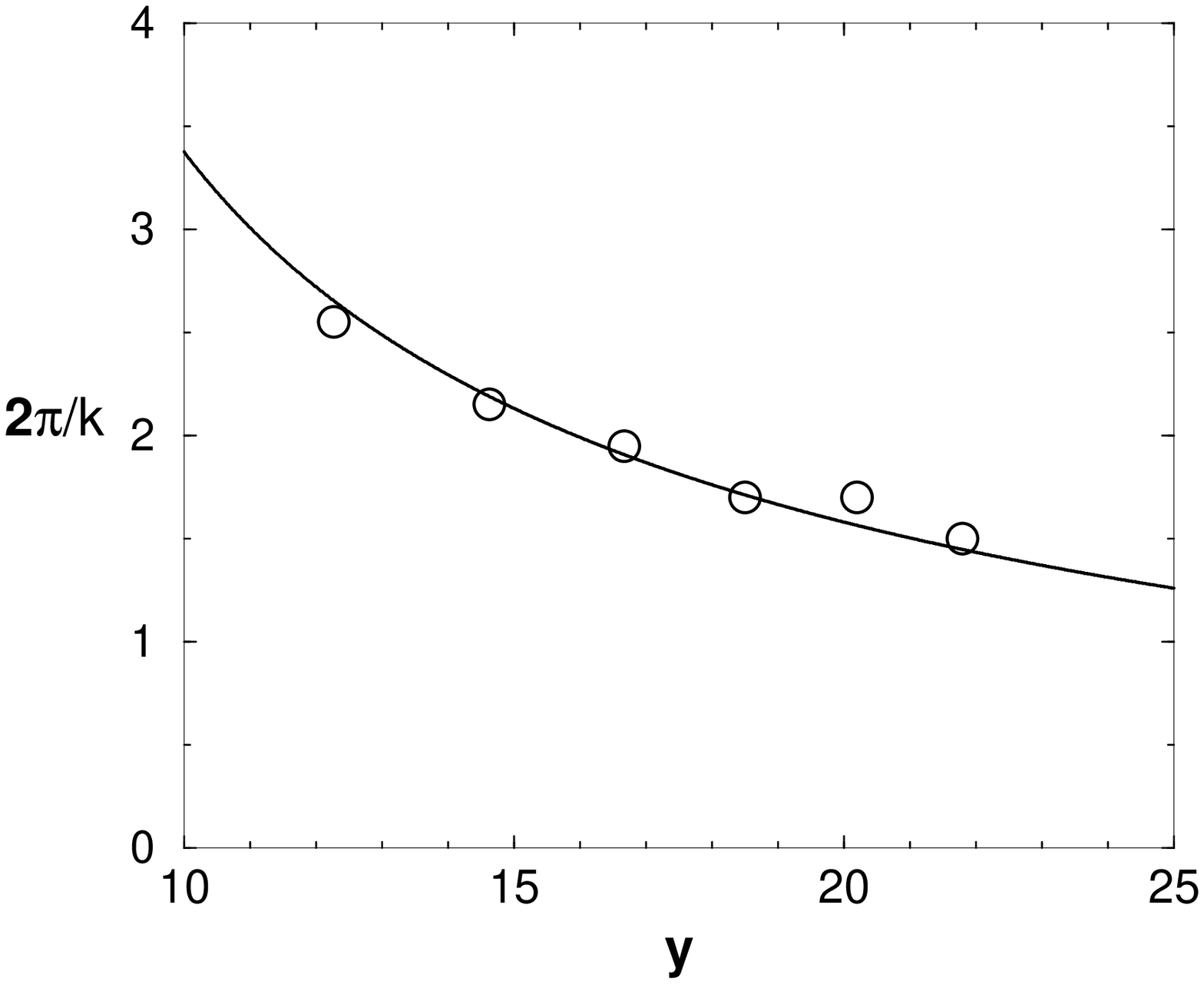}}
\caption{Comparisons for the asymptotic ($(x/M) \gg 1$) amplitude $a$ (left)
and wavelength $2\pi/k$ (right) distributions in the front DSW generated by
the wing with a parabolic profile (\ref{Fpar}) ($L=100$,  $\alpha= 0.15$)
placed in the supersonic NLS flow with the Mach number $M=10$. The comparisons
between the asymptotic modulation solution (\ref{frontas})
(solid line) and numerical solution (circles) are made for a fixed $x=50$.}
\label{fig18}
\end{figure}

The amplitude distribution $a(x,y)$ in (\ref{frontas}), on the contrary, depends,
via the function $A(\la_4)$, on the wing profile. We stress that in spite of
the fact that  the asymptotic distribution $a(x,y)$ satisfies the amplitude
equation of the linear modulation theory (see \cite{wh74}), the determination
of the amplitude dependence on the boundary conditions (i.e. the determination
of the function $A(\la_4)$) has required full nonlinear modulation analysis.
To explicitly evaluate the function $A(\la_4)$ for the parabolic profile
(\ref{Fpar}) we just need to know the function $w(z)$ entering the integral
in (\ref{A}).
Since $w(z)$ is the inverse of $\la_+(Y,0)$ on the interval $[Y_0, 0]$ it
is readily obtained as
\begin{equation}\label{wz}
w(z)=-\frac{l}{2\alpha M}(\alpha M+1-z)(z-\frac{\alpha
M}{2})\, .
\end{equation}
We recall that $l=L/M$, where $L$ is the length of the ``wing''.
Then the integral in (\ref{A}) is evaluated explicitly to give:
\begin{equation}\label{A1}
\begin{split}
A(\la_4) =
4\sqrt{\frac{\la_4(\la_4 +1)}{\pi(2\la_4^2+1)}\ } \
(\la_4 - 1)^{1/4}\left(\frac{l}{15 \alpha
M}(1+\alpha M -\la_4)^{3/2}(8\la_4-3\alpha M + 2) \right)^{1/2}.
\end{split}
\end{equation}
The comparisons of the asymptotic distributions (\ref{frontas}), (\ref{A1})
with the distributions of $a$ and $k$ obtained from the 2D numerical solution
are shown in Fig.~\ref{fig18}.
One can see a very good agreement for both distributions.

\subsubsection{Trailing edge}

The leading (outer) edge $y^+(x)$ of the DSW is determined by  formula (\ref{slopelead}).
To complete the modulation solution we
need to determine the trailing (inner) edge $y^-(x)$ defined by  the soliton
condition $m=1$. As in the straight corner case, we shall mainly be concerned
with the amplitude of the trailing soliton and its slope as its actual position
might differ significantly from
the curve $y^-(x)$ obtained from the modulation theory due to the loss of the
initial phase (see the relevant discussion and comparisons with the numerical
solution in Section V.B).

In the piston problem terms, we are going to find the curve $Y=Y^-(T)$,
where $\la_3=\la_2=1$.
Remarkably, the equation for $Y=Y^-(T)$ and, as a result, the parameters
of the trailing soliton can be found directly, without using the full modulation solution
obtained in the previous subsection. We use the fact that
in the front DSW one has  $\la_2=1$ (see the previous subsection) so the matching
condition (\ref{match1}) at the trailing edge
 assumes the form:
\begin{equation}\label{traila}
\hbox{At} \ \ Y=Y^-(T):\qquad \la_3=\la_2=1\, , \ \  \la_4 = \la_+ , \  \   \la_1 = -1 \, , \\
\end{equation}
where $\la_+(Y,T)$ obeys the simple-wave equation (\ref{simple}).
On the other hand,  the curve $Y=Y^-(T)$ is specified by the
kinematic condition (\ref{mult}) in which we set the values of $\la_j$'s
from (\ref{traila}).
As a result we get a closed system
\begin{equation}\label{trailb}
 Y-{\tfrac{1}{2}(3\la_+-1)}T= w(\la_+) \, , \qquad \frac{dY}{dT}=\frac{1}{2}(1+\la_+)
\end{equation}
along the trailing edge.
We introduce  in (\ref{trailb}) $Y=Y_s(\la^*)$, $T=T_s(\la^*)$, $\la_+=\la^*$,
where $\la^*$  is the parameter along the trailing edge curve
so that $Y_s(A^+)=0$,  $T_s(A^+)=0$ (since $\la^+(0,0)=A^+$---see Fig.~14).
Next, eliminating $Y_s'$ we obtain a single ordinary differential
equation for $T_s(\la^*)$
\begin{equation}\label{tdiff}
(\la^*-1)T_s ' + \frac{3}{2}T_s+w'(\la^*)=0 \, , \quad T_s(A^+)=0\, ,
\end{equation}
which is readily  integrated to give
\begin{equation}\label{T0}
T_s=\frac{1}{(\la^*-1)^{3/2}}\int \limits_ {\la^*}^{A^+}(z- 1)^{1/2}w'(z) d z \, .
\end{equation}
Next, substituting (\ref{T0}) into the first equation (\ref{trailb}) we obtain
the function $Y_0(\la^*)$ in the form
\begin{equation}\label{Y}
Y_s=\frac{1}{2}(3\la^* - 1)T_s(\la^*) + w(\la^*)\, .
\end{equation}
Thus, equations (\ref{T0})--(\ref{Y}) specify the DSW trailing edge
$\{Y=Y^-(T): \ Y=Y_s(\la^*), T=T_s(\la^*)\}$. Correspondingly, the geometric
location $y^-(x)$ of this edge  in the physical $x,y$-plane is given by
\begin{equation}\label{xy}
y=y^-(x): \quad   y=Y_s(\la^*), \ x=MT_s(\la^*) \, .
\end{equation}
Within the NLS modulation theory the position of the trailing edge determines,
up to an inherent phase shift,  the location of the trailing dark soliton.
Thus, equations (\ref{T0})--(\ref{xy}) define the geometric shape of this
spatial trailing dark soliton.
We note that, unlike recently found oblique dark solitons generated in the
2D supersonic NLS flow past small obstacles \cite{egk06}, \cite{kp08},
and stretching along straight lines, the trailing dark soliton in the front
DSW has a curved contour in the $xy$-plane. In fact, the ``bending'' of this
2D soliton has the same nature  as the speed variations of a 1D soliton
propagating through a non-uniform medium. Here the non-uniformity is due
to the large-scale density variations
in the flow past extended obstacle.

Using (\ref{T0}) we obtain an implicit expression for the variations of
the trailing dark soliton amplitude $a^-=2(\la^* - 1)$ along the wave crest
line $y^-(x)$
\begin{equation}\label{xa}
x= \frac{2^{3/2}M }{(a^-)^{3/2}}\int \limits_ {1+a^-/2}^{1+\alpha M}
(z-1)^{1/2} w'(z) d z \, .
\end{equation}
The relationship between the local slope $s^-$ of the trailing dark
soliton and  its amplitude is given by (see (\ref{trailb}))
\begin{equation}\label{str}
s^-=\frac{dy^-}{dx}=\frac{1}{M}(1+a^-/4) \, .
\end{equation}
Since close to the the origin, $(x, y) \to (0,0)$, we have $\la_3 \to\la_2=1$,
$\la_4 \to A^+=1+\alpha M$ we get for the soliton amplitude
$a^-(0,0)=2(\la_4 - \la_3)=2 \alpha M$, so the initial slope of the trailing
edge is $s^-(0)=1/M + \alpha/2$, i.e. it coincides
with the slope of the dark soliton in the DSW generated in the flow past
straight corner (see (\ref{s-}) as one can expect (note that this result
does not have much practical significance as the modulation theory performs
rather poorly for small $x,y$).

Next, since the  integral in (\ref{xa}) is $O(1)$ we conclude that
$a^- \sim x^{-2/3}\to 0$ as $x \to \infty$ i.e. the trailing dark soliton
amplitude vanishes along the  line
$y^-(x)$ while its slope asymptotically approaches the Mach line of the
highly supersonic undisturbed flow: $y^- \to x/M$ as $x \to 1$.

There still remains an issue of the transition from the amplitude decay
$a \sim x^{-1/2}$ (\ref{frontas}) for the major part of the the DSW
to the decay $a \sim x^{-2/3}$ (\ref{xa}) for the trailing dark soliton
at trailing edge. This matching requires a detailed analysis of the
asymptotic behaviour of the hodograph solution (\ref{Ts11}) in the small
vicinity of the singular point $\la_3=\la_4=1$. Such an analysis, whilst
being relatively straightforward, is beyond the scope of the present paper.

For the parabolic wing profile (\ref{Fpar}) the function $y^-(x)$ defining
the location of the trailing dark soliton is given by
(\ref{xy}), where for $T_s(\la^*)$ we obtain from (\ref{T0}) by using formula
(\ref{wz}) for the inverse function  $w(z)$:
\begin{equation}\label{T01}
T_s =\frac{l}{30\alpha
M}\left((10-3 \alpha M)\left(\frac{\alpha M}{\la^*-1}\right)^{3/2}-12\la^*
+15 \alpha M + 2 \right)\, ,
\end{equation}
and for $Y_s(\la^*)$ we have (\ref{Y}).

And, finally, for the trailing soliton amplitude $a^-(x)$ at $x \gg 1$ we
obtain from (\ref{xa}) (or directly from (\ref{T01})) a simple implicit formula
\begin{equation}\label{xa1}
x= \frac{l}{30 \alpha}\left((10-3 \alpha M)\left(\frac{2 \alpha M}{a^-}\right)^{3/2}
- 6a^-   +15 \alpha M - 10\right)\, .
\end{equation}
One can see that at $x=0$ one has $a^-=2\alpha M$ and $a^- \to 0$ as $x \to \infty$
as predicted by the general theory. In particular for $x \gg 1$ we have the
asymptotic behaviour of the amplitude along the soliton wavecrest,
\begin{equation}\label{xa2}
a^- \simeq  \left(\frac{l(10 - 3 \alpha M)}{30 \alpha}\right)^{2/3}
\frac{2\alpha M}{x^{2/3}}
\end{equation}
\begin{figure}[ht]
\centerline{\includegraphics[width=5cm]{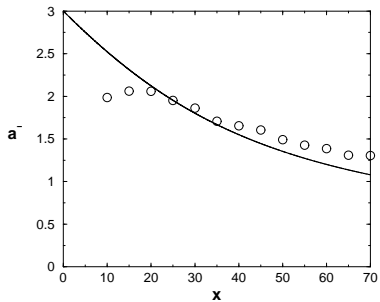}}
\caption{Comparison for the amplitude decay along the trailing dark soliton.
Solid line: asymptotic modulation solution (\ref{xa1}), circles:
the amplitude values from the direct numerical simulation. }
\label{fig19}
\end{figure}
The comparison of the amplitude behaviour (\ref{xa1}) along the DSW trailing
edge with the numerical simulation data is shown in Fig.~\ref{fig19}.
One can see a good agreement between the analytical curve and numerical
solutions for large enough  (as expected from the range of validity of
Eq.~(\ref{xa1})) values of $x$.

We note that dependencies (\ref{xa}) and (\ref{str}) have been derived under
an implicit assumption that the ``time'' $x/M$ of the establishment of the
trailing dark soliton in the DSW is much less than the typical modulation
``time'' scale  $\sim L/M$, i.e. under the assumption that the DSW is fully established.
This assumption works quite well for the straight wedge type profiles studied
in Section VI but it may fail for the wing-type profiles with sufficiently rapidly
decaying derivative so that
condition $|f''(\xi)|\ll 1$ is not satisfied for a significant part of the
profile (or, in terms of asymptotically equivalent initial conditions
(\ref{eq27}) the inequality being $|d\la_+(Y,0)/dY| \ll 1$).
In that case, the trailing soliton establishes itself very slowly and realizes
only asymptotically for $x \gg L$ (see \cite{gkm89} for the analysis of a similar
issue in the context of the KdV equation). Taking into account that the amplitude
of this  ``slowly developing soliton''  decreases with $x$ on the scale $\sim L$,
its behaviour for finite $x/L$ could actually be rather well approximated by
the linear theory (see the next section).  At the same time, one should stress
that the dependence  (\ref{xa}) of the values  of the trailing soliton amplitude
on the obstacle size and shape cannot be found within the linear theory.
Determination of this dependence requires full nonlinear analysis
(either modulation or IST-based) -- see the discussion in section VII A.2.

\subsubsection{Extension of the modulation solution: the ``ship wave'' pattern.}

The modulation solution obtained in section VII A.2 is defined within the domain
$y^-(x) \le y  \le  y^+(x)$ and implies that  the wave amplitude vanishes at
the outer (leading) DSW edge $y^+(x)$ and outside of the DSW region flow is
assumed to be constant. At the same time, the boundary $y=y^+(x)$, associated
with the linear group velocity, is not a wavecrest line so it is clear that
one should be able to extend the wave crests beyond the DSW boundary. Indeed,
it is clearly seen from the results of numerical simulations (see the density
plots in Figs.~11--13) that the wavecrests do not stop at the external
boundary of the modulation solution and the small (but quite noticeable)
oscillations are present outside the DSW. To resolve this apparent contradiction
one can notice that  the vanishing of the amplitude at $y=y^+(x)$ for the DSW
modulation solution does not necessarily imply that the actual wave amplitude
turns into zero, this simply means that the oscillations are linear.
To capture these linear oscillations occurring for $y> y^+(x)$ we introduce
a small-amplitude wave packet as a natural extension of the DSW, and will
use the {\it linear} modulation theory for its description.

In linear modulation theory the equation for the wave amplitude is decoupled
from the equation for the wavenumber (see \cite{wh74}) so one can put $a=0$
and consider the ``wave conservation'' law separately.  We note that such an
extension, whilst being automatically consistent with the DSW modulation
solution at $y=y^+(x)$, is not quite trivial as the linear modulation theory
is not valid {\it inside} the DSW region, even in a small neighborhood of the
zero-amplitude leading edge $y^+(x)$---see \cite{gke91}. We note that the
modulation solution for the wavenumber in the linear wave packet has already
been  obtained, this is equation (\ref{group}) (see the explanation after
formula (105)), so we simply postulate that this solution describes the wave
distribution for $y>y^+(x)$.

In effect, modulation solution (\ref{group}) enables one
to derive the two-dimensional ``ship-wave'' pattern generated by the front edge
of the obstacle. To this end, we notice that, up to an arbitrary initial
phase $\Theta_0 \in [0, 2\pi]$,
the local angular phase of the two-dimensional ``travelling'' wave is given
by (see (\ref{eq016})),
\begin{equation}\label{1a}
    \Theta= k_y \theta = k_y\left(y-U\frac{x}M\right)=k_y\cdot y-
    \frac{k_y}M\sqrt{1+\frac{k_y^2}4}\cdot x,
\end{equation}
hence the wave vector of the modulated linear wave is equal to
\begin{equation}\label{2a}
    \mathbf{k}=\left(-\frac{k_y}M\sqrt{1+\frac{k_y^2}4}, k_y\right).
\end{equation}
As in the 2D theory of ship waves produced by a point-like obstacle in
the supersonic NLS flow
\cite{ship1, ship2}, we introduce the angle $\chi$ between the radius-vector
$\mathbf{r}$ and the $x$ axis, i.e., the flow direction, and the angle
$\eta$ between the wave-vector $\mathbf{k}$ and $-x$ axis (see Fig.~20, left)):
\begin{equation}\label{3a}
    \mathbf{r}=(r\cos\chi,r\sin\chi),\quad \mathbf{k}=(-|{\bf k}|\cos\eta, |{\bf k}|\sin\eta).
\end{equation}
Then Eq.~(\ref{2a}) leads to the following expression for the wave vector
length
\begin{equation}\label{4a}
    |{\bf k} |=\frac{2\sqrt{M^2\cot^2\eta-1}}{\sin\eta}
\end{equation}
in the hypersonic approximation. One should emphasize that the  wavenumber
$k$ defined by Eq. (\ref{kU}) and occuring in the zero-amplitude limit
(\ref{group}) of the one-dimensional piston approximation of the DSW
modulation solution  is consistent with the $y$-component $k_y$ of the
full two-dimensional vector ${\bf k}$ (\ref{2a}). Hence the substitution of
\begin{equation}\label{5a}
    \la_4=\sqrt{1+k_y^2/4}=\sqrt{1+|{\bf k}|^2\sin^2\eta/4}=M\cot\eta
\end{equation}
into Eq.~(102) yields the relationship between $\chi$ and $\eta$
\begin{equation}\label{6a}
    \tan\chi=2\cot\eta-\tan\eta/M^2.
\end{equation}
\begin{figure}[ht]
\centerline{\includegraphics[width=5cm]{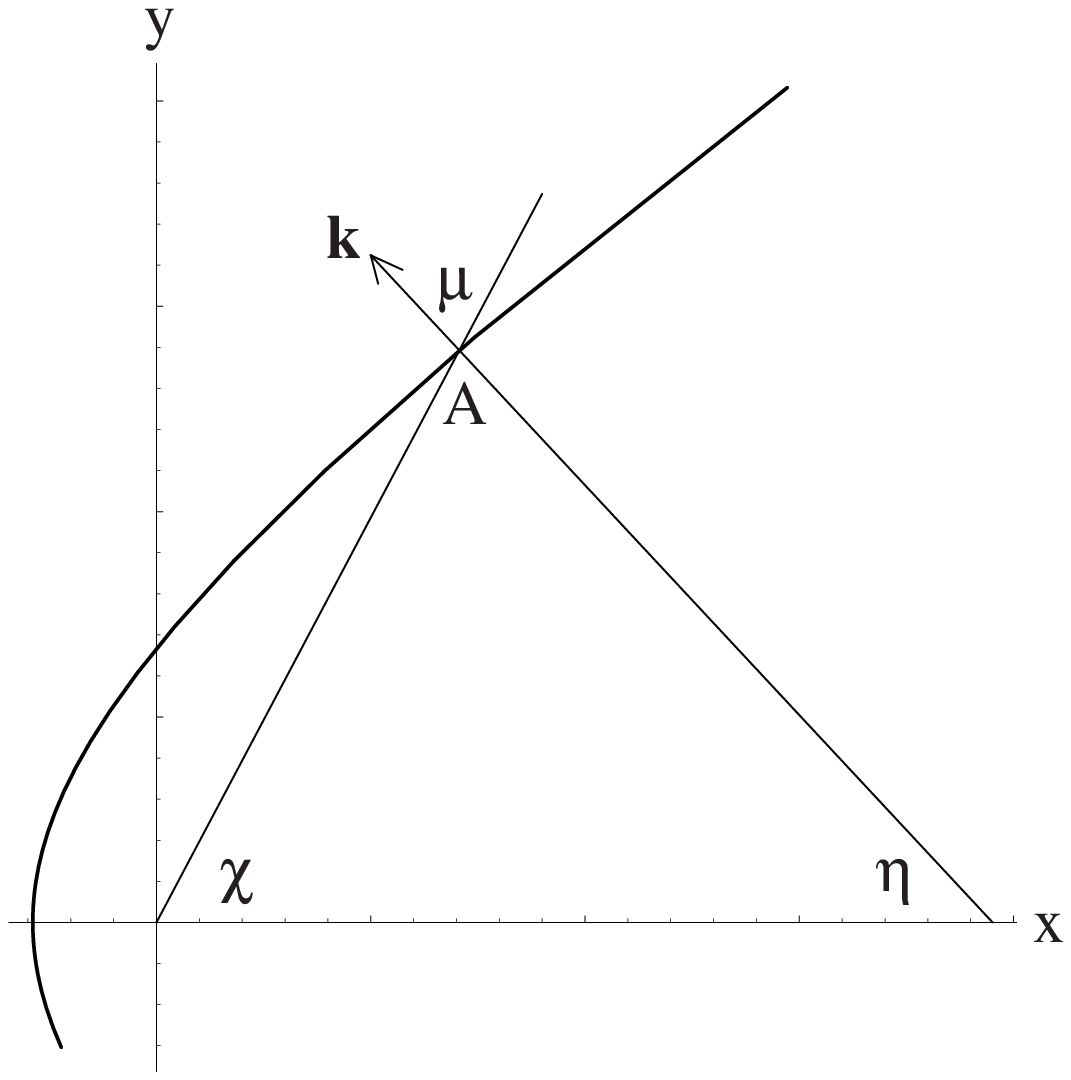} \qquad  \qquad \qquad  \includegraphics[width=6cm]{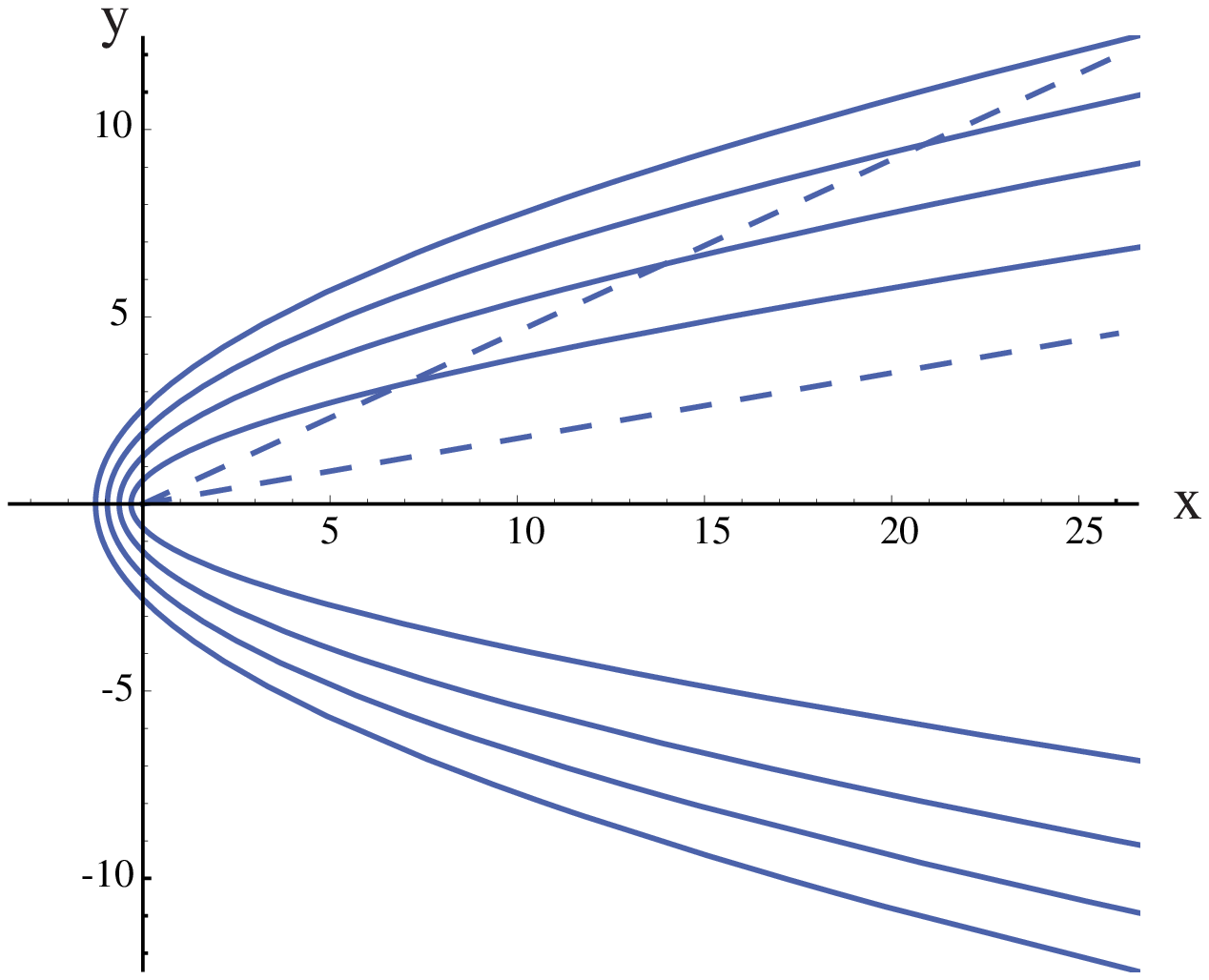}}
\caption{Left: The wave crest geometry in the NLS ship wave pattern.
The wave vector $\mathbf{k}$ is normal to the wave crest line
which is shown schematically by a curve.
Right:  the theoretical wave crest lines in the ship-wave pattern
(solid lines); the DSW boundaries determined by the modulation solution
are shown by the dashed line.} \label{fig20}
\end{figure}

Now we notice that Eqs.~(\ref{4a}) and (\ref{6a}) are nothing but the
highly supersonic approximation of the ``ship wave'' theory developed earlier
\cite{ship1, ship2} for the case of a localized point-like obstacle.
Indeed, in this theory the length of the wave vector is given by the
expression
\begin{equation}\label{1c}
    |\bk|=2\sqrt{M^2\cos^2\eta-1}
\end{equation}
which in our hypersonic approximation can be easily transformed to
\begin{equation}\label{2c}
    |\bk|=2\sin\eta\sqrt{M^2\cot^2\eta-1}.
\end{equation}
This expression is approximately equal to (\ref{4a}) if $\sin\eta\cong1$,
that is
\begin{equation}\label{3c}
    \tan\eta\gg1,\quad \text{or}\quad |k_y/k_x|\gg1.
\end{equation}
In a similar way, the relation
\begin{equation}\label{4c}
    \tan\chi=\frac{(1+|\bk|^2/2)\tan\eta}{M^2-(1+|\bk|^2/2)}
\end{equation}
between the angles $\chi$ and $\eta$ for the point-like obstacle case
in the hypersonic limit can be cast into the form
\begin{equation}\label{5c}
    \tan\chi\cong\frac{2\tan\eta}{\tan^2\eta-1}-\frac{\tan^3\eta}{M^2(\tan^2\eta-1)},
\end{equation}
and again this formula is reduced to (\ref{6a}) under condition (\ref{3c})
which means that the flow parameters change much slower in the
$x$-direction than in the $y$-direction what is assumed in our approach.
Thus, we have arrived at a remarkable result: the  solution of the Whitham
equations describing the DSW region, turns out to coincide, for $x, y \gg 1 $,
with the corresponding approximation of the linear ship wave theory
describing the waves outside the DSW. Thus, the far-field asymptotic solution (\ref{frontas})
is not restricted to the DSW region and can be used for the description of
the whole flow at the distances sufficiently far from the front edge of the wing.

This observation permits us to extend the the wave crest lines to the
whole region outside the Mach cone. It remains only to show that the
ship wave pattern produced by a slender body can be approximated by
the pattern produced by a point-like obstacle. To this end we turn to
the formula for oscillations of density in the ship-wave theory
(see Eq.~(20) in \cite{ship2}),
\begin{equation}\label{6c}
    \delta n=\int\frac{V(\bk)|\bk|^2e^{i\bk\cdot\br}}{(\bk\cdot\mathbf{U})^2
    -|\bk|^2(1+|\bk|^2/4)}\frac{d\bk}{(2\pi)^2}
\end{equation}
where $\mathbf{U}=(M,0)$ and
\begin{equation}\label{7c}
    V(\bk)=\int V(\br)e^{-i\bk\cdot\br}d\br
\end{equation}
is the Fourier image of the potential $V(\br)$ created by the obstacle.
Integral over wave vector length $|\bk|$ can be estimated as contribution
of the poles in (\ref{6c}) which depends on the dispersion relation only.
Moreover, for obstacles with a sharp form their Fourier images must
include wide range of harmonics and hence they are smooth functions
of $\bk$. Therefore in the integration over directions of $\bk$
performed for $|\bk\cdot\br|\gg1$ by the stationary phase method
(see \cite{ship2}) the main contribution is given by a stationary point
of the phase $\bk\cdot\br$ which, again, does not depend  on the
function $V(\bk)$. This yields the relation (\ref{4c}) and,
subsequently, the parametric formulae for the wave crest lines,
\begin{equation}\label{8c}
    \begin{split}
    x&=\frac{4\Theta}{|\bk|^3}\cos\eta(1-M^2\cos2\eta),\\
    y&=\frac{4\Theta}{|\bk|^3}\sin\eta(2M^2\cos^2\eta-1),
    \end{split}
\end{equation}
where $\Theta=2\pi,\,4\pi,\,6\pi,\ldots$, the wavevector $|\bk|$ is given by Eq.~(\ref{1c})
and $\eta$ changes formally in the range
\begin{equation}\label{9c}
    -\arccos(1/M)\leq\eta\leq\arccos(1/M).
\end{equation}

In the limit $\cos\eta\gg1/M$ the crest lines take a parabolic form
\begin{equation}\label{12a}
    x(y)\cong -\frac{\Theta}{2M}+\frac{M}{2\Theta}y^2.
\end{equation}
This limit corresponds to the region located not too far from the front edge
of the obstacle.

In the opposite limit, when $\cos\eta\to1/M$ the crest lines
converge asymptotically to the straight lines parallel to the Mach cone lines:
\begin{equation}\label{13a}
    y\cong\frac{x}M.
\end{equation}
\begin{figure}[ht]
\centerline{\includegraphics[width=10cm]{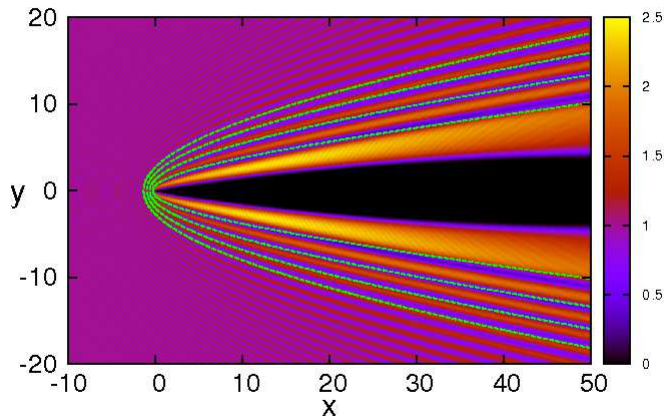}}
\caption{(color online) Comparison of the universal  ``ship-wave'' pattern (\ref{8c})
for a slender body (dashed lines) with the wavecrest lines in the
supersonic NLS flow past the front edge of the parabolic wing profile
(\ref{Fpar}) with $\alpha=0.15$, $L=100$. The oncoming flow speed is $M=10$.}
\label{fig21}
\end{figure}

\bigskip

It is important to note that the obtained expressions for the geometry of the wave crest lines
do not depend on the opening angle $\alpha$ of the obstacle and, hence,
on the slope of the outer edge $y^+(x)$ of the DSW. 
Indeed, as we have shown, although in the
Whitham approximation the amplitude of the wave vanishes at $y^+(x)$, the curves of the wave crest lines can be 
continued outside  this DSW boundary where they represent the spatial distribution of the small-amplitude linear waves.
For this reason the linear ship wave pattern can be viewed as a natural continuation
of the wave crest distribution in the oblique spatial DSW. However, when we go inside
the DSW region along the wave crest line, the amplitude of
the DSW gradually increases and the linear approximation loses its
applicability. In an established (strongly nonlinear) DSW, the shape of the wave crests is determined by the shape of the obstacle
(see section VI.A.3).
But for the profiles with sufficiently rapidly decaying derivative the DSW establishment ``time'' $x/L$ is rather large (see the discussion in the end of section VI.A.3)
so the linear approximation works quite well in a wide region
around the front tip of the obstacle including the neighborhood of the
outer boundary of the DSW.  This is illustrated in Fig.~\ref{fig21} by the comparison of
the analytical predictions given by Eq.~(\ref{8c}) with the results of full 2D numerical simulations.

\subsection{Flow past rear edge of a wing}

\begin{figure}[ht]
\centerline{\includegraphics[width=6cm]{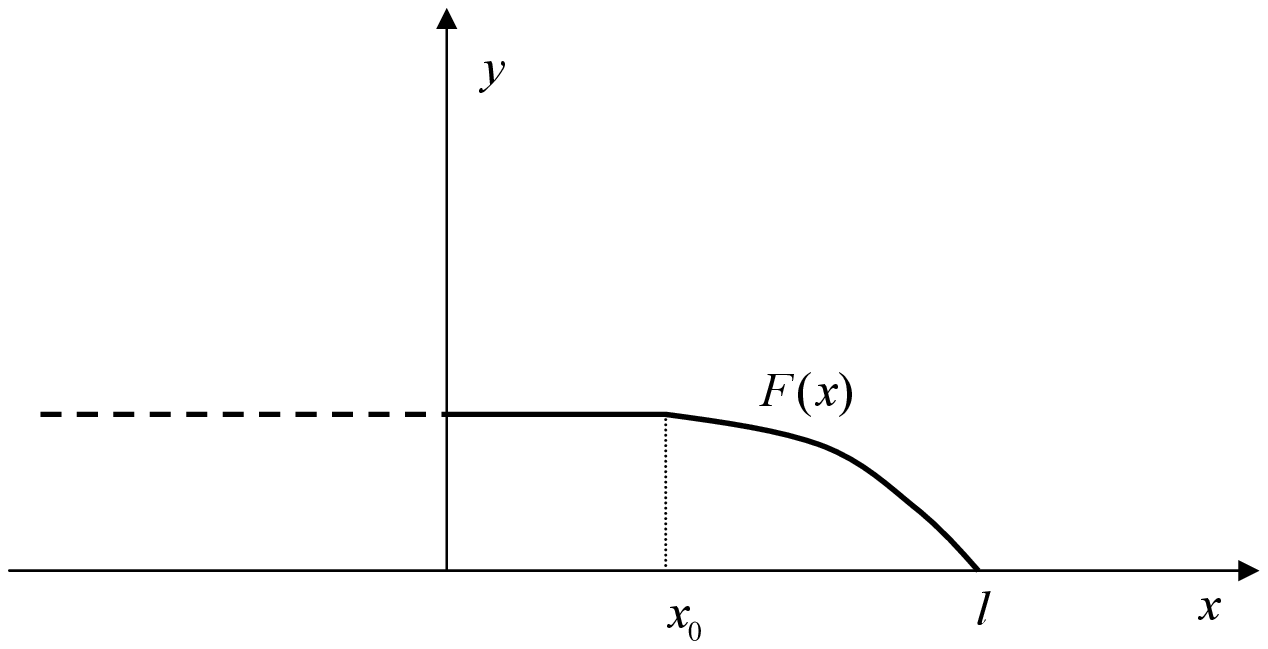} \qquad \includegraphics[width=6.0cm]{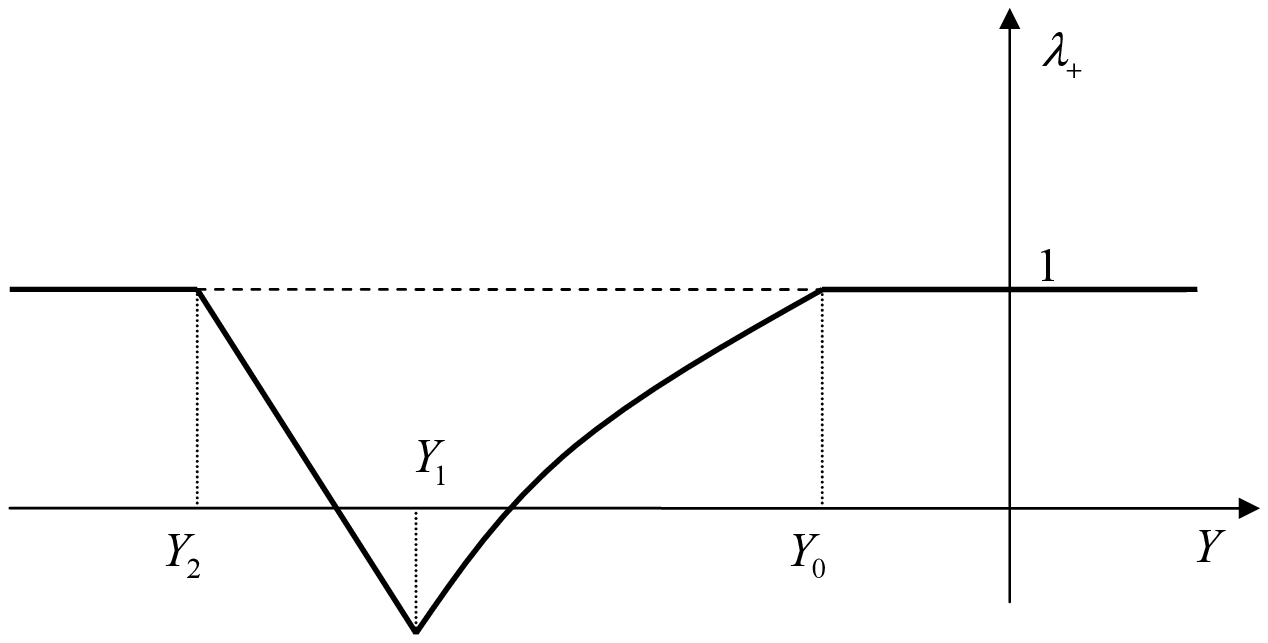}}
\caption{Left: profile of the rear edge of a wing in the upper half-plane.
Right: asymptotically equivalent initial condition for $\lambda_+$.}
\label{fig22}
\end{figure}

Now we consider the DSW generated by the flow past the rear edge of the wing.
The corresponding  initial profile of the Riemann invariant $\la_+$ is
given by Eq.~(\ref{eq27}) for $Y_1 < Y < Y_0$
and by (\ref{lapar2}) for $Y_2<Y<Y_1$. We recall that $Y_0=f(x_0)-x_0$, $Y_1=-(\tfrac{3}{2}f'(l-0)+1)l$,
$Y_2=-l$ and $ \lambda_+(Y_1,0)=1+f'(l-0)$ (see (\ref{Y2}), (\ref{Y1})).
The second invariant is constant, $\la_-=-1$. A typical form of the function
$\la_+(Y,0)$
is schematically shown in Fig.~\ref{fig17} (right).  The evolution of the
``potential well'' $\la_+(Y,T)$ leads to the wave breaking at
\begin{equation}\label{}
T_b=\frac{Y_1-Y_2}{V_+(1,-1) - V_+(\la_+(Y_1,0), -1)} =l \, ,
\qquad Y_b=Y_2 + V_+(1,-1)T_b= 0\, ,
\end{equation}
which simply means that the rear DSW spreads directly from the rear endpoint
of the wing  (see Fig.~\ref{fig4}). A typical profile of $\la_+(Y,T_b)$
is shown in Fig.~\ref{fig22} (left).

Thus, for $T>T_b$ one has a DSW forming behind the body (see Region IV in
Fig.~\ref{fig4}). This DSW can be described by the modulated travelling wave
solution analogous to that constructed in the previous section for the front DSW.
The main difference is that now one has the Riemann invariants $\la_3$ and
$\la_2$ varying within the modulation solution while $\la_1=-1$, $\la_4=1$
(see Fig.~\ref{fig18} (right)). As a result, the modulation solution yields
that as $T \to \infty$ one has  $\la_2 \to \la_3$ (i.e. $m \to 1$) everywhere
except some small vicinity of the leading edge where $\la_3=1$ and $m \to 0$.
That means that the rear DSW  for $x \gg 1$, $y \gg 1$ asymptotically transforms
into a soliton train (a fan of oblique dark solitons). Of course, such a
behaviour is to be expected as the initial profile of $\la_{+}$
(see Fig.~\ref{fig22} (right)) corresponds to a large-scale `potential well'
in the associated scattering problem in the Zakharov-Shabat IST formalism for
the 1D NLS equation, and, therefore, leads to a semi-classical distribution
of the bound states, each linked to a dark soliton in the NLS equation solution
\cite{jl99,kku02}.
\begin{figure}[ht]
\centerline{\includegraphics[width=6cm]{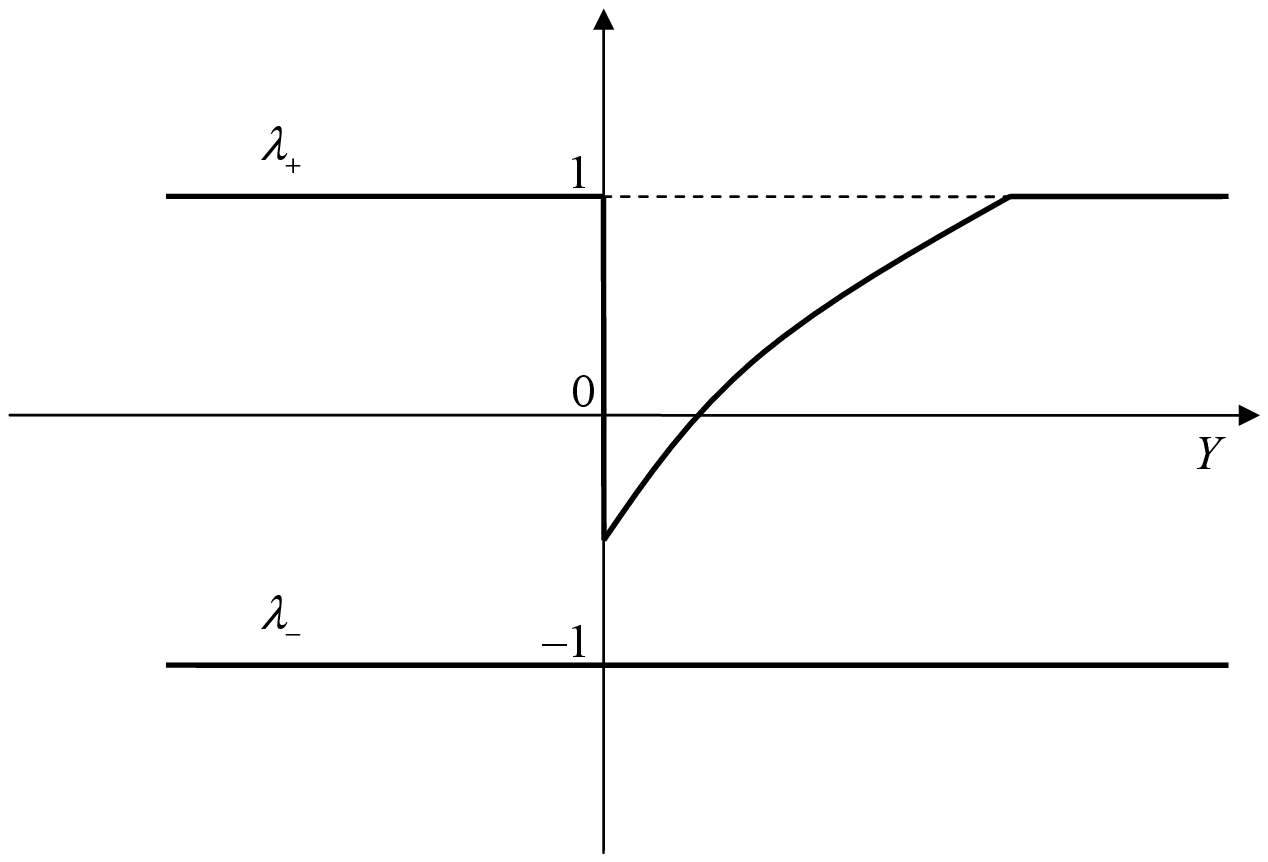}
\qquad \includegraphics[width=6.5cm]{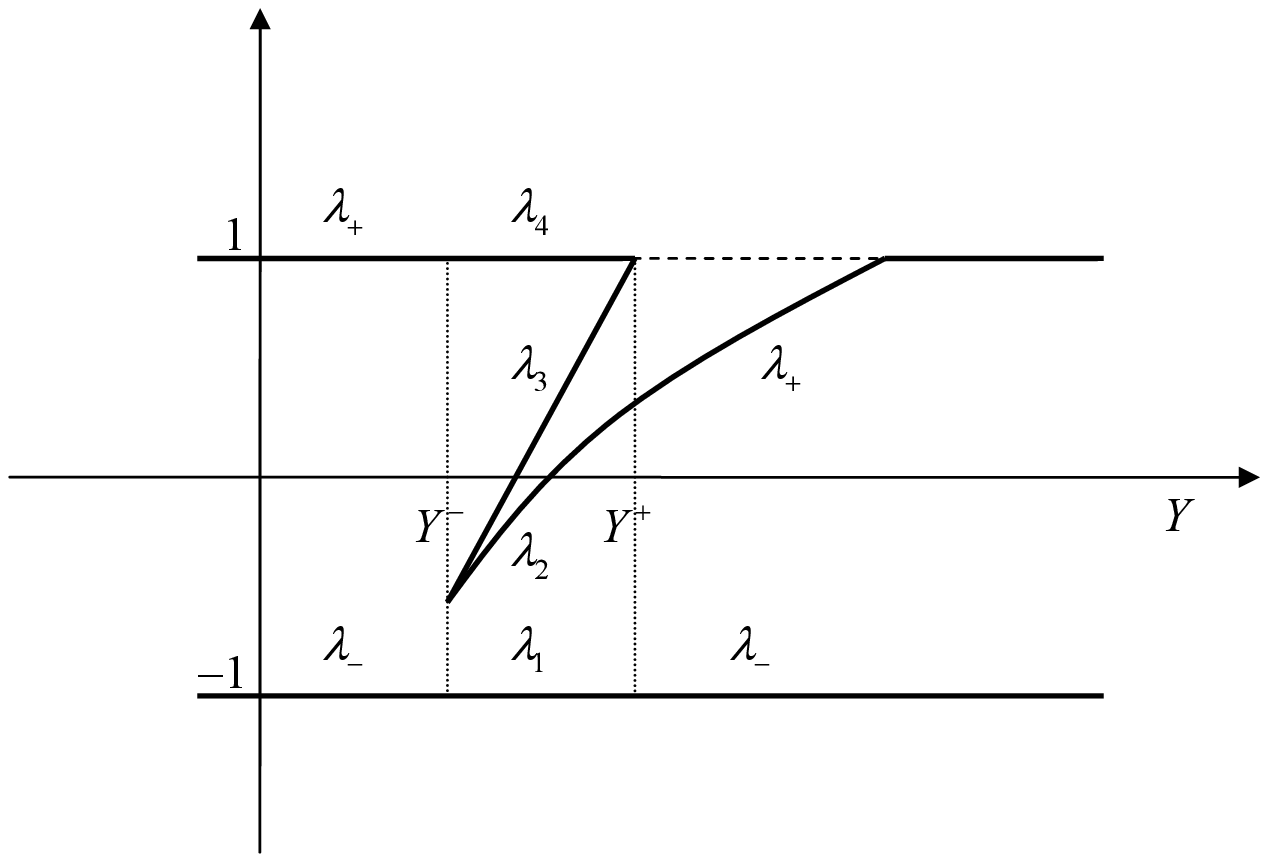}}
\caption{Left: profile of the Riemann invariants $\la_{\pm}$ at the point
of wave breaking, $T=T_b$.
Right: schematic behaviour of the Riemann invariants in the modulation
solution for rear DSW, $T>T_b$}
\label{fig23}
\end{figure}

Thus, if one is interested in the asymptotic structure of the
flow in the region far enough from the body
where the rear DSW transforms into a ``fan'' of spatial solitons well
separated from each other, there is no need to derive the full modulation solution.
As was shown in Refs.~\cite{jl99,kku02},
each soliton in the soliton train evolving from the initial ``well''
is parameterized by the eigenvalue $\la=\la_k$ found from the generalized
Bohr-Sommerfeld quantization rule, consistent with the Whitham approximation
used before,
\begin{equation}\label{bs}
\begin{split}
   \oint\sqrt{(\la-\la_+)(\la-\la_-)}\,dY=2\pi(k+\tfrac12),\quad
   k=0,1,\ldots,K,
   \end{split}
\end{equation}
where in our case $\la_+=\la_+(Y,0)$ is given by (\ref{eq27}), (\ref{lapar2}),
$\la_-=-1$, and the integration is taken over the cycle around two
turning points defined by $\la=\la_+(Y,0)$.
The $k$-th soliton amplitude $a_k$ is related with the eigenvalue $\la_k$ by
\begin{equation}\label{}
a_k = 1-\la_k^2
\end{equation}
Returning to
the spatial coordinates (\ref{eq15}), we find the profile of the
$\la_k$-soliton in the train as (see \cite{kku02})
\begin{equation}\label{eq29}
   n_k(x,y)=1-\frac{1-\la_k^2}{\cosh^2[\sqrt{1-\la_k^2}(y-(\la_k/M)x)]},
\end{equation}
that is the ``fan'' of spatial dark solitons  is made of
soliton ``feathers'' lying asymptotically along the lines
\begin{equation}\label{eq30}
   y=({\la_k}/M) x,\quad k=0,1,\ldots,K,
\end{equation}
in the upper half-plane and symmetric ``fan'' of solitons is
generated in the lower half-plane.

Remarkably, the distribution (\ref{bs}) is invariant with respect to the evolution,
up to a breaking point at $T=T_b$, of the Riemann invariant $\la_+$,
described by the simple-wave equation (\ref{eq24}) (which is consistent with
the dispersionless limit of the NLS equation (\ref{nls-1D})). Indeed,
it is not difficult to show that (\ref{eq24}) implies that
\begin{equation}\label{inv}
\frac{\partial}{\partial T} \oint\sqrt{(\la-\la_+)(\la+1)}\,dY = 0\, .
\end{equation}
The property (\ref{inv}) can be viewed as a semi-classical analog of
isospectrality of the 1D NLS evolution (see \cite{jl99}).
Thus, the initial profile of $\la_+$ is defined up to the deformation
(\ref{eq24}) and, thus, should not necessarily be a single-valued function
as in Fig.~\ref{fig23} (see also the discussion in  Section V).

For the parabolic profile (\ref{Fpar}), the function $\la_+(Y,0)$ corresponding
to the rear part of the wing
is specified by formulae (\ref{lapar1}), (\ref{lalin}) in the interval
$Y_2<Y < Y_0$ and $\la_+=1$ outside of this interval.
It has its minimum at $Y_1$: $\la_+(Y_1,0)=1-\alpha M$.

Now the integral in (\ref{bs}) is evaluated in a closed form giving
the equation for the  bound states $\la=\la_k$:
\begin{equation}\label{2-1}
\frac{l\sqrt{\la+1}}{15 \alpha M}(\la-1+\alpha M)^{3/2} (3\alpha M + 8 \la + 2) = 2 \pi(k+\tfrac12),\quad
    k=0,1,\ldots,N.
\end{equation}
The physically meaningful roots $\la_k$ lie in the interval $ 1-\alpha M < \la_k<1$.
It immediately follows
from the requirement $\la_+ > \la_-=-1$ that one must also impose a restriction
that $\alpha M <2$ (for $\alpha M>2$ the description should be modified
as the vacuum point
appears at $y=0$).
The greatest root $\la_N$ has the value close to unity
so that the number of solitons in the fan can be estimated by putting
$\la_k=1$, $k=N$
in (\ref{2-1}), i.e
\begin{equation}\label{N}
N \approx \frac{1}{2\pi}\frac{\sqrt{2} l}{15}(\alpha M)^{1/2} (10+3\alpha M)
\end{equation}
The semi-classical formula (\ref{2-1}), strictly speaking, is asymptotically
valid as long as $N \gg 1$, which, by (\ref{N}) presumes rough general criterion $l(\alpha M)^{1/2} \gg 1$.
However,  as is often the case with the Bohr-Sommerfeld type distributions,
formula (\ref{2-1}) works reasonably well for a much broader range of parameters.
Say, for $l=10$, $\alpha=0.15$, $M=10$  one has just 3 physical roots of the
equation  (\ref{2-1}), which agrees with three dark solitons observed in numerical
solution (see Fig.~\ref{fig15}). The comparisons between the predictions of
(\ref{2-1}) and the numerical simulations data for the amplitudes $a_k$ and
slopes $s_k$ of the oblique dark solitons are presented in the Table below.

\medskip
\begin{center}
\begin{tabular}{|l|l| l | l|l| l|}
   \hline
   $k$  & $\la_k$  & $a_k=1-\la_k^2$ &  $a_k$ (num) &$s_k=\la_k/M$ &  $s_k$ (num) \\
   \hline
   0  & 0.2915 & 0.9150 & 0.9170 & 0.0291 & 0.02\\
   1  & 0.7101 & 0.4957 & 0.5689 & 0.0710 & 0.06\\
   2  & 0.9649 & 0.0688 & 0.1903 & 0.0964 & 0.09\\
   \hline
 \end{tabular}
\end{center}

\medskip
One can see that, taking into account the inherent in the hypersonic
approximation error $O(1/M)$ for the soliton amplitude and $O(1/M^2)$ for
the slope, the comparison should be viewed as quite favorable.

\section{Discussion}

In this paper, we have constructed an asymptotic theory of the supersonic
flow of a superfluid past slender  bodies. The theory is constructed in the framework
of the 2D defocusing NLS equation with the impenetrability condition at the
body surface and the condition of an equilibrium steady flow with Mach number
$M$ at infinity.
The description is made under the following assumptions: $M \gg 1$, $\alpha \ll 1$,
$M\alpha = O(1)$, where $\alpha$  is the body slenderness parameter (e.g.
the opening angle of a ``wing'' or a wedge).
Under these assumptions we have asymptotically (with respect to the small
parameter $1/M$) reduced the original two-dimensional stationary boundary-value
problem for the
time-independent 2D  NLS equation in the $x,y$-plane with the oncoming flow
along the $x$-axis to the  dispersive piston problem for 1D defocusing NLS equation,
in which the role of time is played by the stretched  $x$-coordinate, $T=x/M$,
and the spatial variable is the transverse coordinate $y$. The flow is globally
described using the  semi-classical approximation of the NLS equation,
when the solution is governed by the dispersionless
limit equations (the shallow-water system) in the regions of non-oscillating
flow and by the Whitham modulation equations in the regions of dispersive shock waves,
representing rapidly oscillating expanding nonlinear wave structures. We use the
so-called Gurevich-Pitaevskii formulation of the problem to match the solutions
of the Whitham equations with the solutions of the shallow-water equations at
free boundaries. The full modulation solutions are constructed and analyzed for
two canonical cases of the supersonic flow past bodies: the flow past infinite
straight corner (a wedge) and the flow past a wing. Our analytical solutions
are supported by direct 2D unsteady numerical simulations.

We now summarize the main results of the paper.
\begin{itemize}
\item{We have shown that the highly supersonic NLS flow past 2D slender bodies
is accompanied by the generation of two DSWs with contrasting asymptotic properties.}
\item{By making the  comparisons of the numerical solutions for the 2D problem
of supersonic flow past infinite wedge with the 1D numerical and analytical
modulation solutions of the
associated dispersive piston problem we have shown that the piston problem
describes the arising 2D wave patterns remarkably well for sufficiently large Mach numbers.
  }
\item{Using the dispersive piston approximation, we have constructed exact
modulation solutions for the problems of the supersonic NLS flow past a
straight infinite wedge and a slender ``wing''.}
\item{By analyzing the asymptotic behaviour of the obtained modulation
solution for the front DSW in the flow past a wing we have derived the
distributions of the amplitude $a$ and the wavenumber $k$
far enough from the front edge of the wing (Eq.~(\ref{frontas})):
\begin{equation}\label{frontas1}
x,y \gg 1: \qquad  a \cong \left(\frac{M}{x} \right)^{1/2} A \left(\frac{\tau+\sqrt{\tau^2+8}}{4}\right),
\quad k \cong \frac{1}{2}\sqrt{\left(\tau+\sqrt{\tau^2+8}\right)^2-16},
\quad \hbox{where} \ \ \tau=M\frac{y}{x} \, ,
\end{equation}
where the function $A(\xi)$ is given by Eq.~(\ref{frontas}). These
distributions describe the Kelvin-Bogoliubov ``ship wave'' pattern
and relate it, via the function $A(\xi)$, with the
geometric parameters of the wing}
\item{The distribution of oblique dark solitons in the rear DSW is obtained using the
generalized semi-classical Bohr-Sommerfeld quantization rule.}
\end{itemize}

The developed in this paper theory could find the applications to the description of
Bose-Einstein condensates behaviour in current experiments on loading of
ultracold quantum gases in
traps, their coherent manipulation and transport. Such processes are now
under intense investigations in atom chips---microfabricated, integrated devices
in which electric, magnetic and optical fields can confine, control and
manipulate cold atoms. Understanding of the interplay of dispersive and nonlinear
properties in Bose-Einstein condensate dynamics is of crucial importance for the effective
use of these devices which have very promising technological applications.

\subsection*{Acknowledgments}
A.M.K. thanks the Royal Society for the financial support
of his visit to Loughborough University and RFBR (grant 09-02-00499) for partial support.
V.V.Kh. thanks the London Mathematical Society for partial support of his visit to Loughborough University.


\newpage
\begin{thebibliography}{99}

\bibitem{cf48} R. Courant and K.O. Friedrichs, {\it Supersonic flow and shock
waves} (Interscience, New York, 1948).

\bibitem{ll} L.D. Landau and E.M. Lifshitz, {\it Fluid Mechanics}
(Pergamon Press, Oxford, 1987).

\bibitem{wh74} G.B. Whitham,
{\it Linear and Nonlinear Waves} (Wiley--Interscience, New York, 1974).

\bibitem{gp74} A.V. Gurevich and L.P. Pitaevskii,
{Sov. Phys. JETP,} {\bf 38,} 291 (1974).

\bibitem{gk87}  A.V. Gurevich, A.L. Krylov, Sov. Phys. JETP {\bf 65,}  944 (1987).

\bibitem{eggk95} G.A. El, V.V. Geogjaev, A.V. Gurevich, and
A.L. Krylov, Physica D {\bf 87,} 186 (1995).

\bibitem{kod99} Y. Kodama, SIAM J. Appl. Math., {\bf 59},  2162 (1999).

\bibitem{bk06} G. Biondini and Y. Kodama, J. Nonlinear Sci., {\bf 16}
435 (2006).

\bibitem{gke92}
A.V. Gurevich, A.L. Krylov and G.A. El,
{Sov. Phys. JETP} {\bf 74,} 957 (1992).

\bibitem{kke92} A.L. Krylov, V.V.  Khodorovskii, and G.A.  El,
{JETP Letters} {\bf 56,} 323 (1992).

\bibitem{tian93}
F.R. Tian,   Comm. Pure Appl. Math.
{\bf 46,}  1093-1129 (1993).

\bibitem{ek95} G.A. El and A.L. Krylov, Phys. Lett. A {\bf 203}, 77 (1995).

\bibitem{kku02} A.M. Kamchatnov, R.A. Kraenkel, and B.A. Umarov,
Phys. Rev. E {\bf 66},  036609 (2002).

\bibitem{ts85} S.P. Tsarev,  Soviet Math. Dokl.  {\bf 31,} 488 (1985).

\bibitem{egkJFM} G.A. El, R.H.J. Grimshaw, and A.M. Kamchatnov,
Journ. Fluid Mech {\bf 585}, 213 (2007).

\bibitem{kgk04} A.M. Kamchatnov, A. Gammal, and R.A. Kraenkel, Phys. Rev. A
{\bf 69,} 063605 (2004).

\bibitem{ha06} M.A. Hoefer, M.J. Ablowitz, I. Coddington,
E.A. Cornell, P. Engels, and V. Schweikhard, Phys. Rev. A {\bf 74,}
023623 (2006).

\bibitem{barrier09} A. M. Leszczyszyn, G. A. El, Yu. G. Gladush, and A. M. Kamchatnov,
Phys. Rev. A {\bf 79}, 063608 (2009).

\bibitem{ea07} P. Engels and C. Atherton, Phys. Rev. Lett. {\bf 99,} 160405 (2007).

\bibitem{gkke} A.V. Gurevich,
A.L. Krylov, V.V. Khodorovskii, and
G.A. El, { JETP,} {\bf 81,} 87 (1995); {\bf 82,} 709 (1996).

\bibitem{karpman} V.I. Karpman, {\it  Nonlinear Waves in Dispersive Media}
(Nauka, Moscow, 1973).

\bibitem{ekt04} G.A. El, V.V. Khodorovskii, and A.V. Tyurina,
Phys. Lett. A {\bf 333,} 334 (2004).

\bibitem{ovs}
L.V. Ovsyannikov,   {\it Lectures on the Foundations of Gas
Dynamics} (Nauka, Moscow, 1981)  [in Russian].

\bibitem{ek06} G.A. El and A.M. Kamchatnov, Phys. Lett.  A {\bf 350,} 192 (2006);
erratum: Phys. Lett. A {\bf 352,} 554 (2006).

\bibitem{cornell05} E.A. Cornell, Talk at the ``Conference on
Nonlinear Waves, Integrable Systems and their Applications",
(Colorado Springs, June 2005);
http://jilawww.colorado.edu/bec/papers.html.

\bibitem{caruso} I. Carusotto, S.X. Hu, L.A. Collins, and A. Smerzi,
Phys. Rev. Lett. {\bf 97,} 260403 (2006).

\bibitem{wmca-99} T. Winiecki, J.F. McCann, and C.S. Adams, Phys. Rev. Lett.
{\bf 82,} 5186 (1999).

\bibitem{ap04} G.E. Astrakharchik and L.P. Pitaevskii, Phys. Rev. A {\bf 70,}
013608 (2004).

\bibitem{egk06} G.A. El, A. Gammal, and A.M. Kamchatnov, Phys. Rev. Lett.
{\bf 97,} 180405 (2006).

\bibitem{kp08} A.M. Kamchatnov and L.P. Pitaevskii,
Phys. Rev. Lett. {\bf 100,} 160402 (2008).

\bibitem{ship1} Yu. G. Gladush, G. A. El, A. Gammal, and A. M. Kamchatnov,
Phys. Rev. A {\bf 75}, 033619 (2007).

\bibitem{ship2} Yu.G. Gladush, L.A. Smirnov, and A.M. Kamchatnov,
J. Phys. B: At. Mol. Opt. Phys. {\bf 41,} 165301 (2008).

\bibitem{ship3D}T.-L. Horng,  S.-C. Gou, T.-C. Lin,  G.A. El, A.P. Itin,
and A.M. Kamchatnov, Phys. Rev. A.
{\bf 79,} 053619 (2009).

\bibitem{sh-08} R. G. Scott and D. A. W. Hutchinson, Phys. Rev. A
{\bf 78,} 063614 (2008).

\bibitem{two-comp} Yu.G. Gladush, A.M. Kamchatnov, Z. Shi, P.G. Kevrekidis,
D.J. Frantzeskakis, B.A. Malomed, Phys. Rev. A {\bf 79,} 033623 (2009).

\bibitem{chip0} R. Folman, P.Kr\"uger, J. Schmiedmayer, J. Denschlag, C. Henkel,
Adv. At. Mol. Opt. Phys. {\bf 48,} 263 (2002).

\bibitem{chip} J. Fort\'agh and C. Zimmermann, Rev. Mod. Phys. {\bf 79,} 235 (2007).



\bibitem{ha08} M.A. Hoefer, M.J. Ablowitz, P.Engels, Phys. Rev. Lett.
{\bf 100},  084504  (2008).

\bibitem{kp-1970} B. B. Kadomtsev and V. I. Petviashvili, Sov. Phys. Doklady,
{\bf 15,} 539 (1970).

\bibitem{zakharov-1975} V. E. Zakharov, JETP Lett, {\bf 22,} 172 (1975).

\bibitem{kuztur} E.A. Kuznetsov and S.K. Turitsyn, Sov. Phys. JETP {\bf 67},
1583 (1988).

\bibitem{anderson} B.P. Anderson, P.C. Haljan, C.A. Regal, D.L. Feder,
L.A. Collins, C.W. Clark, and E.A. Cornell, Phys. Rev. Lett. {\bf 86},
2926 (2001).

\bibitem{jl99} S. Jin, C.D. Levermore, D.W. McLaughlin, Comm. Pure Appl. Math.,
{\bf 52,} 613 (1999).

\bibitem{kamch2000}  A.M. Kamchatnov, {\it  Nonlinear Periodic Waves and
Their Modulations} (World Scientific, Singapore, 2000).


\bibitem{fl86} M.G. Forest and J.E. Lee, Geometry and modulation theory for
periodic nonlinear Schr\"odinger equation, in {\it Oscillation Theory,
Computation, and Methods of Compensated Compactness,} Eds. C. Dafermos et al,
IMA Volumes on Mathematics and its Applications {\bf 2,} (Springer, N.Y., 1987).

\bibitem{pavlov87} M.V. Pavlov, Theor. Math. Phys. {\bf 71}, 351 (1987).

\bibitem{tricomi}
F. Tricomi, F. {\it  Differential equations} (Blackie and Sons, Boston, 1961).

\bibitem{ha07} M.A. Hoefer, M.J. Ablowitz, Physica D {\bf 236}, 44 (2007).

\bibitem{el05} G.A. El, Chaos, {\bf 15,} 037103 (2005).

\bibitem{egk07} G.A. El, Yu.G. Gladush, and A.M. Kamchatnov,
J. Phys. A: Math. Theor. {\bf 40,} 611 (2007).

\bibitem{abramowitz} M. Abramowitz and I.A. Stegun, {\it Handbook of Mathematical
Functions with Formulas, Graphs, and Mathematical Tables}
(New York, Dover Publications, 1972).

\bibitem{gkm89}
A.V. Gurevich, A.L. Krylov and N.G. Mazur, Sov. Phys. JETP,
{\bf 68}, 966 (1989); A.V. Gurevich, A.L. Krylov, N.G. Mazur and G.A. El,
Sov. Phys. Doklady, {\bf 37}, 198 (1992).

\bibitem{gke91} A.V. Gurevich, A.L. Krylov, and G.A. El,
Sov. Phys. JETP {\bf 71,} 899 (1991).

\end{thebibliography}
\end{document}